\documentclass[aps,showpacs,pra,twocolumn]{revtex4-1}

\usepackage{hyperref}
\hypersetup{
    colorlinks=true,
    linkcolor=black,
    filecolor=black,      
    urlcolor=black,
    citecolor=black,
}

\usepackage{exscale}
\usepackage{graphicx}
\usepackage{amsmath}
\usepackage{latexsym}
\usepackage{epstopdf}
\usepackage{amsfonts}
\usepackage{amssymb}
\usepackage{bbm}
\usepackage{times}
\usepackage[T1]{fontenc}
\usepackage{lipsum}
\usepackage{amsthm}
\usepackage{fancyhdr,txfonts,bbm}
\usepackage{appendix}
\usepackage{soul}

\usepackage{epsfig,pstricks}
\usepackage{changes}

\newtheorem{definition}{Definition}

\newcommand{{\Cd}}{{\mathbb{C}^d}}
\newcommand{{\C}}{{\mathbb{C}}}

\newcommand{\sign}{\text{sign}}

\newcommand{\tr}[1]{\mbox{Tr}\left[ #1\right]}
\newcommand{\trE}[1]{\mbox{Tr}_E\left[ #1\right]}

\newcommand{\expp}[1]{e^{#1}}

\begin{document}

\title{Measuring  non-Markovianity  via incoherent mixing with Markovian dynamics}

\author{Dario De Santis$^{1}$ and Vittorio Giovannetti$^{2}$}
\affiliation{$^1$ICFO-Institut de Ciencies Fotoniques, The Barcelona Institute of Science and Technology, 08860 Castelldefels (Barcelona), Spain \\ 
$^2$ NEST, Scuola Normale Superiore and Istituto Nanoscienze-CNR, I-56127 Pisa, Italy\\
}

\date{\today}
\begin{abstract}

We introduce a measure of non-Markovianity based on the minimal amount of extra
 Markovian noise we have to add to the process  via incoherent mixing, in order to 
make the resulting transformation Markovian too at all times. 
 We show how to evaluate  this measure by considering the set of depolarizing evolutions in arbitrary dimension and the set of dephasing evolutions for qubits.
\end{abstract}

\maketitle

\section{Introduction}

In open quantum system dynamics~\cite{OPENBOOK} {Markovian} evolutions 
are  characterized 
by the existence of a one-way flow of information from  the system  to its environment. While approximatively valid in many contexts of physical relevance
(in particular under system-environment  weak-coupling conditions), 
in the vast majority of settings the Markovianity of the dynamical evolution 
is  lost and one witnesses  
 \textit{backflows} of information from the environment to the system~\cite{BREUERREV,RIVASREV,REVMOD,REVPIILO}.
The study of  these non-Markovian effects is a central topic of  quantum information theory
both because they arise almost everywhere, but also because, when  
 properly exploited, they may 
 show advantages in different quantum information processing tasks, such as quantum metrology \cite{added1}, quantum key distribution \cite{added2}, quantum teleportation \cite{added3}, entanglement generation \cite{added4}, quantum communication \cite{added5} and quantum thermodynamics~\cite{T1,T2,T3,ABIUSO}.

The standard procedure to characterize and possibly measure the non-Markovianity of a given evolution is to
target 
functionals that are guaranteed to be 
 monotonic   under arbitrary Markovian evolutions and to
 check for violations of such behaviour.  
Many quantities have been studied in this framework: the distance between pair of states \cite{BLP,LAINE}, channel capacities \cite{added5},  the guessing probability of evolving ensembles of states \cite{BD}, the volume of the accessible states \cite{volume} and correlation measures \cite{NMMI,DDS}. In the present work 
we introduce a conceptually different approach 
to the problem which tries to quantify   non-Markovian
character of a dynamical  evolution
by computing the minimal amount of extra noise  that 
one has to {\it inject} into the system dynamics  in order to stop
the information backflow at all times.
Specifically we consider 
the minimum
value of the probability needed to introduce  Markovianity  for the entire temporal evolution of the system
by incoherently mixing it  with an arbitrary extra  process which is already Markovian. 
Our measure has a clear operational meaning due to the fact 
that  creating stochastic convolutions of processes is a well defined physical procedure.
We remark however that since
neither the set of Markovian evolutions, nor its complementary counterpart, are convex~\cite{assessing}  the explicit evaluation of the proposed measure is typically hard to comply.  At variance with the 
approaches presented in Refs.~\cite{epsilon,epsilonres} which discuss similar ideas 
focusing on
infinitesimal Markovian evolutions~\cite{Ko,Li,Go},
the lack of convexity also prevents us from framing our proposal in the context of a conventional (convex) resource theory of evolutions where Markovian trajectories constitute the resource-free set \cite{oneshot,FGSL}.
After introducing the procedure in the general case of arbitrary 
open quantum evolutions  we focus on 
the special subset of depolarizing transformations of arbitrary dimension and for qubit dephasing channels~\cite{HOLEVO,WILDE,KING} which, thanks to their highly symmetric character, allow for an explicit analytical treatment.
Depolarizing 
channels represent an important error model in quantum information theory. Indeed by pre- and post- processing and classical communication via twirling~\cite{TW}, any other open quantum dynamics can be mapped into a depolarizing channel  whose efficiency in 
protecting the information stored into the system is lower than or equal to the corresponding one of the original process. Accordingly the study of the non-Markovian character of this special set of open quantum evolutions 
is an important task in its own. 

 The manuscript is organized as follows. We start in Sec.~\ref{secII} by defining Markovian and non-Markovian evolutions. In Sec.~\ref{secdep} we introduce the depolarizing evolutions set. In addition, we describe its Markovian and non-Markovian subsets  (Sec.~\ref{secMNMdep}), we discuss some geometrical properties of these subsets (Sec.~\ref{example222}) and we characterize  continuous depolarizing evolutions (Sec.~\ref{secMNMdepcont}).
 In Sec.~\ref{sectionmeasure} we present the measure of non-Markovianity that we study throughout this work and we describe how to apply it to non-Markovian depolarizing evolutions (Sec.~\ref{secmeasDep}). 
We follow in Sec.~\ref{measureofnmc} by evaluating this measure of non-Markovianity for continuous  depolarizing evolutions. 
Sec.~\ref{contnoncont} is dedicated to show that, considering the task of making continuous  depolarizing evolutions Markovian by mixing them with Markovian evolutions, non-continuous Markovian evolutions are less efficient than continuous Markovian evolutions. 
From Sec.~\ref{sec:inter} we start to study non-continuous non-Markovian depolarizing evolutions. In particular, we show that in some particular cases the approaches considered for continuous non-Markovian evolutions are still valid to evaluate the degree of non-Markovianity of these evolutions. 
In Sec.~\ref{SECV} we consider our measure of non-Markovianity applied to generic non-continuous non-Markovian depolarizing evolutions. We start by noticing some features of these evolutions that imply an ambiguity for the identification of the optimal Markovian evolution that makes a generic  non-Markovian depolarizing evolution Markovian (Sec.~\ref{AMB}).  Hence, in  Sec.~\ref{SECVI}, we propose a strategy to calculate our measure of non-Markovianity for any non-continuous depolarizing evolutions.
Finally, in Sec.~\ref{DEPHASING} we extend the analysis to the case of dephasing channels for  qubits. The paper ends in Sec.~\ref{CONCL} with the conclusions. Technical material is presented in the appendices.

\section{Markovian and non-Markovian evolutions}\label{secII}

 Let $\mathcal{S}(\mathcal{H})$ be
 the set of density matrices on a $d$-dimensional Hilbert space $\mathcal{H}$. Any time evolution on $\mathcal{S}(\mathcal{H})$ is defined by a one-parameter family $\Lambda = \{\Lambda_t\}_{t\geq 0}$ of  superoperators called
  \textit{dynamical maps}  $\Lambda_t$. These are completely positive, trace preserving (CPTP) transformations
  which   induce the evolution of a generic initial state $\rho$ at time $t\geq 0$ via  the relation 
  $\rho(t)= \Lambda_t (\rho)$~\cite{HOLEVO,WILDE,WAT}.
The CPTP requirement can  be enforced via the
Stinespring-Kraus representation theorem \cite{stine,kraus}, which allows us to describe
the action of $\Lambda_t$ 
 in terms of a Hamiltonian interaction with an initially uncorelated  external environment $E$ via the expression 
\begin{equation}\label{CPTPmap}
\Lambda_t(\rho) = \trE{U_t \left( \rho\otimes \sigma_E\right) {U_t^\dagger} }\;,
\end{equation}
 with $\sigma_E \in \mathcal{S}(\mathcal{H}_E)$ the initial state of $E$,  $U_t$ a unitary operator on 
 the compound system, and $\trE{\cdot}$ the partial trace over the environment.

In what follows we shall impose
that for $t=0$, $\Lambda_{t}$ should correspond to the identity map, i.e., 
\begin{eqnarray} \label{CONDINI} 
\Lambda_0 = \mbox{id} (\cdot)\;,\end{eqnarray}  and require 
the family $\Lambda$ to be continuous and differentiable almost everywhere,
allowing at most a countable set of discontinuity points.
These assumptions are physically well motivated when considering that 
the partial trace in Eq.~(\ref{CPTPmap}) is a continuous operation and that 
$U_t$  should be 
 the solution of a Schr\"{o}dinger equation, hence
continuous and differentiable in $t$ apart from the presence of abrupt Hamiltonian quenches possibly induced
 by external controls. 
We hence define  $\mathcal{E}\equiv \{ \Lambda\}$ 
to be the set of all the evolutions  on $\mathcal{S}(\mathcal{H})$ that obey the above constraints.
One can easily verify that such set is closed under convex combination meaning that \begin{equation}
p \Lambda+ (1-p) \Omega \in \mathcal{E}\;,\qquad \forall p\in[0,1], \; \forall \Lambda,\Omega\in \mathcal{E}\;.
\end{equation} 

Following~\cite{RIVAS,assessing,HOU,SAB,HALL} 
we now identify 
 Markovian and non-Markovian evolutions  of the system by linking it directly to the
 divisibility condition of the quantum trajectory, i.e., 
\begin{definition}\label{defM}
An evolution $\Lambda =\{\Lambda_t\}_{t\geq 0}\in {\mathcal E}$ is CP-divisible if and only if for any $0\leq s \leq t$ there exists a linear CPTP super-operator $V_{t,s}$ such that 
\begin{equation} \Lambda_t(\cdot)  = (V_{t,s}   \circ \Lambda_s)(\cdot) \equiv V_{t,s}  (\Lambda_s(\cdot))
\;. \label{intermap} \end{equation} We also
 call $V_{t,s}$ the \textit{intermediate map} of  $\Lambda$ between the times $s$ and $t$.
\end{definition} 

Accordingly we identify the Markovian subset $\mathcal{E}^M$ of $\mathcal{E}$  by
the collection of  all CP-divisible evolutions, i.e., 
\begin{equation}
\mathcal{E}^M \equiv \{ \Lambda \in \mathcal{E} \, | \,  \Lambda  \mbox{ is CP-divisible} \} \, ,
\end{equation}
and  define the complement to $\mathcal{E}$ of $\mathcal{E}^M$ as the 
set of non-Markovian evolutions of the system, i.e.
\begin{equation}
\mathcal{E}^{NM} \equiv \mathcal{E} \setminus \mathcal{E}^M \, .
\end{equation}
As already mentioned in the introduction neither  $\mathcal{E}^M$ nor  $\mathcal{E}^{NM}$
are closed under  convex convolutions~\cite{assessing}. 

\section{Depolarizing evolutions}\label{secdep}

Depolarizing evolutions $\mathcal{D}$ form a closed convex subset of~$\mathcal{E}$~\cite{HOLEVO,WILDE,KING}.  
An evolution $D=\{D_t\}_t$ belongs to $\mathcal{D}$ if and only if at any time $t\geq 0$ the corresponding dynamical map $D_t$ can be written as a linear combination of the identity transformation $\mbox{id} (\cdot)$  and 
the map that sends every inputs into the completely mixed state. Specifically we have 
\begin{equation}\label{D}
D_t (\cdot) = f(t) \, \mbox{id} (\cdot) + (1-f(t)) \tr{ \cdot }\frac{\mathbbm{1}}{d} \, ,
\end{equation}
with $\mathbbm{1}$  the identity operator on $\mathcal{H}$
and $f(t)$ a real quantity  belonging to the interval 
\begin{equation}\label{intervalI}
 I_{\mathcal{D}}\equiv\left[-\frac{1}{d^2-1}, 1 \right] \, ,
 \end{equation}
 this last property being necessary and sufficient to ensure $D_t$ to be CPTP~\cite{KING}.
 From Eq.~(\ref{D}) it is clear that we can use the function $f(t)$ to uniquely characterize the elements of 
 $\mathcal{D}$. In order to comply with the structural requirements we imposed on
 $\mathcal E$  in the previous section, 
we focus on the collection of functions $f(t): \mathbbm{R}^+ \rightarrow I_{\mathcal{D}}$ that  \begin{enumerate}
\item are continuous for almost-all $t$; 
\item admit  right and left time derivatives  ($\dot f(t^\pm)  \equiv \lim_{\epsilon \rightarrow 0^\pm} \frac{f(t+\epsilon)-f(t)}{\epsilon} $);
\item satisfy $f(0)=1$;
\end{enumerate} 
the last property being introduced  to enforce  Eq.~(\ref{CONDINI}). 
We define $\mathfrak{F}$ to be the set of \textit{characteristic functions}  $f(t)$ that satisfy the above conditions and use {Eq.}~(\ref{D})
to
establishing  a one-to-one relation between such set and $\mathcal{D}$. We also
introduce the special subset of 
continuous depolarizing evolutions $\mathcal{D}_C$ as the collection of depolaring evolutions~(\ref{D}) whose  $f(t)$  belong to the subset 
$\mathfrak{F}_{C} \subset \mathfrak{F}$ formed by continuous characteristic functions.

To fix the notation, if $\{t_i\}_i$ is the discrete collection of times when $f(t)$ is discontinuous, we have that $f(t_i^+)\equiv \lim_{\epsilon \rightarrow 0^+} f(t_i+\epsilon)$ is different from $f(t_i^-)\equiv \lim_{\epsilon \rightarrow 0^+} f(t_i-\epsilon)$. To describe  the discontinuous behavior of $f(t)$ we hence introduce the quantity 
\begin{equation}\label{xi}
\xi(f(t))\equiv  \frac{f(t^+)}{f(t^-)} \, ,
\end{equation}
which assumes values in $[-\infty,+\infty]$, where we fix $\xi(f(t))= \pm \infty$ when $\sign(f(t^+))=\pm 1$ and $f(t^-)=0$.  Moreover, when $f(t^+)=f(t^-)=0$ we define $\xi(f(t))=1$. From Eq.~(\ref{xi}) it follows that $f(t)$ is continuous at time $t$ if $\xi(f(t))=1$ and  that $f(t)\in \mathfrak{F}_{C}$ if and only if $\xi(f(t))=1$ for any $t\geq 0$.
On the contrary from Eq.~(\ref{xi})  it also follows that a discontinuity distances $f(t)$ from zero preserving its sign if $\xi(f(t))>1$, it makes $f(t)$ change its sign if $\xi(f(t))<0$, and finally that $\xi(f(t))=0$ if and only if $f(t^+)=0$ and $f(t^-)\neq 0$.

\subsection{Markovian and non-Markovian depolarizing evolutions}\label{secMNMdep}

In view of the one-to-one correspondence between $\mathcal{D}$ and $\mathfrak{F}$,
we define the Markovian  and non-Markovian depolarizing 
subsets $\mathcal{D}^M\equiv \mathcal{D}\cap \mathcal{E}^M$ and 
$\mathcal{D}^{NM}\equiv \mathcal{D}\cap \mathcal{E}^{NM}=\mathcal{D}\setminus\mathcal{D}^{M}$
by assigning the corresponding sets of the associated characteristic functions $\mathfrak{F}^M$ and 
$\mathfrak{F}^{NM}$.

\begin{figure}
\includegraphics[width=0.49\textwidth]{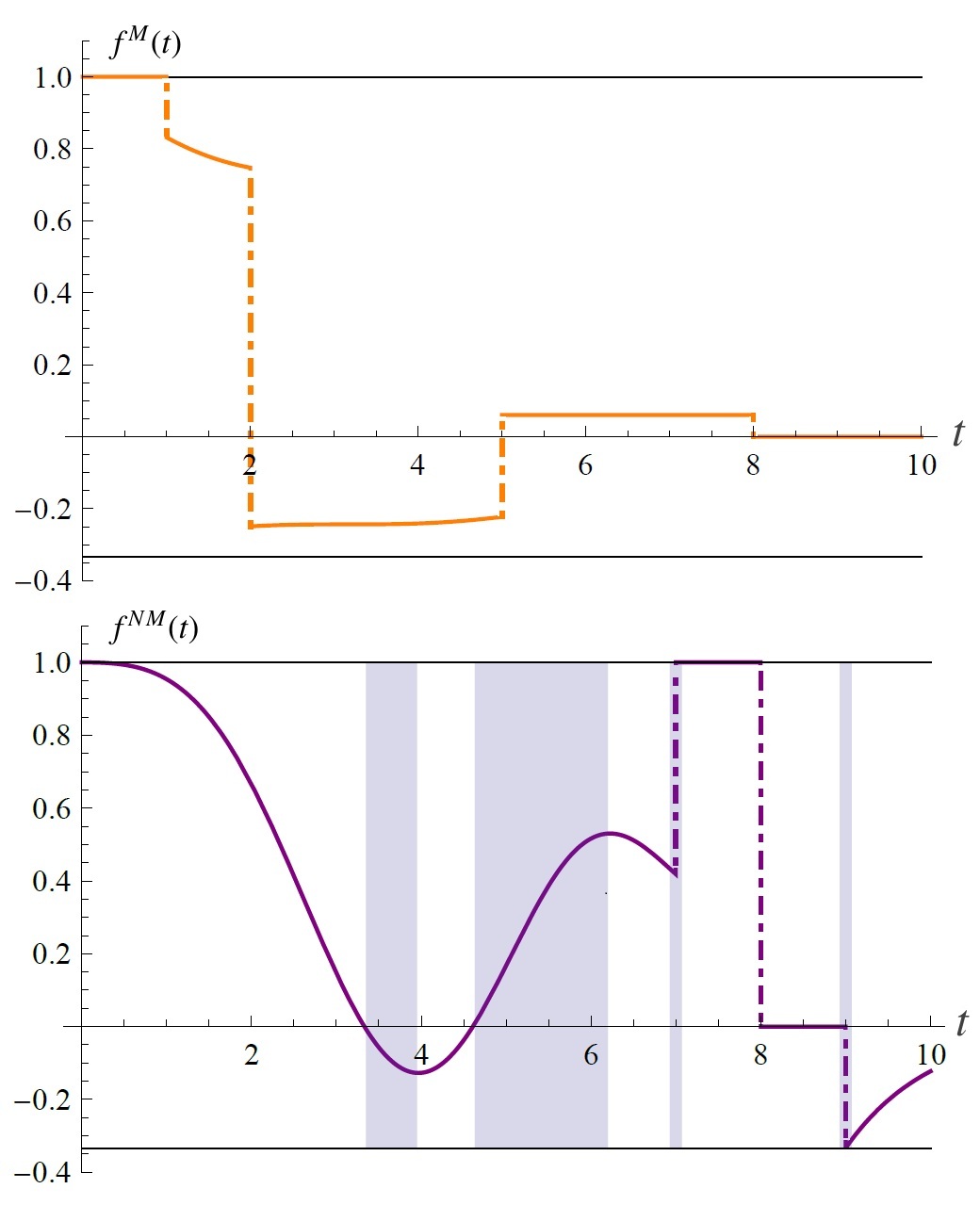}
\caption{Example of a non-continuous Markovian characteristic function (above) $f^M(t)\in \mathfrak{F}^M$ and a non-continuous non-Markovian characteristic function (below) $f^{NM}(t)\in \mathfrak{F}^{NM} $ for $d=2$.  Given Eq.~(\ref{intervalI}),  any characteristic function has to assume values in $I_\mathcal{D}=[-1/3,1]$. Discontinuities are underlined by dotted dashed lines. $f^M(t)$, when continuous, satisfies $\mathbf{CM_1}(\tau)$, i.e., it does not increase its distance from zero.  When $f^M(t)$ is not continuous it satisfies $\mathbf{CM_2}(\tau)$: for the times $\tau=1,2,5$ and $8$, we have $\xi(f^{M}(1))=0.83$, $\xi(f^{M}(2))=-0.33$, $\xi(f^{M}(5))=-0.27$ and $\xi(f^{M}(8))=0$. Since $f^M(8^+)=0$, $f^M(t)$ has to be equal to 0 for any $t>8$. The times when $f^{NM}_C(t)$   has a non-Markovian behavior are colored in purple. This characteristic function shows both time intervals and times of discontinuity when, respectively, $\mathbf{CM_1}(\tau)$ and $\mathbf{CM_2}(\tau)$ are violated. Indeed, for $\tau=7$ and 9 we have non-Markovian discontinuities $\xi(f^{NM}(7))=2.39$ and $\xi(f^{NM}(t))=-\infty$, while at $\tau=8$ we have $\xi(f^{NM}(8))=0$, i.e., a  Markovian discontinuity. The temporal parameter $t$ in the plots is expressed in arbitrary unit.}\label{fncj}
\end{figure}

{We} start by observing that if the characteristic function  of an element $D$ of  $\mathcal{D}$
assumes zero value at $s$ (namely $f(s)=0$) then $D_s$ becomes the complete depolarizing channel $\tr{ \cdot }\frac{\mathbbm{1}}{d}$,  loosing  memory 
of the input state of the system. Accordingly the only possibility we have to 
fulfil  the constraint~(\ref{intermap}) needed for Markovianity 
is that $D_t$ correspond to $\tr{ \cdot }\frac{\mathbbm{1}}{d}$ too, 
i.e., 
\begin{eqnarray} \label{effe0} 
f(s) =0 \quad  \Longrightarrow \quad f(t)=0 \;, \qquad \forall t\geq s\;.
\end{eqnarray} 
On the contrary
 if $f(s)\neq 0$,  Eq.~(\ref{intermap}) can be enforced by observing that the intermediate map $V_{t,s}$ assumes the  same form of Eq.~(\ref{D}), i.e., 
\begin{equation}\label{V}
V_{t,s} (\cdot) =
 \frac{f(t)}{f(s)} \, \mbox{id} (\cdot) + \left(1-\frac{f(t)}{f(s)} \right) \tr{ \cdot }\frac{\mathbbm{1}}{d} \; ,
\end{equation}
which  is CPTP if and only if 
\begin{eqnarray} \frac{f(t)}{f(s)}  \in I_{\mathcal{D}} \;, \label{COND444}  \end{eqnarray}  with $I_{\mathcal{D}}$ the interval defined  in   Eq.~(\ref{intervalI}). This includes also the case (\ref{effe0}) by noticing that only with $f(t)=0$ we prevent   ${f(t)}/{f(s)}$ from diverging 
 when $f(s)=0$.
As shown in  Appendix~\ref{DERIAPP}, Eq.~(\ref{COND444}) can be conveniently casted in the following inequality that in
some case is easier to handle, i.e., 
\begin{equation}\label{COND4441}
 C(t,s)\equiv  \left| 2(d^2-1)  f(t) -  ({d^2-2}) f(s) \right| - d^2 |f(s)|\leq 0\,.
\end{equation} 
From {Definition}~\ref{defM} we have hence that $D \in \mathcal{D}^M$ if and only if 
its characteristic function $f(t)$ is such that (\ref{COND444}) (or equivalently (\ref{COND4441}))
holds true  for any $t \geq s\geq 0$, i.e., 
\begin{equation}\label{MarkDsetnew}
\mathfrak{F}^M \equiv
 \left\{  f(t) \in \mathfrak{F} \, | \,  C(t,s)\leq 0\;, \;   \forall t \geq s\geq 0 
 \right\} \;.
\end{equation}
 Considering the property (\ref{effe0})  and that for $f(t) \in \mathfrak{F}$ we must have $f(0)=1$, it is easy to verify that
 all continuous elements
 of $\mathfrak{F}^M$ are  non-negative and non-increasing (more on this in Sec.~\ref{secMNMdepcont}).
 Markovian characteristic functions can however change their sign through discontinuities.
  Indeed according to (\ref{COND444}) a
   non continuous element   
$f(t)$ of  $\mathfrak{F}^M$
can jump either to a value $f(t^+)$ with the same sign and $|f(t^+)|< |f(t^-)|$, namely $\xi(f(t))\in[0,1)$, or to a value with opposite sign and $|f(t^+)|\leq|f(t^-)|/(d^2-1)$, namely $\xi(f(t))\in[-1/(d^2-1),0]$. These facts can be formalized by saying that a generic
$f(t) \in \mathfrak{F}$ exhibits a {\it Markovian behaviour} at time $\tau
\geq 0$ if one of the two conditions applies
\begin{eqnarray} 
\begin{array}{lclr} 
{\mathbf{CM_1}(\tau):  } && { \mbox{$\xi(f(\tau))=1$ and  $\frac{d}{d\tau}|f(\tau)| \leq 0$}; }  \\
\mathbf{CM_2}(\tau):
&& \mbox{$\xi(f(\tau)) \in I_{\mathcal{D}} \setminus 1$} ; 
\end{array} \label{MarkD}
\end{eqnarray} 
{where $\mathbf{CM_1}(\tau)$ has to be  replaced by $\dot{f}(\tau^{\pm}) f(\tau) \leq 0$ when $\dot{f}(\tau)$ is non-continuous, i.e., $\dot f(\tau^-)\neq \dot f(\tau^+)$.}
Notice  that the conditions given in Eq.~(\ref{MarkD}) do not explicitly exclude the cases for which $\dot f(t)\neq 0$ and $f(t)=0$. Nonetheless, the properties of $\mathfrak{F}$ would imply that $\exists \delta >0$ such that $\dot f(t+\delta) f(t+\delta) >0$, which would exclude $f(t)$ from $\mathfrak{F}^M$.
It is worth stressing that imposing (\ref{MarkD}) for all $\tau\geq 0$ is equivalent to enforce (\ref{COND444}) (or (\ref{COND4441}))
for all couples $0\leq s \leq t$. Hence, 
 Eq.~(\ref{MarkDsetnew}) can be casted in the form 
 \begin{equation}\label{MarkDset}
\mathfrak{F}^M = \left\{  f(t) \in \mathfrak{F} \, | \, 
\mathbf{CM_1}(\tau)\; \mbox{or}\; \mathbf{CM_2}(\tau) = {\mbox{TRUE}, \forall\tau\geq 0 }
 \right\} \;,
\end{equation}
which  involves only local properties of  $f(t)$. 
 By construction any $f(t)\in\mathfrak{F}$ that fails to fulfil both the constraints of Eq.~(\ref{MarkD}) at least for  one $\tau$,
 or the inequality (\ref{COND4441}) for some couple $s$ and $t$, 
  defines an element of the non-Markovian characteristic function set $\mathfrak{F}^{NM}\equiv \mathfrak{F}\setminus\mathfrak{F}^M$ which describes  the non-Markovian depolarizing evolutions  $\mathcal{D}^{NM}$.
  At variance with the elements of $\mathfrak{F}^M$ 
a  characteristic function $f(t)$ which is non-Markovian 
 can show any increasing or decreasing continuous  behaviour and discontinuities with  $\xi(f(t))\in [-\infty,+ \infty ]$. In Fig. \ref{fncj} we show the typical behavior of characteristic functions in $\mathfrak{F}^{M}$ and $\mathfrak{F}^{NM}$.

We notice that any element of $\mathfrak{F}^{NM}$
 can still  obey  the constraints (\ref{MarkD}) on some 
part of the real axis. 
 In particular we say that  $f(t)\in\mathfrak{F}^{NM}$ has a Markovian behaviour 
  in $(t_1,t_2)$ if the function satisfies at least one of the conditions of Eq.~(\ref{MarkD}) for any $\tau\in (t_1,t_2)$. Finally, we say that $\tau$ is a time when $f(t)\in\mathfrak{F}$ shows a \textit{Markovian discontinuity}  if $\xi(f(\tau))\in { I_\mathcal{D}  \setminus 1}$. Instead, if $\xi(f(\tau))\notin I_\mathcal{D}$, we say that $\tau$ is a time when $f(t)$ shows a \textit{non-Markovian discontinuity}.

\subsection{Border and geometry of the Markovian depolarizing set}\label{example222}

It is possible to show that the following properties hold:
\begin{itemize}
\item $\mathcal{D}$ is convex,
\item $\mathcal{D}^M$ is closed, non-convex, and $\mbox{\bf{border}}(\mathcal{D}^M)= \mathcal{D}^M$,
\item $\mathcal{D}^{NM}$ is open, non-convex, and dense. 
\end{itemize} 
The non convexity of $\mathcal{D}^{M}$  and $\mathcal{D}^{NM}$  (and hence $\mathfrak{F}^M$ and $\mathfrak{F}^{NM}$)  can be easily proven  by presenting some explicit counter-examples (see Appendix~\ref{NONCONV}).
 To show instead that
$\mathcal{D}^M$ coincides with its border we can proceed as follows: 
given a generic Markovian depolarizing evolution $D^M\in \mathcal{D}^M$, consider
a time $s>0$ where the associated 
characteristic function $f^M(t)$  is continuous, namely $\xi(f^{M}(s))=1$ (of course such $s$ can alway be found since the  set of discontinuity points for 
a generic element of  $\mathfrak{F}$ is at most countable).
Take then a non-Markovian depolarizing evolution $D^{NM}\in \mathcal{D}^{NM}$ with 
characterstic function $f^{NM}(t)$ which instead has  $\xi(f^{NM}(s))>1$ and  $\sign(f^{NM}(s^-))=\sign(f^{M}(s^+))$  (such an element can always be identified). 
It is then straightforward to verify  that the whole family of elements of $\mathcal{D}$ defined as
{ $D^{(p)} = (1-p) D^{NM} + p D^{M}$ }
 for  {$p\in [ 0,1) $}  is non-Markovian: indeed for all such values, at $t=s$ the characteristic function
 \begin{eqnarray}f^{(p)}(t)=(1-p) f^{NM}(t) + p f^M(t)  \;,  \label{DEFFPPP} \end{eqnarray} 
 of $D^{(p)}$ has a non-Markovian discontinuity ($\xi(f^{(p)}(s))>1$). 
Notice also that as {$p\rightarrow 1$}, $D^{(p)}$ gets arbitrarily close to $D^M$ in
any conceivable norm one can introduce on $\mathcal{E}$ or $\mathcal{D}$
(indeed $\| D^{(p)} - D^M\|={(1- p)} \| D^{NM}- D^M\|$). 
The above argument shows that any neighbour of a Markovian depolarizing trajectory contains non-Markovian processes, i.e., that $\mathcal{D}^{M}$ is a  set of measure zero, or equivalently, that  almost-all depolarizing evolutions are non-Markovian. 
On the contrary, for any non-Markovian depolarizing evolution $D^{NM}$ one can show that there exists no Markovian $D^M$ such that the convex combination {$D^{(p)}= (1-p) D^{NM} + p D^{M}$} is Markovian for any {$p\in(0,1]$}.
More precisely it is possible to
identify a  probability value
{$p^*(D^{NM})\in (0,1]$}  such that,  irrespectively from the choice of $D^M$, we have 
 \begin{eqnarray} 
 D^{(p)} \in \mathcal{D}^{NM} \qquad \forall {p < p^*(D^{NM})}\;. \label{IMPORTANTnew} 
\end{eqnarray} 
Indeed, since $D^{NM}$ is explicitly non-Markovian, there must exist $t\geq s\geq 0$ such that 
its the characteristic function violate the constraint (\ref{COND4441}) which we rewrite here as
 \begin{equation} \label{ASSUMPTION} 
 A^{NM}(t,s) \equiv \left| 2(d^2-1)  f^{NM}(t) -  ({d^2-2}) f^{NM}(s) \right| > d^2  |f^{NM}
 (s)|\;.
 \end{equation} 
On the contrary, if $D^{(p)}$ is Markovian, its characteristic function  must fulfil (\ref{COND4441}), i.e.
 \begin{eqnarray}  \label{PPP1} 
 \left| 2(d^2-1)  f^{(p)}(t) -  ({d^2-2}) f^{(p)}(s) \right| \leq  d^2 |f^{(p)}
 (s)|\;. 
 \end{eqnarray} 
Using~(\ref{DEFFPPP}) we notice however that the left-hand-side of the above expression
can be lower bounded as follows 
\begin{eqnarray} 
&& \left| 2(d^2-1)  f^{(p)}(t) -  ({d^2-2}) f^{(p)}(s) \right|   \nonumber  \\ 
&&\qquad   \geq {(1-p)} A^{NM}(t,s) - 
{p} \left| 2(d^2-1)  f^{M}(t) -  ({d^2-2}) f^{M}(s) \right| \nonumber \\
 &&\qquad  \geq {(1-p)} A^{NM}(t,s) - 
{p} (3 d^2 -4) \;,
  \end{eqnarray} 
  where in the last inequality we exploit the fact that all characteristic functions must have modulus smaller or equal to $1$.
  Similarly the right-hand-side of (\ref{PPP1}) can be upper bounded as  
   \begin{equation} 
    \left| f^{(p)}(s) \right|  \leq   {(1-p)}\left| f^{NM}(s) \right|  + {p} \left| f^{M}(s) \right| \leq  {(1-p)} \left| f^{NM}(s) \right|  + {p} \;. 
  \end{equation} 
  Hence a necessary condition for  (\ref{PPP1}) is to have 
 \begin{equation}\label{IMPORTANTnew1} 
{4 p (d^2  -1 ) \geq   (1-p) C^{NM}(t,s)  }\;, \end{equation} 
where $   C^{NM}(t,s)\equiv A^{NM}(t,s) - d^2 |f^{NM}(s)| $. Due to the strict positivity of the rightmost term of Eq.~(\ref{IMPORTANTnew1})  (see~(\ref{ASSUMPTION})), it
 cannot be fulfilled for all {$p\in (0,1]$}. 
 Equation~(\ref{IMPORTANTnew})
  finally follows from~(\ref{IMPORTANTnew1}) e.g. by setting
  \begin{eqnarray}\label{pstar}
p^*(D^{NM}) = {\frac{C^{NM}(t,s)}{C^{NM}(t,s) + 4(d^2-1) } }\;. 
  \end{eqnarray} 
It is easy to show that this value of $p^*(D^{NM}) $  belongs to {$(0,1]$} if and only if $C^{NM}(t,s)$ violates Eq.~(\ref{COND4441}).

\subsection{Markovian and non-Markovian continuous depolarizing evolutions}\label{secMNMdepcont}

\begin{figure}
\includegraphics[width=0.49\textwidth]{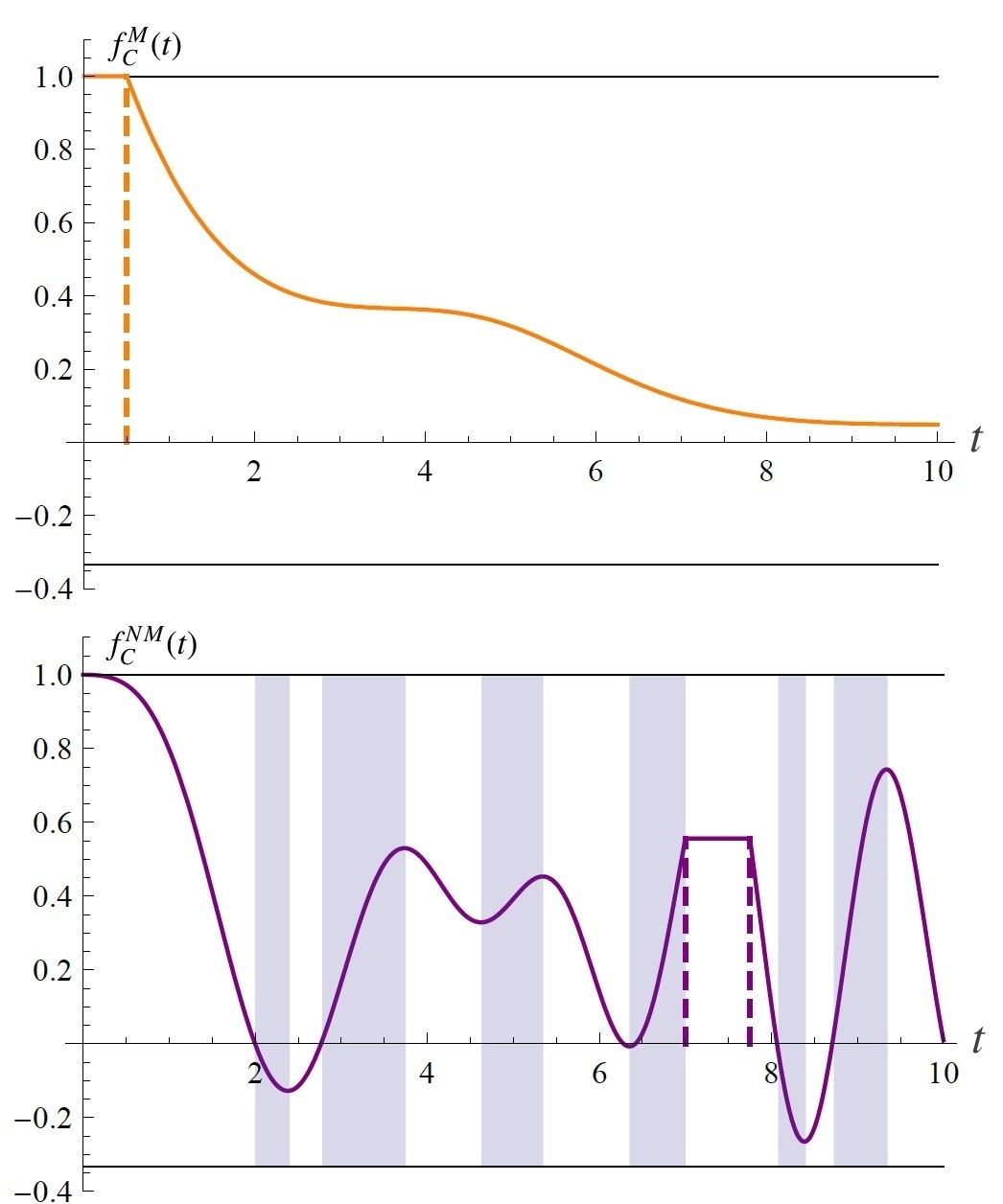}
\caption{Example of a continuous Markovian characteristic function (above) $f^M_C(t)\in \mathfrak{F}^M_C$ and a continuous non-Markovian characteristic function (below) $f^{NM}_C(t) \in \mathfrak{F}^{NM}_C$  for $d=2$.   Given Eq.~(\ref{intervalI}), any characteristic function has to assume values in $I_\mathcal{D}=[-1/3,1]$. $f^M_C(t)$ is non-increasing and assumes values in $[0,1]$. $f^{NM}_C(t)$ assumes values in $I_\mathcal{D}=[-1/3,1]$ (horizontal lines) and violates the Markovian condition $\mathbf{CM_1}(\tau)$ in the time intervals colored in purple, i.e., when it increases its distance from zero. Dahsed lines underline the times when the respective time derivatives are non-continuous. The temporal parameter $t$ in the plots is expressed in arbitrary unit.}\label{fcj}
\end{figure}

 Important subsets of $\mathcal{D}^M$ and $\mathcal{D}^{NM}$ are obtained
 by considering their intersections  with the continuous subset $\mathcal{D}_C$ of $\mathcal{D}$, i.e., 
 \begin{eqnarray}
\mathcal{D}_C^M
\equiv  \mathcal{D}_C \cap \mathcal{D}^M \;, \qquad \mathcal{D}_C^{NM}
\equiv  \mathcal{D}_C \cap \mathcal{D}^{NM}\;. 
 \end{eqnarray}  
By construction $\mathcal{D}_C^M$ and $\mathcal{D}_C^{NM}$ are composed by depolarizing process 
whose associated characteristic functions 
$f(t)$ belong respectively to the intersections  $\mathfrak{F}^{M}_{C}
\equiv  \mathfrak{F}_{C} \cap \mathfrak{F}^M$ and $\mathfrak{F}^{NM}_{C}
\equiv \mathfrak{F}_{C} \cap  \mathfrak{F}^{MN}$. 
From Eq.~(\ref{MarkD}) we  deduce that the elements of $\mathfrak{F}_C^M$ are 
 {\it monotonically non increasing}, continuous functions $f^M_C(t)\in[0,1]$.
In particular, since any convex combination of two continuous functions
{in $\mathfrak{F}^M_C$ belongs to $\mathfrak{F}^M_C$, we have }
\begin{itemize}
\item $\mathcal{D}_C$ is convex,
\item $\mathcal{D}^M_C$ is closed and convex,
\item $\mathcal{D}^{NM}_C$ is open and non-convex. 
\end{itemize}
Furthermore, if  $f_C^{M}(t')=0$ for  some time $t'$, the time derivative of $f_C^{M}(t)$ cannot be different from zero for any $t> t'$ without violating the first condition of Eq.~(\ref{MarkD}).  
Instead the elements of $\mathfrak{F}_C^{NM}$ are  continuous functions $f^{NM}_C(t) $ that can assume any value in $I_\mathcal{D}$ such that $f_C^{NM}(0)=1$.  In Fig. \ref{fcj} we show the typical behavior of continuous characteristic functions in $\mathfrak{F}^{M}_C$ and $\mathfrak{F}^{NM}_C$.

In Appendix~\ref{deppos} we introduce another convex subset of $\mathcal{D}$ given by the positive depolarizing evolutions, namely defined by, in general non-continuous, positive characteristic functions. The Markovian subset of these evolutions is convex and, as we show, it contains the set of continuous Markovian evolutions.

\section{A measure of non-Markovianity by noise addition}\label{sectionmeasure}

In this section we introduce our 
 measure of non-Markovianity. Given  $\Lambda \in \mathcal{E}$ the quantum process we are
 interested in, consider the quantum trajectories 
$\Lambda^{(p)} \in \mathcal{E}$ defined by the convex sums
\begin{equation}\label{Lambdap}
\Lambda^{(p)}={(1-p)  \Lambda + p \Lambda^{M} }  , \qquad  p\in[0,1] \, ,
\end{equation}
one get by incoherently mixing 
the original evolution with an element $\Lambda^{M}$ of the Markovian subset $\mathcal{E}^M$
with  time-independent weights {$1-p$ and $p$}. It is worth stressing that the dynamical evolution 
~(\ref{Lambdap}) can be physically implemented, at least in principle, by a simple
random event taking place at time $t=0$ which decides wether to transform the state of the system
under the action of $\Lambda$ or under the action of $\Lambda^M$.  
{We introduce a measure of non-Markovianity $p(\Lambda)$ by considering} the  {smallest} $p$ that enables us to make $\Lambda^{(p)}$ Markovian 
for some~$\Lambda^{M}$, i.e.
\begin{equation}\label{DEFPNM} 
p(\Lambda) \equiv {\min_p} \{ p \, | \,  \exists \Lambda^M \in \mathcal{E}^M \mbox{ s.t. }  \Lambda^{(p)} \in \mathcal{E}^M \} \, ,
\end{equation} 
 and call \textit{optimal} 
a Markovian evolution $\Lambda^M$ that allows us to attain such value. In other contexts, e.g. resource theories \cite{RES1,RES2}, the measure of non-Markovianity $p(\Lambda)$ is ofter referred to as a \textit{robustness} measure. 
 { $p(\Lambda)$} is always well defined since the set of
$p$  entering the optimization contains at least the  {point~$1$}.
The rational of this choice is that, the {greater} is $p$, the {stronger} is the perturbation we add into 
the system by the mixing operation~(\ref{Lambdap}): indeed, for fixed $\Lambda^M$,
the distance between $\Lambda^{(p)}$ and the original trajectory $\Lambda$ is always
proportional to {$p$}. For instance,  at any given time  $t$ we can write 
$\| \Lambda^{(p)}_t- \Lambda_t \| = {p} \| \Lambda_t^{M} - \Lambda_t \|$
where $\| \cdot\|$ stands for (say)  the diamond norm for super-operators~\cite{DIAMOND}.
{ As a consequence, $p(\Lambda)$ is the minimum perturbation one needs to introduce via the mixing procedure (\ref{Lambdap}) to enforce Markovianity  into the system evolution.} 
The maximum value of this quantity has a precise meaning: {$p(\Lambda)=1$}
 implies that $\Lambda$  cannot be made Markovian by any non-trivial mixture (\ref{Lambdap}). On the contrary, since  {$p(\Lambda)=0$} if and only if $\Lambda\in \mathcal{E}^M$, it is clear that {(\ref{DEFPNM})} is a faithful measure of non-Markovianity. 
 
{We can} consider the case where in Eq.~(\ref{Lambdap}) 
 $\Lambda^{(p)}$ is asked to belong to a specific Markovian target subset $\mathcal{T}^M$ of $\mathcal{E}^M$, while at same time $\Lambda^{M}$ belongs to a particular set $\mathcal{A}^M$ of $\mathcal{T}^M$
(namely $\mathcal{A}^{M} \subseteq \mathcal{T}^M \subseteq \mathcal{E}^M$). This leads to the functional
\begin{eqnarray}\label{pRT}
\!\!\!\!\!\! p(\Lambda | \, \mathcal{A}^{M}, \mathcal{T}^{M}) &\equiv& {\min_p} \{ p \,| \, \exists \Lambda^M \in \mathcal{A}^M \mbox{ s.t. } \Lambda^{(p)} \in \mathcal{T}^M \}\, ,
\end{eqnarray}
 which by construction provides a bound for~(\ref{DEFPNM}) 
 \begin{eqnarray} \label{GAP} 
 p(\Lambda | \, \mathcal{A}^{M}, \mathcal{T}^{M})  \geq
 p(\Lambda | \, \mathcal{A}^{M}, \mathcal{E}^{M}) \geq  
  p(\Lambda)\;, 
  \end{eqnarray} 
{A typical situation where $p(\Lambda | \, \mathcal{A}^{M}, \mathcal{T}^{M})$ can be considered is given when} $\mathcal{A}^M$ {represents} the accessible Markovian evolutions that we are able to reproduce in our laboratory and mix with $\Lambda$, while  $\mathcal{T}^M$ {represents} a particular subset of $\mathcal{E}^M$ for which Markovianity is easy to certify, or which possesses some additional features that we demand. 
From this perspective Eq.~{(\ref{GAP})}, besides  being an upper bound for Eq.~{(\ref{DEFPNM})} can also be seen as a different approach to quantify the degree of non-Markovianity of  the process $\Lambda$. 
A case of special interest  is provided 
by the scenario where the subsets $\mathcal{A}^{M}$ and $\mathcal{T}^{M}$ entering~(\ref{pRT}) coincide and correspond to the Markovian part of  a convex subset of the system  evolutions $\mathcal{B}\subset \mathcal{E}$, i.e.
 $\mathcal{A}^{M}=\mathcal{T}^{M}=\mathcal{B}^{M}\equiv \mathcal{B} \cap 
 \mathcal{E}^M$.
Under these conditions from~(\ref{Lambdap}) it follows that we can write 
\begin{equation}\label{psimple}
p(\Lambda | \mathcal{B}^M) \equiv p(\Lambda | \mathcal{B}^M,\mathcal{B}^M) =
p(\Lambda | \mathcal{B}^M,\mathcal{E}^M)\;,  \qquad  \forall \Lambda \in \mathcal{B}\;, 
\end{equation}
showing that for the elements of $\mathcal{B}$, at least  the first
 of the inequalities in (\ref{GAP}) closes
(of course this does not necessarily hold if $\mathcal{B}$ is not 
convex, as in this case  there could be maps $\Lambda^{(p)}$ in 
$\mathcal{E}^M$ which are not necessarily in $\mathcal{B}^M$).
Furthermore, while we have no explicit evidence in support of this claim, if  $\mathcal{B}$ is a sufficiently "structured" set as in the case of the depolarizing evolutions addressed in the following
subsection, it is also tempting to 
conjecture that  the second gap  in (\ref{GAP}) should collapse too, implying that in this case
$p(\Lambda | \mathcal{B})$ should coincide with $p(\Lambda)$ for all $
\Lambda \in \mathcal{B}$, or equivalently that 
\begin{equation}\label{conj} 
\mbox{\small\bf{(CONJECT.)}}\quad 
 { p(\Lambda) = p(\Lambda | \mathcal{B}^M)   \;,  \qquad  \forall \Lambda \in \mathcal{B} }\;.  
\end{equation} 

\subsection{Measuring  the non-Markovianity of depolarizing evolutions}\label{secmeasDep}

To study the non-Markovian behaviour of   depolarizing evolutions  $D\in {\mathcal D}$ we shall focus on the case where the set  $\mathcal{B}$ entering in Eq.~(\ref{psimple})  corresponds to
$\mathcal{D}$ itself, i.e., the quantity $p(D|\mathcal{D}^M)$.
While for elements  of the Markovian subset $p(D|\mathcal{D}^M)$ is { clearly equal to  $0$}, 
in the case $D^{NM}\in {\mathcal D}^{NM}$ we can invoke (\ref{IMPORTANTnew}) to claim the following
{lower} bound 
\begin{eqnarray} 
p(D^{NM}|\mathcal{D}^M) {\geq } p^*(D^{NM}) \;,
\end{eqnarray} 
 which is non trivial due to the fact that $p^*(D^{NM})$ is strictly {larger than $0$}. 
Since $\mathcal{D}_C$ is a proper subset of $\mathcal{D}$, it is  also clear that in general the following ordering holds
 \begin{eqnarray}\label{INEQ1} 
 p(D|\mathcal{D}_C^M,\mathcal{D}^M) {\geq} p(D|\mathcal{D}^M)
 \;, \qquad \forall D\in \mathcal{D}\;.
 \end{eqnarray} 
In particular
  if  the channel we test is an element of the continuous subset 
of $\mathcal{D}$,  the inequality in Eq.~(\ref{INEQ1}) closes, leading to 
\begin{eqnarray} 
p(D_C|\mathcal{D}_C^M)=p(D_C| \mathcal{D}^M)\;, \qquad \forall D_C\in \mathcal{D}_C\; .\label{IMPORTANT}  
\end{eqnarray} 
Notice that we used the fact that, 
due to the 
 convexity of  $\mathcal{D}_C$, one has that  
 $p(D_C|\mathcal{D}_C^M,\mathcal{D}^M)$ corresponds to 
 $p(D_C|\mathcal{D}_C^M)\equiv p(D_C|\mathcal{D}_C^M,\mathcal{D}_C^M)$ 
 when evaluated on $D_C\in  \mathcal{D}_C$). 
 The proof of {Eq.~(\ref{IMPORTANT})} is rather cumbersome and we posticipate it to
 Sec.~\ref{contnoncont}, focusing first on the explicit   computation of {$p(D_C|\mathcal{D}_C^M)$}, 
 which we present in Sec.~\ref{measureofnmc}.

\section{Measure of non-Markovianity for continuous depolarizing evolutions}\label{measureofnmc}

In this section we  evaluate our measure of non-Markovianity  
\begin{eqnarray} 
 p(D_C|\mathcal{D}_C^M)\;, \label{NMC} 
\end{eqnarray} 
for the cases where $D_C$ is an arbitrary element of the continuous subset $\mathcal{D}_C$  of the depolarizing 
evolutions, under the assumption that also the transformations $D^M$ of (\ref{depp}) 
are elements of $\mathcal{D}_C$. 
Before entering into the details of the analysis it is worth clarifying  that in computing $p(D_C| \mathcal{D}_C^M)$ the map
 $\Lambda^{(p)}$ of Eq.~(\ref{Lambdap})  has the form
\begin{equation}\label{depp}
D_C^{(p)}={(1-p) D_C + p } D_C^M , 
\end{equation}
where $D_C^M \in \mathcal{D}^{M}_C$ and  $D_C \in \mathcal{D}_C$.
Thus, since $\mathcal{D}_C$ is convex, for any $p$, $D_C$ and $D_C^M$, we have that $D_C^{(p)} \in \mathcal{D}_C$ 
with characteristic function $f_C^{(p)}(t)\in \mathfrak{F}_C$ given by the convex sum of the characteristic functions $f_C(t)$ and $f_C^M(t)$ associated with $D_C$ and $D_C^M$ respectively, i.e.
\begin{equation}\label{fp}
f_C^{(p)}(t)={(1-p)} f_C(t) + {p} f_C^{M}(t) \, .
\end{equation} 
In order to evaluate $p(D_C|\mathcal{D}^M)$   our goal is hence to obtain the optimal choice
of $f_C^M(t)\in \mathfrak{F}_C^M$  that allows the {minimum} value of $p$ such that $f_C^{(p)}(t) \in \mathfrak{F}_C^M$. 

{As notice before}, if $D_C$ is an element of $\mathcal{D}^M_C$ then we can simply take
 {$p=0$, i.e.,  $p(D^M_C|\mathcal{D}_C^M)=0$}. For the depolarizing evolutions  which instead have a continuous characteristic function $f^{NM}_C(t)$ that possesses some degree of  non-Markovianity, the computation of
 (\ref{NMC}) requires instead some non trivial work.
 In this case Eq.~(\ref{fp}) becomes 
\begin{equation}\label{fpcont}
f_C^{(p)}(t)={(1-p) f^{NM}_C(t) + p }f_C^{M}(t) \, .
\end{equation}
 While the continuity of $f_C^{(p)}(t)$ is automatically ensured by construction,
 finding the {minimum} $p$ that forces this function into $\mathfrak{F}^{M}_C$ (namely
 that allows it to be also positive and non-increasing) 
 is not a simple task. 
 In order to tackle this problem we start by  first illustrating the relatively simple
  case of 
  non-Markovian depolorazing 
evolutions with 
 positive $f^{NM}_C(t)\in\mathfrak{F}_C^{NM}$ (see Sec.~\ref{onepos}).
  Next we discuss the slightly more complex scenario of $f^{NM}_C(t)\in\mathfrak{F}_C^{NM}$ having a non definite sign, but  which  exhibit
 their  non-Markovian character exclusively on the time intervals where they are negative (Section~\ref{nneg}). Finally we conclude by addressing the general case of 
a non-Markovian continuous characteristic functions $f^{NM}_C(t)\in\mathfrak{F}_C^{NM}$ 
 in Sec.~\ref{nposnegtech}.

\subsection{Positive non-Markovian continuous characteristic functions}  
\label{onepos}

In this section  we consider depolorazing processes  $D_C^{NM}$ characterized 
by   $f^{NM}_C(t)\in \mathfrak{F}^{NM}_C$ which are positive and which have 
 a number $L>0$  of intervals $T_k^+ \equiv (t_{k}^{(in)},t_{k}^{(fin)})$ of non-Markovianity where $\dot f^{NM}(t^{\pm})> 0$, i.e.,  
\begin{equation}\label{fnmc1}
\left\{
\begin{array}{ccc}
f_C^{NM}(t)\geq 0,\dot{f}_C^{NM}(t^{\pm})\leq  0, \xi(f_C^{NM}(t))=1&  t\notin  T^{NM}\;, \\\\
f_C^{NM}(t)\geq 0 , \dot{f}_C^{NM}(t^{\pm}) >0, \xi(f_C^{NM}(t))=1&  t\in T^{NM} \;,
\end{array}
\right.
\end{equation}
with 
$T^{NM} \equiv\bigcup_{k=1}^L T_k^+$ being the collection of the  intervals~$T_k^+$.
As we shall see, in this case the quantity~(\ref{NMC}) is 
a monotonically increasing function of the  gaps 
\begin{eqnarray}\label{Deltask1}
&\Delta_k^{NM} &\equiv  f_C^{NM}(t_k^{(fin)}) - f_C^{NM}(t_k^{(in)})>0 \, ,
\end{eqnarray}
which certify the non-Markovian character of $f_C^{NM}(t)$ on the intervals
$T_k^+$. 
Specifically, given 
 \begin{eqnarray}  \label{DELTANM} 
\Delta^{NM}\equiv \sum_{k=1}^L\Delta_k^{NM}\;,\end{eqnarray}
we have 
 \begin{equation}\label{N1}
{p(D_C^{NM}|\mathcal{D}_C^M)=\frac{\Delta^{NM}}{1+\Delta^{NM}} }\, ,
\end{equation}
which saturates to its upper bound $1$ in the case where $\Delta^{NM}$ diverges, e.g. when $f^{NM}_C(t)$ exhibit infinite, not 
properly dumped,
 oscillations.
In order to derive (\ref{N1}) we first address the simple case of a single non-Markovian
interval ($L=1$), and then generalize it to the case of arbitrary (possibly infinite) $L$.

\subsubsection{One time interval of non-Markovianity for positive characteristic functions ($L=1$)} \label{Seconetime}

Let $D^{NM}_C$ be an element of  $\mathcal{D}^{NM}_C$ with 
characteristic function 
$f_C^{NM}(t) \in \mathfrak{F}_C^{NM}$ that is always positive and which has positive derivative (hence non-Markovian character) in a single 
time interval $T_1^+=(t_1^{(in)},t_1^{(fin)})$ ($t_1^{(fin)}$ being possibly infinite),
i.e,
\begin{equation}\label{fnmc1}
\left\{
\begin{array}{ccc}
f_C^{NM}(t)\geq 0,\dot{f}_C^{NM}(t^{\pm})\leq  0, \xi(f_C^{NM}(t))=1&  t \notin T_1^+ \;, \\ \\
f_C^{NM}(t)\geq 0 , \dot{f}_C^{NM}(t^{\pm}) >0, \xi(f_C^{NM}(t))=1&  t\in T_1^+ \;. 
\end{array}
\right.
\end{equation}
Our goal is to determine the {minimum} value of $p$ which allows $f_C^{(p)}(t)$ of 
(\ref{fpcont}) to be an element of $\mathfrak{F}^{M}_C$, i.e., 
to obey to the first of the constraints~(\ref{MarkD}) -- the function being already continuous by construction. 
Since both $f^{NM}_C(t)$ and $f^{M}_C(t)$ are non-negative, this is equivalent to impose  
\begin{equation}\label{dotfp}
\dot f_C^{(p)}(t^{\pm})= {(1-p) \dot f_C^{NM}(t^{\pm}) + p} \dot f_C^{M}(t^{\pm})\leq 0 \, ,
\end{equation}
which is automatically verified  for $t\notin T_1^+$. 
A necessary condition for (\ref{dotfp}) can  then be obtained by imposing that
$f_C^{(p)}(t)$ experiences a negative gap at the extremal points of 
$T_1^+$, i.e.,  
\begin{eqnarray}
\label{Deltas3}
&\Delta_1^{(p)}&\equiv  f_C^{(p)}(t_1^{(fin)}) - f_C^{(p)}(t_1^{(in)}) \leq 0 \;. 
\end{eqnarray}
From (\ref{fpcont}) we can cast this into the  condition 
\begin{eqnarray}
\label{Deltas3p}
&\Delta_1^{(p)}&= {(1-p) \Delta_1^{NM}+p}\Delta_1^{M}\leq 0 \;,
\end{eqnarray}
where $\Delta_1^{NM}$ is the positive gap defined as in Eq.~(\ref{Deltask1}) and
\begin{eqnarray}
\label{Deltas2}
&\Delta_1^{M}&\equiv f_C^{M}(t_1^{(fin)}) - f_C^{M}(t_1^{(in)}) \, , 
\end{eqnarray}
is the associated gap of $f_C^{M}(t)$.
Notice that from the properties of $f_C^{M}(t)$ it follows that the latter quantity is non-negative and  larger than $-1$ (which is the minimum allowed gap for an element of $\mathfrak{F}_C^{M}$), i.e.
\begin{eqnarray}
\label{Deltas2lim0}
{\Delta_1^{M} \in[-1,0]} \, \quad \Longrightarrow \quad | \Delta_1^{M}| \leq 1\;. 
\end{eqnarray}
From Eq.~(\ref{Deltas3p}) it follows that a necessary condition for $p$ is 
\begin{eqnarray} {p \geq  
\frac{\Delta_1^{NM}}{|\Delta_1^{M}|+\Delta_1^{NM}}\geq \frac{\Delta_1^{NM}}{1+\Delta_1^{NM}} \equiv p_1 \;,}  \label{nece1} 
\end{eqnarray} 
where the last inequality follows from (\ref{Deltas2lim0}).  
To show that (\ref{nece1}) is also a sufficient condition for (\ref{dotfp}), we provide a particular example of $f_C^{M}(t)$ such that $\dot f_C^{(p)}(t)\leq 0$ for {$p\geq p_1$}. For this purpose consider $g_C^{M}(t) \in \mathfrak{F}^{M}_C$ such that
\begin{equation}\label{ofM}
g_C^{M}(t) = \left\{
\begin{array}{cc}
1 & \,\,\, t\leq t_1^{(in)}\;, \\ 
1- \left( f^{NM}(t)-f^{NM}(t_1^{(in)}) \right)/ \Delta_1^{NM}& t\in T_1^+\;, \\ 
0 & \,\,\,\, t\geq t_1^{(fin)}\;.
\end{array}
\right. 
\end{equation}
This function, for $t\in T_1^+$, is a linear manipulation of $f_C^{NM}(t)$, where its slope is stretched and inverted. Moreover, in this case $\Delta_1^{M}=-1$ and $\Delta_1^{(p)} \leq 0$ for {$p\geq p_1$}. Finally, if we consider $g_C^{M}(t) $ in $f^{(p)}(t)$, for $p=p_1$, we obtain 
\begin{equation}\label{fp1}
f^{(p_1)}(t)= \frac{f^{NM}(t_1^{(fin)})}{1+\Delta_1^{NM}}\;,  \hspace{0.7cm} \mbox{ for } 
t\in T_1^+\;, 
\end{equation}
which is a constant. Hence, in this case $\dot f^{(p_1)}(t) \leq 0$ for any $t\geq 0$. 
Putting all together we can hence claim that 
\begin{equation}\label{N11}
{ p(D_C^{NM}|\mathcal{D}_C^M)=p_1  =\frac{\Delta_1^{NM}}{1+\Delta_1^{NM}} }   \, ,
\end{equation}
 which proves the validity of (\ref{N1}) at least for the functions we are considering here, namely when $L=1$.

\subsubsection{Multiple time intervals of non-Markovianity for positive characteristic functions}\label{npos}
Here we extend the previous construction to address the general case of functions of the 
form (\ref{fnmc1}), i.e.,  which are 
positive and which have 
an arbitrary (possibly infinite) number $L>0$  of intervals $T_k^+ \equiv (t_{k}^{(in)},t_{k}^{(fin)})$ of non-Markovianity.
As in the previous section for each of the intervals $T_k^+$ we introduce the gaps
\begin{eqnarray}
\label{Deltask2}
&\Delta_k^{M}&\equiv f_C^{M}(t_k^{(fin)}) - f_C^{M}(t_k^{(in)}) \, , \\ 
\label{Deltask3}
&\Delta_k^{(p)}&\equiv f_C^{(p)}(t_k^{(fin)}) - f_C^{(p)}(t_k^{(in)})=
{(1-p) \Delta_k^{NM}+p} \Delta_k^{M} \, , 
\end{eqnarray}
with $\Delta_k^{NM}$ the positive quantities defined in (\ref{Deltask1}). 
Observe then due to the fact that $f_C^M(t)$ is in $\mathfrak{F}^{M}_C$, the $\Delta_k^{M}$ are all non-positive while 
their global sum is larger than $-1$, i.e.
\begin{eqnarray}
\label{Deltas2lim}
{ \Delta_k^{M} \in[-1,0] \, \;, \qquad 
\Delta^{M}\equiv  \sum_{k=1}^L \Delta_k^{M} \in[-1,0] }\; .
\end{eqnarray}
This  is just a consequence of the fact that the maximum gap of a continuous
Markovian characteristic function is at most equal to $-1$.
A necessary condition for the Markovianity of $f^{(p)}_M(t)$  can then be obtained by imposing that $\Delta_k^{(p)}\leq 0$ for all $k$, which in turn implies 
\begin{eqnarray}
0\geq \sum_{k=1}^L\Delta_k^{(p)}&= & {(1-p) \Delta^{M} +p } \Delta^{NM} \\  
&&\!\!\!\!\!\!\!\! \Longrightarrow {p \geq \frac{\Delta^{NM}}{|\Delta^{M}| + \Delta^{NM} } \geq  \frac{\Delta^{NM} }{1+\Delta^{NM} }\equiv p_L} \label{abrakadabra} \;, 
\end{eqnarray} 
where \ref{abrakadabra} we used (\ref{DELTANM}) and (\ref{Deltas2lim}). 
%
%
Now we show that a $ g_C^{M}(t)\in \mathfrak{F}^M_C$ that makes $f_C^{(p)}(t)$ Markovian for any ${p\geq p_L}$ exists. We consider the following monotonically decreasing function

\begin{equation}\label{ofMgen}
g_C^{M}(t) = \left\{
\begin{array}{cc}
1 & \,\,\, t\leq t_1^{(in)} \\ 
1- \left( f^{NM}(t)-f^{NM}(t_1^{(in)}) \right)/ \Delta^{NM}& t\in T_1^+ \\  
g_C^{M}(t_{1}^{(fin)})  - \left( f^{NM}(t)-f^{NM}(t_2^{(in)}) \right)/ \Delta^{NM}& t\in T_2^+ \\  
\cdots \\ 
g_C^{M} (t_{k-1}^{(fin)}) -\left( f^{NM}(t) - f^{NM}(t_{k}^{(in)}) \right)/{\Delta^{NM}} & t\in T_k^+ \\ 
\cdots 
\end{array}
\right. \!\! ,
\end{equation}
that we define constant and equal to $g_C^M (t_{k-1}^{(fin)})$ in the time intervals $[t_{k-1}^{(fin)},t_k^{(in)}]$, for $k=1,\dots,L$. Therefore, the temporal derivative of $g_C^M(t)$ is particularly simple
\begin{equation}\label{derivwow}
\dot{{g}}_C^M(t^{\pm}) = \left\{
\begin{array}{ccc}
-\dot{f}^{NM}_C(t^{\pm}) /  |\Delta^{NM}| & t\in  {T}_k^+ \;, \\\\
0 & \mbox{ otherwise. } 
\end{array} \right. 
\end{equation}
As a consequence, for $t\in T_k^+$, the function $g_C^{M}(t)$ decreases by a factor proportional to the increase of $f^{NM}(t)$ in the same time interval, namely $\Delta_k^{M} =-\Delta^{NM}_k /\Delta^{NM} <0$. 
An intuitive explanation for the form of $g_C^{M}(t)$ is the following. The ``resource'' of a continuous Markovian characteristic function to contrast the non-Markovianity of $f^{NM}_C(t)$ is its distance from zero. Once that $f_C^{NM}(t)$ decreases, it cannot increase again. Therefore, to efficiently use the maximum available gap allowed for Markovian characteristic functions, namely $\Delta^M=-1$, $g_C^M(t)$ is constant whenever $f^{NM}_C(t)$ behaves as a Markovian characteristic function. Instead, when this behavior is non-Markovian, $g_C^{NM}(t)$ decreases accordingly to the increase of $f^{NM}_C(t)$ in order to make their convex sum $f_C^{(p)}(t)={(1-p) f^{NM}_C(t)+ p } g_C^M(t)$ constant for the {smallest} value of $p$.
This  proves  that, for the continuous  depolarizing evolutions defined as in Eq.~(\ref{fnmc1}), $p(D_C^{NM}|\mathcal{D}^M_C)=p_L$.
Therefore, the corresponding measure of non-Markovianity (\ref{NMC}) is equal to
\begin{equation}\label{manypos}
{p(D_C^{NM}|\mathcal{D}_C^M)=  p_L= \frac{\Delta^{NM}}{1+\Delta^{NM}}} \, ,
\end{equation}
which corresponds to Eq.~(\ref{N1}).

\subsection{Characteristic functions with  non definite sign that exhibit 
non-Markovianity only when negative}\label{nneg}

Here we consider elements of $\mathcal{D}_C^{NM}$ with 
$f^{NM}_C(t)$ such that their non-Markovian nature is shown only in a number $m>0$ of time intervals $T_j^-\equiv(t_{j}^{(in)},t_j^{(fin)})$ where it assumes negative values while being strictly decreasing, namely violating $\mathbf{CM_1}(\tau)$ while being negative, as notified by the following 
negative gaps 
\begin{equation}\label{NMneggaps}
\Theta_j^{NM}\equiv f_C^{NM}(t^{(fin)}_j) - f_C^{NM}(t^{(in)}_j)<0\;.
\end{equation}
It is worth observing that under the above assumption 
$f^{NM}_C(t)$ cannot be positive after that it becomes negative for the first time. Otherwise, {for some time we would have} $f^{NM}_C(t)\geq 0$ and $\dot f^{NM}_C(t^{+})>0$, which contradicts our premise. Therefore, we have that 
\begin{eqnarray}
f^{NM}_C(t) \leq 0 \;, \qquad \forall t\geq t_1^{(in)}\;.
\end{eqnarray} 
We shall see that in this scenario the the measure of non-Markovianity~(\ref{NMC}) reduces to 
\begin{equation}\label{venga2}
{p(D^{NM}_C|\mathcal{D}_C^M)= \frac{|\Theta^{NM}|}{1+|\Theta^{NM}|}  }\, ,
\end{equation}
with 
\begin{eqnarray}\label{DEFTHETA}  \Theta^{NM} \equiv \sum_{j=1}^m \Theta_j^{NM}\;. \end{eqnarray} 
As in the previous section, to derive the above identity   first we obtain a necessary condition for $f^{(p)}(t)$ to belong to $\mathfrak{F}^M_C$ and then we provide an explicit example that saturates this value.  In this 
case however we find it useful to treat separately the case of finite $m$ from those where $m$ is unbounded 
which introduce some technicalities which have to be dealt carefully.

\subsubsection{The  $m$ finite case}

If  $m$ is finite the function $f^{NM}_C(t)$ cannot exhibit infinite oscillations.
Therefore its $t\rightarrow \infty$ limit exists finite, i.e.
\begin{eqnarray} \label{INFINM} 
\lim_{t\rightarrow \infty} f^{NM}_C(t) =  f^{NM}_C(\infty) \leq 0\;.
\end{eqnarray} 
 Define now $\overline{T}_j=(\overline{t}_j^{(in)},\overline{t}_j^{(fin)})$ to be the time intervals when $f^{NM}_C(t)\leq 0$ and $\dot{f}_C^{NM}(t^{\pm})\geq 0$,  namely the times when the Markovian condition $\mathbf{CM_1}(\tau)$ is satisfied while $f^{NM}_C(t)$ is negative. We notice that, since $f^{NM}_C(t)$ is continuous, for any $T_j^-$ there exists a $\overline{T}_j$ such that $t^{(fin)}_j=\overline{t}^{(in)}_j$, the only case when it does not happen is for $t_j^{(fin)}=\infty$: accordingly the total number $\overline{m}$ of the intervals $\overline{T}_j$ is either equal to $m$ or to $m-1$ and is hence also
 finite by assumption. 
We consider now the associated gaps of the functions $f^{NM}_C(t)$, $f^{M}_C(t)$,
and $f^{(p)}_C(t)$, i.e., the quantities 
\begin{eqnarray}\label{deltanmj}
\delta^{NM}_j&\equiv&f^{NM}_C(\overline{t}^{(fin)}_j) - f^{NM}_C(\overline{t}^{(in)}_j) \, ,\\ 
\label{deltamj}
\delta^{M}_j&\equiv&f^{M}_C(\overline{t}^{(fin)}_j) - f^{M}_C(\overline{t}^{(in)}_j) \, ,\\ 
\label{deltapj}
\delta^{(p)}_j & \equiv& f^{(p)}_C(\overline{t}^{(fin)}_j) - f^{(p)}_C(\overline{t}^{(in)}_j)=
{(1-p) \delta^{NM}_j + p} \delta^M_j  \,. 
\end{eqnarray}
By definition we have that the $\delta^{NM}_j$ must be non-negative, while the
$\delta^{M}_j$ must be non-positive, i.e., 
\begin{eqnarray}\label{deltanmj>}
\delta^{NM}_j\geq  0 \, \;, \qquad 
\delta^{M}_j \leq 0 \, \;, \qquad \forall j.
\end{eqnarray}
If $f^{(p)}_C(t)$ is Markovian it has to be positive and non-increasing. Therefore, we should also have 
\begin{eqnarray}\label{deltanmj>>}
\delta^{(p)}_j \leq 0 \,  \;, \qquad \forall j.
\end{eqnarray}
Therefore 
a necessary condition for the Markovianity of $f^{(p)}_C(t)$ is given by 
the following inequality 
\begin{eqnarray}\delta^{(p)}\equiv \sum_{j=1}^{\overline{m}}  \delta_j^{(p)}={(1-p) \delta^{NM}- p } |\delta^M|\leq 0\;,\label{vangava0} 
\end{eqnarray}  
where  $\delta^M\equiv \sum_{j=1}^{\overline{m}} \delta_j^M\leq 0$ and $\delta^{NM}\equiv \sum_{j=1}^{\overline{m}} \delta_j^{NM}\geq 0$.
Observe also that since  $f_C^M(t)$ and $f^{(p)}_C(t)$  are  both elements of $\mathfrak{F}^M_C$ 
their limiting values for $t\rightarrow \infty$ exist  and fulfil the following constraints 
 \begin{eqnarray}
 f_C^M(t)\geq  f_C^M(\infty)\geq 0\;, 
 \qquad 
  f^{(p)}_C(t)\geq  f^{(p)}_C(\infty)\geq 0\;,
  \end{eqnarray} 
  for all $t\geq 0$. Notice finally that since $f_C^M(t)$ is non increasing and upper bounded by $1$,  its limiting value must fulfil the constraint
  \begin{eqnarray} \label{bound1new} 
  1 \geq  f_C^M(\infty) + | \delta^M| \;. 
  \end{eqnarray} 
  Accordingly from (\ref{INFINM})  we can write 
\begin{equation}\label{vengava}
 f^{(p)}_C(\infty) = {(1-p) f_C^{NM}(\infty)  + p } f_C^M(\infty) \geq 0 \, ,
\end{equation}
or equivalently 
 \begin{equation}\label{vangava1}
-{(1-p)  (\delta^{NM} + \Theta^{NM})  -  p} f_C^M(\infty) \leq 0 \, ,
\end{equation}
where we used 
\begin{eqnarray} \label{CONSTR1new}
 f_C^{NM}(\infty)  = \delta^{NM} + \Theta^{NM} \;,
\end{eqnarray} 
with  $\Theta^{NM}$ as in Eq.~(\ref{DEFTHETA}). 
Summing up  (\ref{vangava1}) with (\ref{vangava0}) term by term,  the following 
necessary constraint for $p$ can finally be obtained 
 \begin{equation}\label{vangava2}
- {(1-p)  \Theta^{NM}  - p } ( f_C^M(\infty) + |\delta^{M}|) \leq 0 \, ,
\end{equation}
which implies 
\begin{eqnarray} {p \geq  \frac{|\Theta^{NM}| }{ f_C^M(\infty) + |\delta^{M}| +|\Theta^{NM}|} \geq   \frac{ |\Theta^{NM}| }{ 1+ |\Theta^{NM}|} \equiv p_m} \;, \label{NEW111} 
\end{eqnarray}  
where in the last passage we used  the inequality~(\ref{bound1new}). 
Accordingly we can conclude that the quantity $p_m$ is {lower} bound for the value $p(D_C^{NM}|\mathcal{D}_C^M)$ associated with 
the evolutions $D_C^{NM}$ we are considering here. 
In order to show that $p_m$ does indeed correspond to  $p(D_C^{NM}|\mathcal{D}_C^M)$ we
now present a example of $f_C^M(t)$ which makes  $f_C^{(p)}(t)$ an element of $\mathfrak{F}^M_C$
for $p=p_m$. 
To do so we define $\overline{g}_C^{M}(t) \in \mathfrak{F}^M_C$ to be equal to
\begin{equation}\label{ofMgenX}
 \left\{\!\!
\begin{array}{cc}
1 & \,\,\, t\leq t_1^{(fin)} \\ 
1- \left( f^{NM}(t)-f^{NM}(\overline{t}_1^{(in)}) \right)/ |\Theta^{NM}|& t\in \overline{T}_1 \\  
1-\delta^{NM}_1/|\Theta^{NM}| & t\in T_2^- \\ 
(1-\delta^{NM}/|\Theta^{NM}|) - \left( f^{NM}(t)-f^{NM}(\overline{t}_2^{(in)}) \right)/ |\Theta^{NM}|& t\in \overline{T}_2 \\  
\dots \\ 
\overline{g}_C^M(\overline{t}^{(fin)}_{j-1}) - \left( f^{NM}(t)-f^{NM}(\overline{t}_j^{(in)}) \right)/ |\Theta^{NM}| & t \in \overline{T}_j\\ 
1-\sum_{i=1}^j \delta^{NM}_i/|\Theta^{NM}| & t\in T_{j+1}^- \\ 
\dots \\
1 - \delta^{NM}/|\Theta^{NM}| & t\rightarrow \infty
\end{array}
\right. \! \!\! .
\end{equation}
The temporal derivative of $\overline{g}_C^M(t)$ assumes the simple form
\begin{equation}
\dot{\overline{g}}_C^M(t^{\pm}) = \left\{
\begin{array}{ccc}
-\dot{f}^{NM}_C(t^{\pm}) /  |\Theta^{NM}| & t\in \overline{T}_j \\
0 & \mbox{ otherwise } 
\end{array} \right. \, .
\end{equation}
It is easy to show that $f^{(p)}_C(t)={(1- p) f^{NM}_C(t) + p } \overline{g}^M_C(t)$ Markovian for {$p\geq p_m$}. Therefore, for any $f^{NM}_C(t)$ that shows a non-Markovian behavior while being negative, we have that 
\begin{equation}\label{venga2pm}
{p(D^{NM}_C|\mathcal{D}_C^M)= p_m=} \frac{|\Theta^{NM}|}{1+|\Theta^{NM}|}  \, ,
\end{equation}
which proves (\ref{venga2}). 

\subsubsection{Removing the  finite $m$ constraint}\label{nneginft}
In the previous paragraph we have assumed $m$ to be explicitly finite, a useful hypothesis which allowed us
to assume the existence of (\ref{INFINM}) and to express its value as in (\ref{CONSTR1new}). 
It turns out however that this assumption is not  fundamental and that Eq.~(\ref{venga2}) holds true also if
we drop it. 
In order to show this, instead of studying the Markovian character of 
$f^{(p)}_C(t)$ for all $t\geq 0$, we 
limit the analysis for just all $t\leq T$ with $T$ being finite quantity. 
Observe then that the number  $m(T)$  
of time intervals  
$T_j^-=(t_{j}^{(in)},t_j^{(fin)})$ contained into  domain $[0,T]$, where the characteristic function $f^{NM}_C(t)$ is negative and decreasing, is by construction finite. 
  Same considerations holds for the total number 
$\overline{m}(T)$ of the time intervals  $\overline{T}_j=(\overline{t}_j^{(in)},\overline{t}_j^{(fin)})$ when 
$f^{NM}_C(t)\leq 0$ and $\dot{f}_C^{NM}(t^{\pm})\geq 0$ and which fit on $[0,T]$.
Following the same reasoning we adopted in the previous section, the following relations  can then
be derived 
\begin{eqnarray}
 \label{CONSTR1newT}
 f_C^{NM}(T)  &=& \delta^{NM}(T) + \Theta^{NM}(T) \;, \\
 1 &\geq&  f_C^M(T) + | \delta^M(T)| \;,   \label{CONSTR1newT1}
\end{eqnarray} 
with
 \begin{eqnarray}  \nonumber 
  & \delta^{M}(T)\equiv \sum_{j=1}^{\overline{m}(T)}  \delta_j^{M} \leq  0 \;, 
   \qquad 
 \delta^{NM}(T)\equiv \sum_{j=1}^{\overline{m}(T)}  \delta_j^{NM} \geq  0\;, &\\
 &  \Theta^{NM}(T) \equiv \sum_{j=1}^{m(T)} \Theta_j^{NM} < 0\;. &
 \end{eqnarray} 
Furthermore Eqs.~(\ref{vangava0}) and (\ref{vangava1}) get replaced by 
\begin{eqnarray}
{(1-p) \delta^{NM}(T)- p } |\delta^M(T)|&\leq& 0 \;, \\
-{(1-p)  (\delta^{NM}(T) + \Theta^{NM}(T))  - p } f_C^M(T) &\leq& 0 \, ,
\end{eqnarray}  
that summed up term by term lead to 
\begin{eqnarray} {p \geq \frac{|\Theta^{NM}(T)|}{ 1+ |\Theta^{NM}(T)|} } \;, \label{NEW111T} 
\end{eqnarray}  
which is a necessary condition to have $f^{(p)}_C(t)$ Markovian at least on $[0,T]$. 
Following then a construction which is analogous to the one given in (\ref{ofMgenX}) we can
also show that indeed the right-hand-side term of (\ref{NEW111T})
is the {minimum} value for $p$ to ensure the   Markovianity of $f^{(p)}_C(t)$ on
$[0,T]$. The final result thus can be derived by taking the limit $T \rightarrow \infty$ which leads to (\ref{venga2}) where
now $\Theta^{NM}$ is properly computed  as  $\Theta^{NM}= \lim_{T\rightarrow \infty}  \Theta^{NM}(T)$.
Notice in particular that having extend (\ref{venga2}) to the case of infinite $m$ it is now possible that
$|\Theta^{NM}|$ will diverge (a case that for instance happen whenever $f^{NM}_C(t)$ has infinitely many
-- not properly dumpted -- oscillations) leading to {the maximum value for the measure of non-Markovianity, namely} {$p(D_C^{NM}|\mathcal{D}_C^M)=1$}.

\begin{figure}
\includegraphics[width=0.51\textwidth]{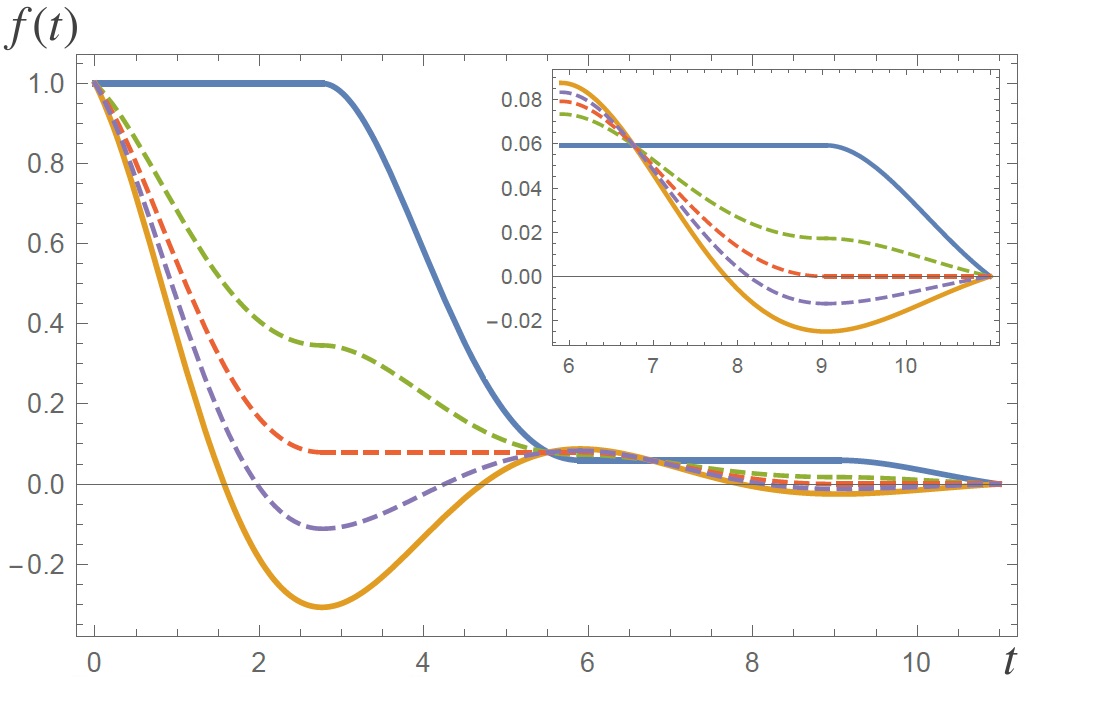}
\caption{Plots of $f_C^{NM}(t)=\expp{-2t/5}\cos (t) $ (yellow), the corresponding optimal Markovian characteristic function $h_C^{M}(t)$ (blue) and $f^{(p)}(t)$ for different values of $p$ (dashed lines)  in the time interval $t\in[0,7\pi/2]$. The inset shows their behavior for $t\geq 5.90$.
In this example $T_1^{-}\simeq(\pi/2,2.76)$, $T_1^{+}\simeq(3\pi/2, 5.90)$ and  $T_2^{-}\simeq(5\pi/2,9.04)$ are the time intervals of non-Markovianity of $f^{NM}(t)$ and  $\Theta_1^{NM}\simeq-0.31$, $\Delta_1^{NM}\simeq 0.09$ and $\Theta_2^{NM}\simeq -0.02$ are the corresponding non-Markovian gaps. The value of the measure of non-Markovianity is $p(D^{NM}_C|\mathcal{D}^M)\simeq 0.30$.
If ${p=0.5 > p(D_C^{NM}|\mathcal{D}^M_C)}$, $f^{(p)}(t) \in \mathfrak{F}^M_C$ is monotonically decreasing (green dashed line). If {$p=p(D_C^{NM}|\mathcal{D}^M_C) \simeq 0.30$},  $f^{(p)}(t)\in \mathfrak{F}^M_C$ is monotonically decreasing and constant when $\dot{f}^{NM}(t)>0$ (red dashed line). If {$p=0.15<p(D_C^{NM}|\mathcal{D}^M_C)$},  $f^{(p)}(t) \in \mathfrak{F}^{NM}_C$ is not monotonic nor positive in more than one time interval (purple dashed line).  }\label{zoomed}
\end{figure}

\subsection{Multiple time intervals of non-Markovianity for continuous  characteristic functions: the general case}\label{nposnegtech}

Building up from the previous sections here we compute {$p(D_C^{NM}|\mathcal{D}_C^M)$} for the general case of  
a non-Markovian depolarizing processes  with continuous 
characteristic function  $f^{NM}_C(t)$. At variance with the examples discussed before, now 
$f^{NM}_C(t)$ may possess both 
 a collection of time intervals $T_k^+ \equiv (t_{k}^{(in)},t_{k}^{(fin)})$  where it is positive and increasing, and also time intervals $T_j^-\equiv(t_{j}^{(in)},t_j^{(fin)})$ where instead it is negative and decreasing (namely it may exhibit 
all the non-Markovian features detailed separately in Sec.~\ref{onepos} and Sec.~\ref{nneg}).

In this case we can show that Eqs.~(\ref{N1}) and (\ref{venga2}) 
get replaced by the more general formula
 \begin{equation}\label{N1final}
{p(D_C^{NM}|\mathcal{D}_C^M)=\frac{\Gamma^{NM}}{1+\Gamma^{NM}} } \, ,
\end{equation}
with $\Gamma^{NM}$ being given by the expression
\begin{eqnarray}  \label{DEFGAMMA} 
\Gamma^{NM} \equiv\Delta^{NM}+ |\Theta^{NM}|\;,
\end{eqnarray} 
where $\Delta^{NM}$ and $\Theta^{NM}$, defined as in Eqs.~(\ref{DELTANM}) and 
(\ref{DEFTHETA}),
are the sums of the non-Markovian increments 
the function $f^{NM}_C(t)$ experiences on the intervals $T_k^+$ and $T_j^-$, respectively.

Since $f_C^{NM}(t)$ may not admit a limiting value for $t\rightarrow \infty$, to prove (\ref{N1final}) we shall proceed as in Section~\ref{nneginft}, determining first the conditions under which the associated $f_C^{(p)}(t)$ is guaranteed to be
Markovian at least on the time interval $[0,T]$ with $T$ finite. 
Under this condition  the numbers  $L(T)$ and $m(T)$ of intervals $T_k^+$ and $T_j^-$ of 
$f_C^{NM}(t)$ that fit on the considered domain, are both finite. 
We introduce also the time intervals $\overline{T}_j
\equiv (\overline{t}_j^{(in)},\overline{t}_j^{(fin)})$ of $[0,T]$ where $f^{NM}_C(t)$ is negative and non decreasing
(their number $\overline{m}(T)$ being finite too), and define the gaps $\Delta^{NM}_k(T)$, $\Delta_k^M(T)$, $\Delta_k^{(p)}(T)$, 
$\Theta_j^{NM}(T)$, $\delta^{NM}_j(T)$, $\delta^{M}_j(T)$ and $\delta^{(p)}_j(T)$
as in 
Eqs. (\ref{Deltask1}), (\ref{Deltask2}), (\ref{Deltask3}), (\ref{NMneggaps}), (\ref{deltanmj}), (\ref{deltamj}), and (\ref{deltapj}).
By construction we have the following conditions
\begin{eqnarray}
&&\Delta^{NM}_k(T) >  0\;,\qquad  \Theta_j^{NM}(T) < 0 \qquad \delta^{NM}_j(T) \geq 0\;,   \nonumber \\
&&\Delta^{M}_k(T) \leq  0\;, \qquad \delta_j^{M}(T) \leq 0\;, \nonumber \\
&& \Delta_k^{(p)}(T)={(1-p) \Delta^{NM}_k(T) + p } \Delta_k^M(T)\;, \\
&& \delta^{(p)}_j(T)={(1-p)  \delta^{NM}_j (T)+  p } \delta_j^M(T)\;, 
\end{eqnarray} 
for all $k$ and $j$. 
A necessary condition for $f^{(p)}(t)$ being Markovian on the considered domain is that all its gaps $\Delta_k^{(p)}(T)$ and 
$\delta_j^{(p)}(T)$ are non-positive,
i.e., 
\begin{eqnarray}
{(1-p) \Delta^{NM}_k(T) + p} \Delta_k^M(T) &\leq& 0\;, \label{FIRST} \\
{(1-p) \delta^{NM}_j (T)+  p} \delta_j^M(T) &\leq& 0\;.  \label{SECOND}
\end{eqnarray} 
By summing up term by term, all contributions from (\ref{FIRST}) and  (\ref{SECOND})  we get  \begin{equation}\label{cond1}
{(1-p) (  \Delta^{NM}(T)+ \delta^{NM}(T) )  -  p (  |\Delta^M(T) | + |\delta^M(T) |) \leq 0 \;, } \end{equation}
where 
\begin{eqnarray} 
&&\Delta^{NM}(T) \equiv \sum_{k=1}^{{L}(T)} \Delta_k^{NM}(T)>0\;, \qquad \Delta^{M}(T)\equiv\sum_{k=1}^{{L}(T)}\Delta_k^M(T)\leq 0\;, \nonumber \\
&& \delta^{NM}(T) \equiv \sum_{k=1}^{\overline{m}(T)} \Delta_k^{NM}(T)>0\;, \qquad \delta^{M}(T)\equiv\sum_{k=1}^{\overline{m}(T)}\Delta_k^M(T)\leq 0\;. \nonumber
 \end{eqnarray}
Suppose now that $f^{NM}_C(T)$ is a non-negative quantity, i.e., $f^{NM}_C(T)\geq 0$.
  Under this condition it is easy to verify that  the total gaps this function experiences
on the interval where it is negative must nullify, i.e., 
  \begin{eqnarray} \delta^{NM}(T)=|\Theta^{NM}(T)|\;, \label{IDEIDE}  \end{eqnarray} 
  with 
  \begin{eqnarray} 
&&\Theta^{NM}(T) \equiv \sum_{j=1}^{{m}(T)} \Theta_j^{NM}(T)< 0
\;. \end{eqnarray} 
Replacing this into (\ref{cond1}) we hence get the condition
\begin{eqnarray}\label{cond1dd}
  p &\geq&  \frac{ \Delta^{NM}(T)+ |\Theta^{NM}(T)| }{ |\Delta^M(T) | + |\delta^M(T) | +   \Delta^{NM}(T)+ |\Theta^{NM}(T)| }  \nonumber \\
&\geq& 
\frac{\Delta^{NM}(T)+ |\Theta^{NM}(T)| }{1+  \Delta^{NM}(T)+ |\Theta^{NM}(T)|}\;,  \label{impo1} 
\end{eqnarray} 
where in the second line we used the fact that the sum over the gaps of a continuous Markovian function cannot 
cannot be larger than 1, i.e.,  $|\Delta^M(T) | + |\delta^M(T) |\leq 1$. 
If $f^{NM}_C(T)$ is negative,  i.e., $f^{NM}_C(T) < 0$, we can still show that (\ref{impo1}) holds, but we need to change
the derivation. In this case we observe that Eq.~(\ref{IDEIDE}) is substituted by the constraint
\begin{eqnarray}
 \label{cCONSTR1newT11}
 f_C^{NM}(T)  &=& \delta^{NM}(T) + \Theta^{NM}(T)  \;,
 \end{eqnarray} 
which allows us to rewrite positivity of $f^{(p)}_M(t)$ for $t=T$ (a necessary condition for $f^{(p)}(t)$ to be Markovian on $[0,T]$) as 
\begin{eqnarray}
{(1-p)   (\delta^{NM}(T) + \Theta^{NM}(T))   + p} f_C^M(T) &\geq& 0 \, .
\end{eqnarray}  
Together with (\ref{cond1}) the above expression finally leads to  
\begin{eqnarray} 
(1-p)  (\Delta^{NM}(T) - \Theta^{NM}(T))  & \leq &  p( |\Delta^M(T) | + |\delta^M(T) | + f_C^M(T) )\nonumber \\
 &\leq& p\;, \label{P11} 
\end{eqnarray}
where in the last passage we used the fact that continuous Markovian characteristic function cannot have drops 
larger than $1$, i.e., $|\Delta^M(T) | + |\delta^M(T) | + f_C^M(T)\leq 1$. 
Equation~(\ref{P11}) coincides with (\ref{impo1}) which hence holds true irrespectively from the sign of $f^{NM}_C(T)$.
Taking the limit $T\rightarrow \infty$ we can finally conclude that a necessary condition for $f^{(p)}_C(t)$ to be Markovian
is 
\begin{eqnarray}p \geq 
\frac{ \Gamma^{NM}}{1+  \Gamma^{NM}}\;, \label{impo1T} 
\end{eqnarray} 
with $\Gamma^{NM}$ as in (\ref{DEFGAMMA}) with $\Delta^{NM}$ and $\Theta^{NM}$ 
formally given by 
\begin{eqnarray}
\Delta^{NM} = \lim_{T\rightarrow \infty}  \Delta^{NM}(T)\;, \qquad   \Theta^{NM}  = \lim_{T\rightarrow \infty}  \Theta^{NM}(T)\;. 
\end{eqnarray} 
To show that the inequality (\ref{impo1T}) is also a sufficient condition for the Markovianity of $f^{(p)}_C(t)$
we now provide an explicit example that saturates it -- in  Appendix~\ref{UNIQUE} we also prove that the solution 
we present here is also unique.

It is intuitive to understand that the function $h_C^{M}(t) \in \mathfrak{F}^M_C$ that we are looking for  must be  a combination of $g_C^M(t)$ (see Eq.~(\ref{ofMgen})) and 
$\overline{g}_C^M(t)$ (see Eq.~(\ref{ofMgenX})).
In order to simplify its complicated formulation, we express $h_C^{M}(t)$ only through its temporal derivative
\begin{equation}
\dot{h}_C^M(t^{\pm}) = \left\{
\begin{array}{ccc}
-\dot{f}^{NM}_C(t^{\pm}) / \Gamma^{NM}  & t\in T_k^+ \\ 
-\dot{f}^{NM}_C(t^{\pm}) / \Gamma^{NM}  & t\in \overline{T}_j \\
0 & \mbox{ otherwise } 
\end{array} \right. , \label{hCM} 
\end{equation}
which can be rewritten in a particularly simple form
\begin{equation}\label{hCMapp}
\dot{h}_C^M(t^{\pm}) = \left\{
\begin{array}{ccc}
-\dot{f}^{NM}_C(t^{\pm}) / \Gamma^{NM}  & \mbox{ if }  \dot{f}^{NM}_C(t)>0 \\ 
0 & \mbox{ otherwise } 
\end{array} \right. ,
\end{equation}
(see Figure \ref{zoomed} for an example).
After a long but straightforward calculation, it is possible to show that  {$f^{(p)}(t)=(1-p) f^{NM}_C(t) + p h_C^M(t)$} belongs to the Markovian set for all $p$ fulfilling (\ref{impo1T}). Therefore, this proves 
that 
\begin{eqnarray} 
p(D^{NM}_C|\mathcal{D}^M)=\frac{ \Gamma^{NM}}{1+  \Gamma^{NM}} \;, \end{eqnarray} 
and therefore (\ref{N1final}).

\section{Optimal Markovian characteristic functions for continuous non-Markovian evolutions are continuous}\label{contnoncont}

In this section we prove the identities  (\ref{IMPORTANT}) showing 
 that in the case of continuous characteristic functions
  $f_C(t)$, non-continuous Markovian characteristic functions $f^M(t)\notin \mathfrak{F}^M_C$ cannot make their
   convex combination   $f^{(p)}(t)$
  Markovian for values of $p$ {smaller} than~$p(D_C| \mathcal{D}_C^M )$.
  This is trivial if $f_C(t)$ is already Markovian as in this case  $p(D_C| \mathcal{D}_C^M )$ saturates to the {minimum} allowed value {$0$}. For characteristic
  functions which are explicitly non-Markovian 
in Sec.~\ref{Seconetime} we analyse the simple scenario of  positive 
   functions  which exhibit non-Markovianity only in a single interval. Then in Sec.~\ref{SECNONM} we discuss the case of functions that have non-Markovian
   behaviour when negative, 
    and conclude in Sec.~\ref{GENERALCASE} with the general case.

\subsection{Single time interval of non-Markovianity with $f_C^{NM}(t)\geq 0$}\label{astro1}

We start by studying the cases discussed in Sec.~\ref{Seconetime}, where $f^{NM}_C(t)$ has a single time interval $(t_1,t_2)$ of non-Markovianity when $f^{NM}_C(t)\geq 0$ and $\dot{f}^{NM}_C(t)>0$. In this case the optimal continuous Markovian function $g_C^M(t)$ which makes the corresponding $f^{(p) }(t)$ Markovian for the {smallest} $p$ is given in  
Eq.~(\ref{ofM}) and leads to
\begin{eqnarray} \label{INE1} p\geq p(D^{NM}_C|\mathcal{D}_C^M)= \frac{ \Delta^{NM}}{1+\Delta^{NM}}\;, 
\end{eqnarray}  where $\Delta^{NM}=f^{NM}_C(t_2) - f^{NM}_C(t_1)>0$. To show that Eq.~(\ref{INE1}) cannot be improved by allowing  $f^M(t)$ to be non continuous, 
we start noticing that in this scenario also
 $f^{(p)}(t)$ will be non-continuous. We distinguish then six possible cases:
  \begin{itemize} 
 \item[(i)] $f^{M}(t_1)> 0$ and $f^{M}(t_2)\geq 0$ with a discontinuity at $T\in(t_1,t_2)$; 
 \item[(ii)] $f^{M}(t_1)\geq 0$ and $f^{M}(t_2)< 0$ with a discontinuity at $T\in(t_1,t_2)$; 
 \item[(iii)] $f^{M}(t_1)<0$ and $f^{M}(t_2)\leq 0$ with $f^M(t)$ continuous in $(t_1,t_2)$;
 \item[(iv)] $f^{M}(t_1)<0$ and $f^{M}(t_2)\leq 0$ with a discontinuity at $T\in(t_1,t_2)$; 
 \item[(v)] $f^{M}(t_1)<0$ and $f^{M}(t_2) > 0$ with a discontinuity at $T\in(t_1,t_2)$; 
 \item[(vi)] $f^{M}(t_1)>0$ and $f^{M}(t_2)\geq 0$ with $f^M(t)$ 
 exhibiting discontinuities before $t_1$.
 \end{itemize}
 Notice that in the cases (iii) and (v) where  $f^M(t_1)<0$ implicitly imply  a discontinuity $\xi(f^M(T_0))\in [-1/(d^2-1),0)$ at  some $T_0< t_1$.

In case (i) we have that  at time $T\in(t_1,t_2)$ a discontinuity is shown such that $f^M(T^+) - f^M(T^-)= - \epsilon <0$, where $\epsilon \in (0,1)$. Notice that $\epsilon =1 $ implies that $f^M(T^-)=1$ and $f^M(T^+)=0$, and therefore this choice does not make sense if our purpose is to make $f^{(p)}(t)$ Markovian. Fixed this $\epsilon$-jump for $f^M(t)$, we build the optimal behavior that makes $f^{(p)}(t)$ Markovian for the {smallest} $p$ possible. Using the same technique used to obtain Eq.~(\ref{hCMapp}), we see that this function is characterized by $f^M(t_1)=1$ and $\dot{f}^M(t)={-\dot{f}^{NM}(t)(1-\overline p)/ \overline p} $ for $t\in (t_1,t_2)$  and the {smallest} value of $\overline p$ for which $f^M(t)$ is Markovian in $(t_1,t_2)$. Indeed, with this structure $f^{(p)}(t)$ is non-increasing for any {$p\geq \overline p$} and $\dot{f}^{(\overline p)}(t) =0$ for $t\in (t_1,t_2)$. By studying the condition of Markovianity $f^M(t_2)\geq 0$, we obtain
$$
\overline p \geq \frac{\Delta^{NM}/(1-\epsilon)}{1 + \Delta^{NM}/(1-\epsilon)}> p(D^M_C| \mathcal{D}_C) \, , 
$$
where the last inequality holds for any $\epsilon \in (0,1)$, i.e., for any discontinuity of this type.

Cases (ii), (iii) and (iv) can be proven to be inefficient to make $f^{(p)}(t)$ Markovian thanks to the following argument. Since $\dot{f}^{NM}(t)>0$ for $t\in (t_1,t_2)$, in order to make $f^{(p)}(t)$ Markovian, we have to require that $f^{(p)}(t_2) \leq  0 $, i.e., it has to assume the same sign of $f^M(t_2)$. It implies that 
\begin{eqnarray}p&\geq&  \frac{f^{M}(t_2) /|f^M(t_2)|}{ 1+ f^{M}(t_2) /|f^M(t_2)| } \geq  \frac{\Delta^{NM} /|f^M(t_2)| }{ 1+\Delta^{NM} /|f^M(t_2)| } \nonumber \\ 
&\geq& \frac{(d^2-1) \Delta^{NM}}{ 1+(d^2-1) \Delta^{NM} } >  p(D_C|\mathcal{D}_C^M) \, ,\end{eqnarray}
where we used $f^{NM}(t_2) \geq \Delta^{NM}$ and $|f^M(t_2)|\leq 1/(d^2-1)$. 

For case (v) we start by noticing that the discontinuity at time $T$ may lead to a non-Markovian discontinuity for $f^{(p)}(t)$. Therefore, we parametrize the discontinuity of $f^M(t)$ as follows: $f^M(T^+)= |f^M(T^-)|\, \lambda/(d^2-1)$, where $\lambda\in [0,1]$. Moreover, in order for $f^M(t)$ to make $f^{(p)}(t)$ Markovian, $f^{(p)}(T^-)<0$. Hence, $f^{(p)}(t)$ shows a Markovian discontinuity at time $t=T$ if and only if $\xi(f^{(p)}(T)) \geq -1/(d^2-1)$. This condition can be written as
\begin{equation}\label{rellambda}
\lambda \leq 1 - \frac{(1-p)d^2}{p} \frac{f_C^{NM}(T)}{|f^M(T^-)|} \, .
\end{equation}
If we consider this bound for $p=p(D^{NM}_C|\mathcal{D}_C^M)$, we have that the difference $h_C^M(T)- f^M(T^+)$ becomes 
\begin{equation}\label{jojo}
h_C^M(T)- f^M(T^+) \geq \frac{1}{\Delta^{NM}} \left(  \frac{f^{NM}_C(T)}{d^2-1}  + f^{NM}_C(t_1)\right) >0 \, ,
\end{equation}
where $h_C^M(T)=1-(f^{NM}(T)-f^{NM}(t_1))/\Delta^{NM}$ (see Eq.~(\ref{ofM})) and we used that in the optimal case $f^M(T^-)=-1/(d^2-1)$. By considering the Markovianity of $f^{(p)}(t)$ in the time interval $(T,t_2)$, the optimal strategy imposes that {$\dot{f}^M(t)=-\dot{f}^{NM}_C(t) (1-\overline p)/\overline p  $} for $t\in (T,t_2)$ and some $\overline p<1$. In analogy to what we found in case (i), Eq.~(\ref{jojo}) implies that $f^M(t)$ cannot make $f^{(p)}(t)$ Markovian for $p= p(D^{NM}_C|\mathcal{D}_C^M)$.

The last case we need to check is (vi), where $f^M(t)$ is continuous (hence non increasing) in $(t_1,t_2)$ but exhibits
some discontinuities before $t_1$. Since by construction  $f^{(p)}(t)$ is 
continuous in $(t_1,t_2)$, it can be Markovian only if it is non increasing in this interval, which in particular implies  
\begin{eqnarray}
0 &\geq& f^{(p)}(t_2^-) - f^{(p)}(t_1^+) \nonumber \\ 
&=&  (1-p) (f^{NM}(t_2)-f^{NM}(t_1)) - p (f^{M}(t_1^+)-f^{M}(t_2^-)) \nonumber \\
&=& (1-p) \Delta^{NM} - p (f^{M}(t_1^+)-f^{M}(t_2^-))\;,
\end{eqnarray} 
that leads to 
\begin{eqnarray}
p \geq  \frac{\Delta^{NM} }{f^{M}(t_1^+)-f^{M}(t_2^-) + \Delta^{NM}} { > }    
p(D^{NM}_C|\mathcal{D}_C^M) \;, \end{eqnarray}
where in the last passage we used the fact that $f^M(t)$ is positive,  continuous
in $(t_1,t_2)$ and, since it shows discontinuities before $t_1$, $f^M(t_1^+)<1$ and therefore $f^{M}(t_1^+)-f^{M}(t_2^-)\in[0,1)$.

\subsection{Single time interval of non-Markovianity with $f^{NM}(t)<0$}\label{SECNONM} 

Let consider a non-Markovian $f^{NM}_C(t)$ such that it has a single time interval of non-Markovianity $(t_1,t_2)$ when $f^{NM}_C(t)< 0$ and $\dot{f}^{NM}_C(t)<0$. An important difference from discontinuous non-Markovian characteristic functions is that $f^{NM}_C(t)$ can become negative if and only if it shows a time interval of non-Markovianity of this type. Indeed, $f^{NM}_C(t_1)=0$. Notice that in the non-continuous case a characteristic function can change its sign without being non-Markovian.

The optimal continuous Markovian characteristic function $h_C^M(t)$ is constant and equal to $1$ for any $t\in[0,t_2]$ and it decreases depending on the behavior of $f^{NM}_C(t)$ (see Eq.~(\ref{ofMgenX}) or (\ref{hCMapp}))  for $t\geq t_2$. It can make the corresponding $f^{(p)}(t)$ Markovian for {$p\geq  p(D^{NM}_C|\mathcal{D}_C^M)=|\Theta^{NM}|/(1+|\Theta^{NM}|)$}, where $\Theta^{NM}=f^{NM}_C(t_2) - f^{NM}_C(t_2) <0$.

Now we consider non-continuous Markovian characteristic functions $f^M(t)$ and we study which scenarios could {potentially} make $f^{(p)}(t)$ Markovian for some {$p<p(D^{NM}_C|\mathcal{D}_C^M)$}. We have to study the following scenarios: 
\begin{itemize}
\item[(i)] $f^M(t_2)\in(0,1)$;
\item[(ii)]
 $f^M(t)$ jumps at time
 $T\leq t_1$ to some negative value and $f^M(t_2)<0$;
 \item[(iii)] $f^M(t)$ jumps at time $T\in(t_1,t_2)$ to some negative value and $f^M(t_2)<0$.
 \end{itemize} 

In case (i) we include all those situations where $f^M(t)$ shows discontinuities with or without changes of sign for one or more times prior to $t_2$ and such that $f^M(t_2)>0$. A necessary condition for $f^M(t)$ to make $f^{(p)}(t)$ Markovian is $f^{(p)}(t_2)\geq 0$. The non-negativity of $f^{(p)}(t_2)$ holds  if and only if
$$
{p\geq  \frac{|\Theta^{NM}|/f^M(t_2)}{1+|\Theta^{NM}|/f^M(t_2)} } \, .
$$
Since $f^M(t_2)=1$ if and only if $f^M(t)=1$ for any $t\in [0,t_2]$ we have that all the $f^M(t)$ with discontinuities of this type cannot perform better than $h_C^M(t)$ in making $f^{(p)}(t)$ Markovian. 

Considering case (ii), we start by noticing that, if $f^M(t_1)<0$ and $f^M(t)$ is continuous for any $t\in(t_1,t_2)$, the optimal $f^M(t)$ of this type can make $f^{(p)}(t)$ Markovian for 
$$
{p \geq  \frac{|\Theta^{NM}|/f^M(t_1)}{1+|\Theta^{NM}|/f^M(t_1)} \geq \frac{(d^2-1)|\Theta^{NM}|}{1+(d^2-1)|\Theta^{NM}|} } {>  p(D^{NM}_C|\mathcal{D}_C^M)} \, ,
$$
{where $ p(D^{NM}_C|\mathcal{D}_C^M)=|\Theta^{NM}|/(1+|\Theta^{NM}|)$}. In the case of a discontinuity of $f^M(t)$ (without change of sign) during the time interval $(t_1,t_2)$, in analogy with case (i) of the previous section, we conclude that $f^M(t)$ cannot make $f^{(p)}(t)$ Markovian for {$p<p(D^{NM}_C|\mathcal{D}_C^M)$} also in this scenario.

In case (iii) $f^M(T^-)>0$ and $f^M(T^+)<0$ for some $T\in(t_1,t_2)$. We have to make $f^{(p)}(t)$ Markovian in $(t_1,t_2)$ and in order to obtain this result we need that $f^{(p)}(t)$ and $f^M(t)$ have the same sign. As a consequence, $f^{(p)}(t)$ shows a discontinuity at time $T$ such that $\xi(f^{(p)}(T))<0$. If we study the condition of Markovianity $\xi(f^{(p)}(T))\geq -1/(d^2-1)$, we obtain
$$
\xi(f^{(p)}(T))=\frac{(1-p) f^{NM}(T) + pf^M(T^+)}{(1-p) f^{NM}(T) + p  f^M(T^-)}
$$
\begin{equation}\label{Ebiii}
{=\frac{-(1-p)|f^{NM}(T)| -  p \lambda f^M(T^-) /(d^2-1) }{-(1-p)|f^{NM}(T)| + p f^M(T^-)  } \geq  \frac{-1}{d^2-1}  } \, ,
\end{equation}
where we used $f^{NM}(T)=-|f^{NM}(T)|$ and $|f^M(T^+)| =   f^M(T^-) \lambda/(d^2-1)$, where $\lambda\in(0,1)$. We can use Eq.~(\ref{Ebiii}) to find a $p$-dependent bound for the values of $\lambda$ that make $\xi(f^{(p)}(T)) \geq -1/(d^2-1)$. By doing so we obtain  {$\lambda \leq 1- (1-p)d^2 |f^{NM}(T)|/(p f^M(T^-))$}. Now we check if the $f^M(t)$ of this case can make $f^{(p)}(t)$ Markovian for {$p=p(D^{NM}_C|\mathcal{D}_C^M)=|\Theta^{NM}|/(1+|\Theta^{NM}|)$}. The optimal scenario is obtained when $f^{M}(T^-)=1$ and therefore we get 
$$f^M(T^+)=\frac{-\lambda}{d^2-1} \geq  \frac{-1}{d^2-1} + \frac{d^2 |f^{NM}(T)|}{(d^2-1) |\Theta^{NM}|} \, , $$
where we used {$(1-p(D^{NM}_C|\mathcal{D}_C^M))/ p(D^{NM}_C|\mathcal{D}_C^M)=1/|\Theta^{NM}|$}. The optimal behavior of $f^M(t)$ that makes the derivative $\dot{f}^{(p)}(t)\geq 0$ for the smallest increase of $f^M(t)$ in $(T,t_2)$ is achieved by considering {$\dot{f}^M(t)= -  \dot{f}^{NM}(t)(1-\overline p)/\overline p$}, for the smallest $\overline p$ that allows a Markovian $f^M(t)$. Therefore, for {$\overline p=p(D^{NM}_C|\mathcal{D}_C^M)=|\Theta^{NM}|/(1+|\Theta^{NM}|)$}, we get $\dot{f}^M(t)= -  \dot{f}^{NM}(t)/|\Theta^{NM}|$. This implies that at time $t_2$ we have
$$
f^M(t_2)\geq  \left( \frac{d^2 |f^{NM}(T)| }{(d^2-1) |\Theta^{NM}| } - \frac{1}{d^2-1} \right) + \frac{|f^{NM}(t_2)| - |f^{NM}(T)|}{|\Theta^{NM}|}
$$
\begin{equation}\label{Ebiii2}
=|f^{NM}(T)|\left( \frac{d^2}{(d^2-1)|\Theta^{NM}|} - \frac{1}{|\Theta^{NM}|} \right) + 1 - \frac{1}{d^2-1} > 0 \, ,
\end{equation}
where we used $f^{NM}(t_2)=\Theta^{NM}<0$. In summary, we proved that a $f^M(t)$ that jumps at  $T\in(t_1,t_2)$ to some negative value such that $f^{(p)}(t)$ does not show a non-Markovian jump at time $t=T$, cannot make $f^{(p)}(t)$ Markovian in the time interval $(T,t_2)$ for $p= p(D^{NM}_C|\mathcal{D}_C^M)$. Indeed, the Markovianity of $f^{p(D^{NM}_C|\mathcal{D}_C^M)}(t)$ in this time interval implies that $f^M(t_2)>0$, i.e., $f^M(t)$ should change sign while being continuous (this behavior is not allowed for Markovian characteristic functions). We underline that Markovian functions  of case (iii) can make $f^{(p)}(t)$ Markovian but only for values of $p$ {larger} than $p(D^{NM}_C|\mathcal{D}_C^M)$, i.e., by imposing {$\dot{f}^M(t)= -  \dot{f}^{NM}(t)(1-\overline{p} )/\overline{p}$} in $(T,t_2)$ with some {$\overline p > p(D^{NM}_C|\mathcal{D}_C^M)$} that allows $f^M(t_2)\leq 0$.

From the results obtained in this section it is clear that, if we add to cases (i), (ii) and (iii) any additional discontinuity in $(t_1,t_2)$, we cannot {reduce} the value of $p$ for which $f^{(p)}(t)$ can be made Markovian with a discontinuous $f^M(t)\in\mathfrak{F}^M(t)$.

\subsection{General case}\label{GENERALCASE} 

In order to prove  (\ref{IMPORTANT}) for any $D_C\in \mathcal{D}^{NM}_C$ represented by a $f^{NM}_C(t)\in \mathfrak{F}^{NM}_C$, we notice that the same technique that we used to derive the optimal continuous solution $h_C^M(t)$ given in  Eq.~(\ref{hCMapp}) can be generalized to the case where we fix the discontinuities that the Markovian characteristic function has to show. Indeed, the rules given in Eq.~(\ref{hCMapp}) can be generalized to the cases where $f^M(t)$ jumps with or without a change of sign and we obtain
\begin{equation}\label{hNC}
{h}_{NC}^M(t)=\left\{
\begin{array}{ccc}
-\dot{f}^{NM}_C(t) / \Gamma'  & \mbox{ if }  \dot{f}^{NM}_C(t)>0 \mbox{ and } h_{NC}^M(t) > 0 \\ 
0 & \mbox{ if }  \dot{f}^{NM}_C(t)\leq  0 \mbox{ and } h_{NC}^M(t) > 0 \\ 
-\dot{f}^{NM}_C(t) / \Gamma'  & \mbox{ if }  \dot{f}^{NM}_C(t)<0 \mbox{ and } h_{NC}^M(t) < 0 \\ 
0 & \mbox{ if }  \dot{f}^{NM}_C(t)\geq   0 \mbox{ and } h_{NC}^M(t) < 0
\end{array} \right. \, ,
\end{equation}
where the sign of $h_{NC}^M(t)$ depends on the discontinuities $\xi(h_{NC}^M(t))\in I_{\mathcal{D}}$ that we impose and $\Gamma'>0$ has to be chosen such that $h_{NC}^M(t)$ is Markovian and $f^{(p)}(t)$ is made Markovian for the {smallest} possible $p$. 

The main difference between $h_C^M(t)$ and $h_{NC}^M(t)$ is that $\Gamma^{NM}$ is replaced by  $\Gamma'$, which in general depends on the particular jumps that $h_{NC}^M(t)$ has to show. Notice that in the previous two sections we used {$\Gamma'=\overline p/(1-\overline p)$}. Our goal is to prove that in every scenario $\Gamma'>\Gamma^{NM}$. Indeed, $h_{NC}^M(t)$ makes $f^{(p)}(t)$ Markovian for {$p\geq \Gamma' /(1+\Gamma')=\overline p$}  {and $\Gamma'>\Gamma^{NM}$ implies that $\overline p>$} ${p(D^{NM}_C|\mathcal{D}_C^M)=\Gamma^{NM}/(1+\Gamma^{NM})}$.

We consider those cases where the discontinuities of $h_{NC}^M(t)$ does not take place during time intervals of non-Markovianity of $f^{NM}_C(t)$.  We show that, even if we ignore possible non-Markovian discontinuities of $f^{(p)}(t)$ caused by the discontinuities of $h_{NC}^M(t)$ (which may {increase the minimum} $p$ for which $f^{(p)}(t)$ can be made Markovian by $h_{NC}^M(t)$), $\Gamma'>\Gamma^{NM}$. We use the following notation for the intervals of non-Markovianity of $f^{NM}_C(t)$: the $i$-th interval $(t_i^{(in)},t_i^{(fin)})$ can either be a time interval where $f^{NM}_C(t)$ shows a non-Markovian behavior while being positive or negative. The $i$-th gap $\Gamma^{NM}_i\equiv |f^{NM}_C(t_i^{(fin)}) - f^{NM}_C(t_i^{(in)}) |>0$ is therefore the non-Markovian gap shown  in the time interval $(t_i^{(in)},t_i^{(fin)})$. Notice that $\Gamma^{NM}=\sum_i \Gamma_i^{NM}$ (see Eq.~(\ref{DEFGAMMA})).
Let start with the case of a  $h_{NC}^M(t)$ that shows a single discontinuity at time $T_1<t_1^{(in)}$, where $\xi_1=\xi(h_{NC}^M(T_1))\in \{I_\mathcal{D} \setminus 1\}$. It is easy to prove that the {minimum} probability $\overline p$ for which $h_{NC}^M(t)$ can make $f^{(p)}(t)$ Markovian satisfies the following {lower} bound  {$\overline p \geq (\Gamma^{NM}/|\xi_1|) /(1+\Gamma^{NM}/|\xi_1|)$}. Therefore, in these cases
\begin{equation}\label{unadiscantes}\Gamma'=\Gamma^{NM}/|\xi_1|>\Gamma^{NM} .
\end{equation}
Now, suppose that a discontinuity characterized by $\xi_1=\xi(h_{NC}^M(T_1))\in \{I_\mathcal{D} \setminus 1\}$ is verified for $t_{k_1}^{(fin)} \leq  T_1 \leq t_{k_1+1}^{(in)}$, i.e., between the $k_1$-th and the $k_1+1$-th non-Markovian time interval. It is easy to show that in this case 
\begin{equation}\label{unadisc} \Gamma'= \sum_{i=1}^{k_1} \Gamma_i^{NM} + \frac{\sum_{i=k_1+1}^N \Gamma_i^{NM} }{|\xi_1|}>\Gamma^{NM}  \, ,
\end{equation}
where $N$ (which may be infinite)  is the number of non-Markovianity intervals  of $f^{NM}_C(t)$.
In the case of an additional discontinuity  $\xi_2=\xi(h_{NC}^M(T_2))\in \{I_\mathcal{D} \setminus 1\}$ that is shown at time $t_{k_2}^{(fin)} \leq  T_2 \leq t_{k_2+1}^{(in)}$, we have
\begin{equation}\label{duedisc} \Gamma'= \sum_{i=1}^{k_1} \Gamma_i^{NM} + \frac{\sum_{i=k_1 +1}^{k_2} \Gamma_i^{NM} }{|\xi_1|} + \frac{\sum_{i=k_2 +1}^{N}  \Gamma_i^{NM} }{|\xi_1 \, \xi_2|  }>\Gamma^{NM} \, .
\end{equation}
We notice that, the presence of two Markovian discontinuities for $h_{NC}^M(t)$ provides a value of $\Gamma'$ that is strictly larger than the $\Gamma'$ obtained with only the first or the second discontinuity (see Eq.~(\ref{unadisc})). The generalization of Eq.~(\ref{duedisc}) to any number of this type of discontinuities is trivial. We conclude that the $h_{NC}^M(t)$ obtained by any number of discontinuities $\{\xi_j\}_j$ of this type are always characterized by $\Gamma' > \Gamma^{NM}$.

In the previous sections we proved that the presence of any discontinuity that takes place during a single time interval of non-Markovianity $(t_1,t_2)$ does not allow to make $f^{(p)}(t)$ Markovian for {$p\leq  p(D^{NM}_C|\mathcal{D}_C^M)$}. It is clear that Eq.~(\ref{hNC}) provides an optimal non-continuous Markovian solution for any set of discontinuities that takes place inside or outside the time intervals $(t_i^{(in)},t_i^{(fin)})$. Moreover, combining the previous results together we obtain  that in every scenario $\Gamma'=\overline p/(1-\overline p)$ is larger than $\Gamma^{NM}=p(D^{NM}_C|\mathcal{D}_C^M) /(1+p(D^{NM}_C|\mathcal{D}_C^M))$ hence proving Eq.~(\ref{IMPORTANT}).

\section{Interlude: a remark on a special subset of non-continuous, non-Markovian depolarizing
evolutions} \label{sec:inter} 

As we shall see in details in the next section, computing our measure of non-Markovianity for depolorazing trajectories which are explicitly non continuous is rather demanding. For this reason we find it
useful to remark that 
the construction presented in Sec.~\ref{measureofnmc}
can however be shown to generalize beyond the  domain $\mathcal{D}_C^{NM}$ allowing us to
compute  $p(D^{NM}|\mathcal{D}_C^M,\mathcal{D}^M)$ at least for some non continuous elements $D^{NM}$.

\subsection{Non-Markovian characteristic functions with Markovian discontinuities}
In particular,  following the same approach we used in Sec.~\ref{Seconetime},  
the function $g_C^{M}(t)$ of  Eq.~(\ref{ofM}) can be shown to provide the optimal choice for the computation of 
 $p(D^{NM}|\mathcal{D}_C^M,\mathcal{D}^M)$ for the whole set of 
 non-Markovian evolutions $D^{NM}\in \mathcal{D}^{NM}$ 
with characteristic functions of the form \begin{equation}
\left\{
\begin{array}{ll}
f^{NM}(t)\geq 0, \dot{f}^{NM}(t^{\pm})\leq  0, \xi(f^{NM}(t))\in [0,1] &  t<t_1^{(in)} \\ \\
f^{NM}(t) \geq 0 , \, \dot{f}^{NM}(t^{\pm}) >0, \xi(f^{NM}(t))=1&  t\in T_1^+ \\ \\
\mbox{\qquad \qquad \qquad  ``Markovian''} & t>t_1^{(fin)}\;.
\end{array}
\right.  \label{EXTENDE1} 
\end{equation}
Notice that differently from the case addressed in Eq.~(\ref{fnmc1}) this new set of functions (i)  can show Markovian discontinuities without changing their sign for any $t<t_1^{(in)}$, and (ii) can follow any behaviour allowed by the Markovian conditions (see Eq.~(\ref{MarkD})), even changing sign, for $t>t_1^{(fin)}$.  
 Since $\mathcal{D}^M$ is non-convex (see Section~\ref{example1}),  the mixture between $f^{NM}(t)$ and $g_C^M(t)$ may in principle make $f^{(p)}(t)$ non-Markovian for one or more times when $f^{NM}(t)$ behaves as a Markovian characteristic function. Nonethelss, this is not the case. Indeed, for $t>t_1^{(fin)}$, we have $g_C^M(t)=0$ and therefore {$f^{(p)}(t)=(1-p) f^{NM}(t)$} is always Markovian. Instead, for $t<t_1^{(in)}$, since $g_C^M(t)$ and $f^{NM}(t)$ are positive,  $f^{(p)}(t)$ cannot behave as a non-Markovian characteristic function.
As a result of this observation one has that for the functions of the form~(\ref{EXTENDE1}) we have 
\begin{eqnarray} 
p(D^{NM}|\mathcal{D}_C^M,\mathcal{D}^M)= \frac{\Delta^{NM}_1}{1+\Delta^{NM}_1} \;, 
\end{eqnarray} 
with $\Delta^{NM}_1$ being the gap associated with the non-Markovian character of the function on $T_1^+$.

Analogously 
 the function $g_C^M(t)$ given in Eq.~(\ref{ofMgen}) 
can be shown to provide the value of $p(D^{NM}|\mathcal{D}_C^M,\mathcal{D}^M)$ 
also for the following class of not necessarily continuous, non-Markovian characteristic functions $f^{NM}(t)$ of the form
\begin{equation}\label{genmulttext}
\left\{
\begin{array}{ll}
f^{NM}(t)\geq 0, \dot{f}^{NM}(t)\leq  0, \xi(f^{NM}(t))\in [0,1] &  t\notin T ^{NM} \\ \\
f^{NM}(t) \geq 0 , \, \dot{f}^{NM}(t) >0, \xi(f^{NM}(t))=1&  t\in T ^{NM} \\ \\
\mbox{\qquad \qquad \qquad   ``Markovian''} & t>t_N^{(fin)}\;, 
\end{array}
\right. 
\end{equation}
where, if $t_N^{(fin)}<\tau$ for some $\tau>0$, the latter of Eq.~(\ref{genmulttext}) is the condition that we consider for $t>t_N^{(fin)}$. Therefore, also for the depolarizing evolutions $D^{NM}$ defined by Eq.~(\ref{genmulttext}), we have 
\begin{eqnarray} 
p(D^{NM}|\mathcal{D}_C^M,\mathcal{D}^M)=\frac{\Delta^{NM}}{1+\Delta^{NM}} \;.
\end{eqnarray}
By the same token  one can show that  $h_C^M(t)$ of Eq.~(\ref{hCM}) yields the measure of non-Markovianity 
{$p(D^{NM}|\mathcal{D}_C^M,\mathcal{D}^M)$}  also for the class of characteristic functions of the form   
\begin{equation}\label{genmult}
\left\{
\begin{array}{lc}
f^{NM}(t)\geq 0, \dot{f}^{NM}(t)\leq  0, \xi(f^{NM}(t))\in [0,1] &  t\notin T ^{NM} \\ \\
f^{NM}(t)\leq 0, \dot{f}^{NM}(t)\geq  0, \xi(f^{NM}(t))=1 &  t\notin T ^{NM} \\ \\
f^{NM}(t) \geq 0 , \, \dot{f}^{NM}(t) >0, \xi(f^{NM}(t))=1 &  t\in T ^{NM} \\ \\
f^{NM}(t) \leq  0 , \, \dot{f}^{NM}(t) < 0, \xi(f^{NM}(t))=1 &  t\in T ^{NM} \\ \\
\mbox{\qquad \qquad \qquad  ``Markovian''} & t>t^{(fin)}\;, 
\end{array}
\right. 
\end{equation}
with $T^{NM}=(\cup_k T_k^+) \cup (\cup_j T_j^-)$ being the same intervals defined in Sec.~\ref{nposnegtech}
and where, if there exits a time $t^{(fin)}$ such that $f^{NM}_C(t)$ does not show any non-Markovian behavior for $t\geq t^{(fin)}$, the last condition replaces the first two for $t\geq t^{(fin)}$. 
In this case we get 
\begin{eqnarray} 
{p(D^{NM}|\mathcal{D}_C^M,\mathcal{D}^M)=\frac{\Gamma^{NM}}{1+\Gamma^{NM}} \;,}
\end{eqnarray} 
where again $\Gamma^{NM}$ is defined as in (\ref{DEFGAMMA}).

\section{Non-continuous depolarizing evolutions}\label{SECV}

Extending the results of the previous sections to the general case
 of non-Markovian depolarizing evolutions $D^{NM}$ which are not necessarily continuous 
 is rather complex. This has to due with the fact that in computing {$p(D^{NM}|\mathcal{D}^M)$} 
 we have to perform an optimization with respect to all the elements  of $\mathcal{D}^M$, 
  which as discussed in 
 Sec.~\ref{example222} is not convex.  As we shall see in Sec.~\ref{AMB} this introduces
 an ambiguity in the definition of the optimal Markovian element which is hard to handle. 
 Nonetheless in Sec.~\ref{SECVI} we propose a solution to the problem which, even though
 does not allow to derive a closed formula for {$p(D^{NM}|\mathcal{D}^M)$}  leads in principle to the exact
 results for any assigned element of $\mathcal{D}^{NM}$.

Before entering into the details of the analysis we define two sets of times: $W_{C}$ is the set of times when $f^{NM}(t)$ is continuous, namely $\xi(f^{NM}(t))=1$ if and only if $t\in  W_C$ and $W_{NC} \equiv \{t_{NC,i}\}_i= \mathbb{R}^+ \setminus  W_C$ is the discrete set of times when $f^{NM}(t)$ is discontinuous, namely $\xi(f^{NM}(t))\neq 1$ if and only if $t\in W_{NC} $. Moreover, we divide $W_{NC} $ in $W_{NC}^{M}$ and $W_{NC}^{NM}$, namely the times when $f^{NM}(t)$ shows Markovian ($\xi(f^{NM}(t_{NC,i}^{M}))\in I_\mathcal{D}$) and non-Markovian ($\xi(f^{NM}(t_{NC,i}^{NM}))\notin I_\mathcal{D}$) discontinuities, respectively.

\subsection{Ambiguity for the choice of the optimal Markovian evolution}\label{AMB} 

In Section~\ref{measureofnmc}, while evaluating {the measure of non-Markovianity} {$p(D_C^{NM}|\mathcal{D}^M)$} for continuous evolutions, we never assumed any particular shape for $f^{NM}_C(t)$ and $\dot f^{NM}_C(t)$ in order to provide the the optimal $f^M_C(t)$ needed to calculate this measure. In the following example, instead, we show that for non-continuous evolutions there is an ambiguity for the choice of the times when the optimal $f^M(t)$ shows discontinuities.
This ambiguity is solved only if we know exactly the shape of $f^{NM}(t)$. Moreover, in these cases the value of the measure of non-Markovianity does not depend solely from $\Gamma^{NM}$.

We consider the non-Markovian characteristic function for qubits  $f_\Theta^{NM}(t) \in\mathfrak{F}^{NM}$ with a single Markovian discontinuity at time $t_{NC}$, i.e., $W_{NC}^{M}=\{t_{NC}\}$, and  a single time interval of non-Markovianity $T^-=(t^{(in)},t^{(fin)})$ when the characteristic function and its time derivative are negative.  More in details
\begin{equation}
f^{NM}_\Theta(t) = \left\{
\begin{array}{cc}
1 & t\in[0,t_{NC}] \\ 
-1/3 & t\rightarrow t_{NC}^+ \\ 
f^{NM}_\Theta(t) \leq 0 , \dot{f}^{NM}_\Theta(t) \geq 0 & t\in [t_{NC},t^{(in)}] \\ 
\Theta- 1/3 & t=t^{(in)} \\ 
f^{NM}_\Theta(t) \leq 0 , \dot{f}^{NM}_\Theta(t) < 0 & t\in(t^{(in)},t^{(fin)}) \\ 
-1/3 & t\geq t^{(fin)} 
\end{array} \right. \, ,
\end{equation}
where $\Theta\in (0,1/3]$. 
It is clear that this function is characterized by a null positive non-Markovian gap $\Delta^{NM}=0$ and a negative non-Markovian gap $\Theta^{NM}=-\Theta$ that is shown in the time interval $T^-=(t^{(in)},t^{(fin)})$. This example can be easily generalized to the qudit case: if we have a $d$-dimensional system, we have to replace the following conditions $f^{NM}_\Theta (t_{NC}^+)=-1/(d^2-1)$, $f_\Theta^{NM}(t)=-1/(d^2-1)$ for any $t\geq t^{(fin)}$, $f_\Theta^{NM}(t^{(in)})=\Theta-1/(d^2-1)$ and $\Theta\in(0,1/(d^2-1)]$.

\begin{figure}
\includegraphics[width=0.5\textwidth]{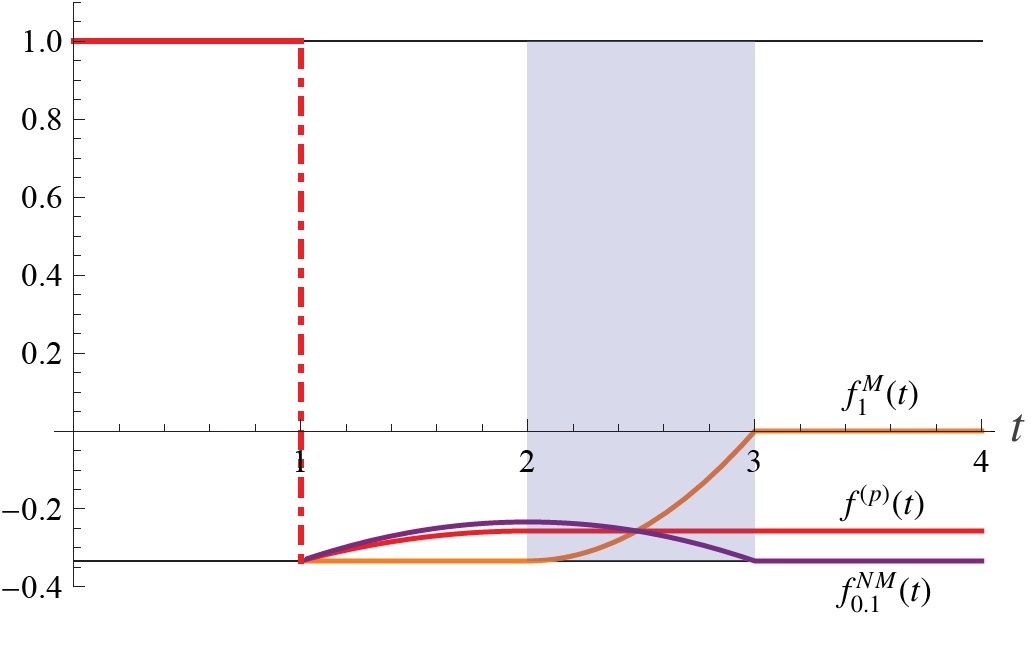}
\caption{Plots of $f^M_1(t)$, $f^{NM}_\Theta(t)$  and $f^{(p)}(t)$ for a non-Markovian gap $\Theta=-\Theta^{NM}=0.1$ and $p=p(D^{NM}_{\Theta}|D)\simeq 0.77$. The time interval of non-Markovianity $T^-=(2,3)$ of $f^{NM}_{0.1}(t)$ is colored in purple. Since $\Theta<1/6$, the optimal Markovian characteristic function is  $f^M_1(t)$. }\label{theta01}
\includegraphics[width=0.5\textwidth]{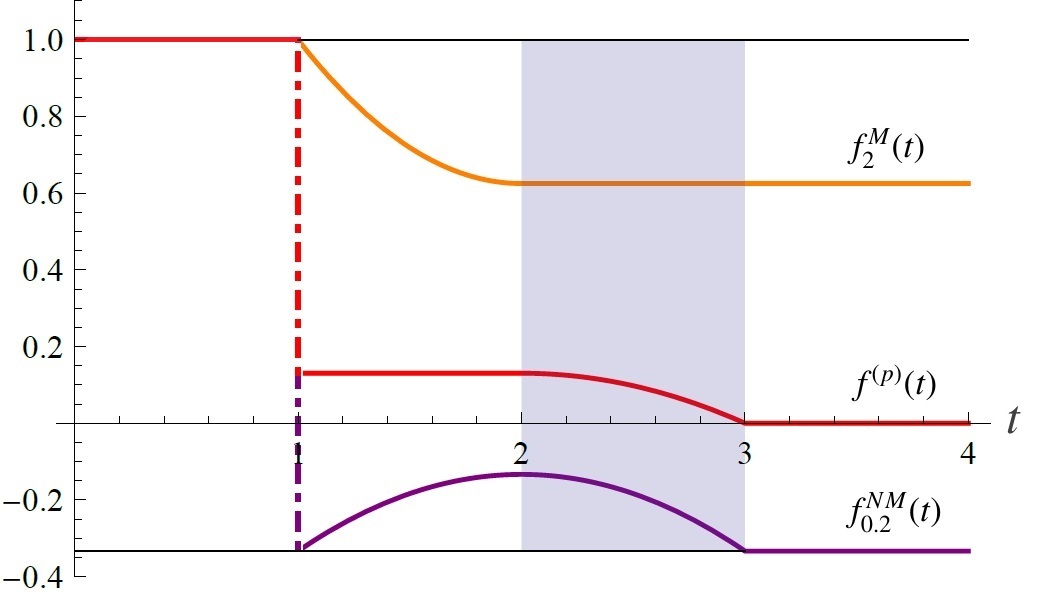}
\caption{Plots of $f^M_2(t)$, $f^{NM}_\Theta(t)$  and $f^{(p)}(t)$ for a non-Markovian gap $\Theta=-\Theta^{NM}=0.2$ and $p=p(D^{NM}_{\Theta}|D)\simeq 0.65$. The time interval of non-Markovianity $T^-=(2,3)$ of $f^{NM}_{0.2}(t)$ is colored in purple. Since $\Theta>1/6$, the optimal Markovian characteristic function is  $f^M_2(t)$.  }\label{theta02}\end{figure}

We can adopt two inequivalent $f^{M,1}(t)$ and $f^{M,2}(t)$ in order to make {$f^{(p)}(t)=(1-p)f^{NM}_\Theta(t) + pf^{M}(t)$} Markovian. We show that the {form} of the optimal Markovian characteristic function needed to evaluate {$p(D^{NM}_\Theta|\mathcal{D}^M)$} depends on the particular value of $\Theta$. Indeed, consider
\begin{equation}
f^{M,1}(t) = \left\{
\begin{array}{cc}
1 & t\in[0,t_{NC}] \\ 
-1/3 & t\in(t_{NC},t^{(in)}] \\ 
f^{M,2} (t) \leq 0 , \dot{f}^{M,2}(t) > 0 & t\in (t^{(in)},t^{(fin)}] \\ 
0 & t \geq  t^{(fin)} 
\end{array} \right. \, ,
\end{equation}
or
\begin{equation}
f^{M,2}(t) = \left\{
\begin{array}{cc}
1 & t\in[0,t^{(in)}] \\ 
f^{M,1} (t) > 0 , \dot{f}^{M,1}(t) < 0 & t\in (t^{(in)},t^{(fin)}] \\ 
f^{M,1} (t) > 0 , \dot{f}^{M,1}(t) = 0& t\geq t^{(fin)} 
\end{array} \right. \, ,
\end{equation}
where, when the time derivative of the characteristic funciton is different from zero, we impose it to be equal to $-f^{NM}_\Theta(t)/\Delta^{eff}_1$ and $-f^{NM}_\Theta(t)/\Delta^{eff}_2$, respectively.
In Fig. \ref{theta01} and \ref{theta02} we provide an example of this situation. We find that $f^{(p)}(t)$ can be made Markovian for
\begin{itemize}
\item {$p\geq \frac{3\Theta}{1+3 \Theta }$}, if we consider $f^{M,1}(t)$ with $\Delta^{eff}_1=3 \Theta$;
\item {$p\geq \frac{1/3+\Theta}{4/3+\Theta}$}, if we consider $f^{M,2}(t)$ with $\Delta^{eff}_2=\Theta+\frac{1}{3}$.
\end{itemize}
It follows that, depending on the value of $\Theta\in(0,1/3]$, the optimal Markovian characteristic function needed to evaluate the measure of non-Markovianity is different, namely it is $f^{M,1}(t)$, if $\Theta \in (0,1/6]$ and $f^{M,2}(t)$, if $\Theta \in [1/6,1/3]$.
As a consequence 
\begin{equation}
p(D_\Theta^{NM}| \mathcal{D}) = \left\{
\begin{array}{cc}
{\frac{3\Theta}{1+3 \Theta}} & \Theta \in (0,\frac{1}{6}] \\
{\frac{1/3+\Theta}{4/3 + \Theta}} & \Theta \in [\frac{1}{6},\frac{1}{3}]
\end{array} \right. \, .
\end{equation}

We notice that, differently from the continuous case, given the signs of $f^{NM}(t)$ and $\dot{f}^{NM}(t)$, it is not possible  to know a priori which are the signs of the optimal $f^{M}(t)$ and $\dot{f}^{M}(t)$  that make $f^{(p)}$ Markovian for the {smallest} value of $p$. Indeed, we have to consider all the possible alternatives for the optimal $f^M(t)$ and evaluate the {minimum} $p$ for which each one make the corresponding $f^{(p)}(t)$ Markovian. This ambiguity is generated by the sign that we decide to assign to $f^M(t)$ during its evolution. Notice that in the continuous case $f^{M}_C(t)$ could not change its sign and we had no ambiguity in the definition of the optimal Markovian characteristic function.
For instance, as we concluded studying $f^{NM}_\Theta(t)$, the difference between $f^{M,1}(t)$ and $f^{M,2}(t)$ is obtained solely by the choice of making the Markovian characteristic function change its sign at time $t_{NC}$ with a discontinuity or not. The remaining part of their definitions are analogous to the optimal solution obtained for continuous evolutions {(see Eq.~(\ref{hNC}))}

In the following, we describe how to evaluate the measure of non-Markovianity for generic non-Markovian depolarizing evolutions, where we pay particular attention to all the possible choices for the signs of the Markovian characteristic function during its evolution.

\subsection{Measure of non-Markovianity for non-continuous depolarizing evolutions}\label{SECVI}

In this section we propose a technique to evaluate the measure of non-Markovianity for any non-Markovian depolarizing channel. For this purpose, we collect the results of the previous sections in order to find a strategy that singles out the optimal $D^{M}$ needed to evaluate this measure.

Given the previous results, we consider two rules
\begin{itemize}
\item If $t' \in W_C$, the $f^{M}(t)$ that are discontinuous at $t=t'$ do not provide larger values of $p$ (if compared with the $f^{M}(t)$ that are continuous for $t=t'$);
\item If $t' \in W_{NC}$, the $f^{M}(t)$ that are discontinuous at $t=t'$ may provide larger values of $p$.
\end{itemize} 
Therefore, the optimal Markovian evolution that we need to evaluate {$p(D^{NM}|\mathcal{D}^M)$} is continuous at least for any $t\in W_C$.

\subsubsection{Vector of signs} 

We define $T_{C,i}=(t_{NC,i-1},t_{NC,i})$ to be the time intervals defined between the times in $W_{NC}=\{t_{NC,i}\}_{i=1}^N$, where we fix $t_{NC,0}=0$ and, if $N$ is finite, $t_{NC,N+1}=\infty$. With this procedure we define $N+1$ time intervals such that $\cup_i T_{C,i}=W_{C}$.

 We  consider a dichotomic variable $\sigma_i\in \{-1,1\}$ that we attach to each time interval $T_{C,i}$. Therefore, we obtain a vector $\boldsymbol{\sigma}=(\sigma_1,\sigma_2,\dots)$ of values equal to +1 or -1. We have a countable number of combinations for this vector. We label each combination $\boldsymbol{\sigma}_a=(\sigma_{a,1},\sigma_{a,2},\dots)$ with a different value of an integer number $a=1,2,\dots$. We impose $\sigma_{a,0}=+1$ for each combination and we fix a labeling scheme,   for instance
\begin{eqnarray*}
\boldsymbol{\sigma}_1=(+1,+1,+1,+1,\dots),& \,\,\, \boldsymbol{\sigma}_5=(+1,+1,+1,-1,\dots) &\, , \nonumber \\ 
\boldsymbol{\sigma}_2=(+1,-1,+1,+1,\dots),&\,\,\, \boldsymbol{\sigma}_6=(+1,-1,+1,-1,\dots)& \, , \nonumber \\ 
\boldsymbol{\sigma}_3=(+1,+1,-1,+1,\dots),&\,\,\, \boldsymbol{\sigma}_7=(+1,+1,-1,-1,\dots)&\, , \nonumber \\ 
 \boldsymbol{\sigma}_4=(+1,-1,-1,+1,\dots),&\,\,\, \boldsymbol{\sigma}_8=(+1,-1,-1,-1,\dots)& \, , \, \dots \nonumber 
\end{eqnarray*}

We call each $\boldsymbol{\sigma}_a$ a \textit{vector of signs} for the following reason. We call $f^{M}_a(t)$ the Markovian characteristic functions such that their sign is defined by $\boldsymbol{\sigma}_{a}$ as follows
\begin{equation}\label{faM}
\sign(f_a^{M}(t))=\left\{
\begin{array}{cc}
\sigma_{a,1}=+1 & t\in[0,t_{NC,1}] \\ 
\sigma_{a,2} & t\in (t_{NC,1},t_{NC,2}] \\ 
\sigma_{a,3} & t\in (t_{NC,2},t_{NC,3}] \\
\dots & \dots  
\end{array} \right. \, .
\end{equation}
We underline that, as noticed in Section~\ref{secMNMdep}, a Markovian characteristic function can change its sign only with discontinuities such that $\xi(f^M(t))\in[-1/(d^2-1),0)$.
Indeed, we imposed that  $f_a^M(t)$ is continuous at least for any $t\in W_C$. Indeed, $f^M_a(t)$  can show a discontinuity only when $f^{NM}(t)$ shows a discontinuity. Therefore,
\begin{itemize}
\item $\sigma_{a,i}=\sigma_{a,i+1}$: $f^M_a(t)$ can either be continuous or show a discontinuity at $t=t_{NC,i}$;
\item $\sigma_{a,i}=- \sigma_{a,i+1}$: $f_a^M(t)$ must show a discontinuity  $\xi(f^M(t_{NC,i}))\in [-1/(d^2-1),0)$ while it changes sign.
\end{itemize}
The Markovian characteristic functions with these features define the set $\mathfrak{F}^M_a$.

Consider the convex sum {$f^{(p)}(t)=(1-p) f^{NM}(t)+ p f^M_a(t)$}. First, it is continuous for any $t\in W_C$. Second, if it is Markovian for some $p$ and $f^M_a(t)$, it also has to belong to $\mathfrak{F}^M_b$ for some vector of signs $\boldsymbol{\sigma}_b$, namely such that $\sign(f^{(p)}(t))=\sigma_{b,i}$ for any $t\in T_{C,i}$. Notice that $\boldsymbol{\sigma}_b$ may be different from $\boldsymbol{\sigma}_a$. Therefore,  in order to obtain {$p(D^{NM}|\mathcal{D}^M)$} we proceed as follows. 
We fix a vector $\boldsymbol{\sigma}_a$ for $f^M_a(t)$ and we make $f^{(p)}(t)\in\mathfrak{F}^M_b$ for the {smallest} $p$
\begin{equation}\label{pab}
p_{a,b}\equiv \min \{p\,|\, \exists f^M_a(t)\in\mathfrak{F}^M_a \mbox{ s.t. } f^{(p)}(t) \in \mathfrak{F}^M_b\} \, ,
\end{equation}
Therefore, we get
\begin{equation}\label{pabfin}
p(D^{NM}|\mathcal{D}^M)=\min_{a,b} p_{a,b} \, .
\end{equation}
The procedure to evaluate $p_{a,a}$ is given in Section~\ref{secaa}, while the evaluation of $p_{a,b}$ for $a\neq b$ is given in Appendix \ref{aneqb}. In both cases, we simplify the {minimization} over a functional space given in Eq.~(\ref{pab}) with a {minimization} over a discrete set of real parameters.

\subsubsection{Optimal Markovian function for a generic vector of signs}\label{secaa}

In this section we evaluate $p_{a,a}$. Therefore, we fix a generic vector of signs $\boldsymbol{\sigma}_a$ that describes the signs of $f^M_a(t)$ and $f^{(p)}(t)$, namely $\sign(f^M_a(t))=\sign(f^{(p)}(t))=\sigma_{a,i}$ for any $t\in T_{C,i}$.

A generic $f^{NM}(t)\in \mathfrak{F}^{NM}$ is characterized by:
\begin{itemize}
\item Time intervals $T_{C,i}=(t_{NC,i-1},t_{NC,i})$ when $f^{NM}(t)$ is continuous, namely $\cup_i T_{C,i}=W_C$. 
\item Discrete set of times $W_{NC}^{M}=\{t_{NC,i}^{M}\}_i$ when $f^{NM}(t)$ shows Markovian discontinuities $\xi(f^{NM}(t))\in I_\mathcal{D}$ for any $t\in W_{NC}^{M}$. We define $W_{NC}=W_{NC}^{NM}\cup W_{NC}^M$.
\item Discrete set of times $W_{NC}^{NM}=\{t_{NC,i}^{NM}\}_i$ when $f^{NM}(t)$ shows non-Markovian discontinuities $\xi(f^{NM}(t))\notin I_\mathcal{D}$ for any $t\in W_{NC}^{NM}$.
\end{itemize}
 Our goal is not only to make $f^{(p)}(t)$ Markovian during the times when $f^{NM}(t)$ behaves as a non-Markovian characteristic function, but we also have to take care of the possible non-Markovianity generated from the convex sum of two characteristic functions, namely $f^{NM}(t)$ and $f^{M}_a(t)$, that for for some times behave as Markovian functions (see the example in Section~\ref{example1}). 

We adopt the following strategy. First, we generalize the technique introduced in Section~\ref{measureofnmc} in order to make $f^{(p)}(t)$ behave as a Markovian characteristic function for any $t\in W_C$  (Section~\ref{ToC}). Second, we make sure not to generate non-Markovianity for those times $t\in W_{NC}^M$ when $f^{NM}(t)$ shows Markovian discontinuities (Section~\ref{secMD}). Finally, we study the cases of those times $t\in W_{NC}^{NM}$ when $f^{NM}(t)$ shows non-Markovian discontinuities (Section~\ref{secNMD}).

\paragraph{Times of continuity:--}\label{ToC}

Consider those times $t\in W_{C}$ when $f^{NM}(t)$ is continuous. Following what we saw in Section~\ref{nposnegtech}, it is straightforward to obtain the behavior of the optimal $f^M_a(t)$ that allows to obtain $p_{a,a}$. The definition of $f^M_a(t)$ has to change depending on (i) the Markovian/non-Markovian behavior of $f^{NM}(t)$ at time $t$, (ii) the sign of $f^{NM}(t)$ at time $t$ and (iii) the sign of $f^{M}_a(t)$ at time $t$.  
Therefore, we focus on a generic $T_{C,i}=(t_{NC,i-1},t_{NC,i})$ when $\sign(f^M_a)=\sigma_{a,i}$. Then, the definition of the time derivative of $f^M_a(t)$ is given in Table \ref{fMaC}. 
The adopted strategy has the following purpose. We have $\dot{f}^M_a(t)=0$ for all those times when a non-zero derivative is not needed to make $f^{(p)}(t)$ Markovian. This strategy cannot be used when the sign of the time derivative of $f^{NM}(t)$ is such that $\sign(\dot{f}^{NM}(t))\sign(f^{{(p)}}_a(t))=+1$. Indeed, if $\dot{f}^M_a(t)=0$, $\sign(\dot{f}^{(p)}(t))\sign(f^{{(p)}}_a(t))=+1$ and $f^{(p)}(t)$ would not satisfy the first Markovian condition (\ref{MarkD}). 
The condition  $\dot{f}^{(p)}(t)=0$ is given in analogy to the continuous case. In order to apply it, we  introduce a parameter $\Delta>0$ as follows: $\dot{f}^M_a(t)=-\dot{f}^{NM}(t)/\Delta$ \cite{foot1}, which indeed makes $f^{(p)}(t)$ Markovian in these time intervals for {$p\geq  \Delta/(1+\Delta)$}. We notice that not all values of $\Delta>0$ are allowed. Indeed, if $\Delta$ is not large enough, $f^M_a(t)$ could violate the Markovian conditions of Eq.~(\ref{MarkD}). The introduction of this parameter imposes to consider $f^M_a(t)$ as a function of $t$ and $\Delta$: 
\begin{equation}\label{afterDelta}
f^M_a(t)=f^M_a(t , \Delta) \, .
\end{equation}
 If not  necessary, we omit this dependence on $\Delta$.

\begin{table}[t]
\begin{tabular}{l || l | l}
$\boldsymbol{t\in T_{C,i} }$ & $t\in T^M$ & $t\in T^{NM}$    \\ \hline  \hline 
$\sign(f^{NM}(t))=\sigma_{a,i}$ & $\dot{f}^M_a(t)=0$  &  $\dot{f}^{(p)}(t)=0 $  \\ \hline
$\sign(f^{NM}(t))=-\sigma_{a,i}$ & $\dot{f}^{(p)}(t)=0$   &  $\dot{f}^M_a(t)=0$   
\end{tabular}
\caption{The conditions for time derivative of the optimal $f^M_a(t)$ for $t\in T_{C,i}$ depends on  $\sigma_{a,i}$, $f^{NM}(t)$ and $\dot{f}^{NM}(t)$. $T^{M}$ ($T^{NM}$) is the set of times when $f^{NM}(t)$ behaves as a Markovian (non-Markovian) characteristic function.}\label{fMaC}
\end{table}

\paragraph{Markovian discontinuities:--}\label{secMD}
In this section we define the behavior of the optimal $f^M_a(t)$ for those times when $f^{NM}(t)$ shows Markovian discontinuities, namely we consider times $t_{NC,i}\in W_{NC}^{M}$ such that $\xi(f^{NM}(t_{NC,i}))\in I_\mathcal{D}$. Having fixed $\boldsymbol{\sigma}_{a}=(\sigma_{a,1},\dots, \sigma_{a,i},\sigma_{a,i+1}, \dots)$, we know the sign of $f^{M}_a(t)$ and $f^{(p)}_a(t)$  before and after $t_{NC,i}$. Moreover, we need to decide what value has to assume $f^M_a(t_{NC,i}^+)$, while we consider $f^M_a(t_{NC,i}^-)$ fixed by its behavior in the time interval $T_{C,i}=(t_{NC,i-1},t_{NC,i})$. 

\begin{table}[t]
\centering
\begin{tabular}{l ||  l | l  }
$\boldsymbol{t\in W_{NC}^M}$ &  $
\begin{array}{c}
\sigma_{a,i}=+1 \\ 
\sigma_{a,i+1}=+1
\end{array}
$ & $
\begin{array}{c}
\sigma_{a,i}=+1 \\ 
\sigma_{a,i+1}=-1
\end{array}
$  \\ \hline \hline
$
\begin{array}{c}
\sign{(f^{NM}(t_{NC,i}^-))}=+1 \\ 
\sign{(f^{NM}(t_{NC,i}^+))}=+1
\end{array}
$ & \hspace{0.6cm} (a) & \hspace{0.6cm} (b) \\ \hline
$
\begin{array}{c}
\sign{(f^{NM}(t_{NC,i}^-))}=+1 \\ 
\sign{(f^{NM}(t_{NC,i}^+))}=-1
\end{array}
$ & \hspace{0.6cm} (a) &  \hspace{0.6cm}  (b) \\ \hline
$
\begin{array}{c}
\sign{(f^{NM}(t_{NC,i}^-))}=-1 \\ 
\sign{(f^{NM}(t_{NC,i}^+))}=+1
\end{array}
$  & \hspace{0.6cm} (c)& \hspace{0.6cm} (d)\\ \hline
$
\begin{array}{c}
\sign{(f^{NM}(t_{NC,i}^-))}=-1 \\ 
\sign{(f^{NM}(t_{NC,i}^+))}=-1
\end{array}
$   &\hspace{0.6cm} (c) & \hspace{0.6cm}  (d)
\end{tabular}
\caption{Discontinuities of $f^M_a(t)$ depending of $\sigma_{a,i}$, $\sigma_{a,i+1}$, $\sign{(f^{NM}(t_{NC,i}^-))}$ and $\sign{(f^{NM}(t_{NC,i}^+))}$ in the case that $t_{NC,i}$ is a Markovian discontinuity for $f^{NM}(t)$. The remaining combinations are obtained by flipping all the signs of this table, where the optimal strategies are the same.}\label{fMaNCM}
\end{table}

If $\sigma_{a,i}=\sigma_{a,i+1}$, for $f^{M}_a(t)$ the time $t=t_{NC,i}$ can  be either (i) a time of continuity $\xi(f^M_a(t))=1$ or (ii)  a time of discontinuity when it does not change its sign, namely $\xi(f^{M}_a(t_{NC,i}))\in[0,1)$ \cite{foot2}.
Instead, if $\sigma_{a,i}=-\sigma_{a,i+1}$, for $f^{M}_a(t)$ the time $t=t_{NC,i}$ is a time of (Markovian) discontinuity $\xi(f^{NM}(t))\in [-1/(d^2-1),0)$ when its sign changes.

Straightforward counts show that, if the starting sign of $f^{NM}(t)$ and $f^M_a(t)$ are the same and they are both showing a Markovian discontinuity, $f^{(p)}(t)$ shows a Markovian discontinuity independently from their final signs. In order to illustrate the discontinuities that $f^M_a(t)$ has to show for any combination of $\sigma_{a,i}$, $\sigma_{a,i+1}$, $\sign{(f^{NM}(t_{NC,i}^-))}$ and $\sign{(f^{NM}(t_{NC,i}^+))}$, we follow the scheme of Table \ref{fMaNCM}.
\begin{itemize}
\item[(a)]  $f^M_a(t)$ preserves its sign and, indipendently from the final value and sign of $f^{NM}(t_{NC,i}^+)$, the time $t_{NC,i}$ is not a non-Markovian discontiuity for $f^{(p)}(t)$. Therefore, the best strategy is to consider  $f^M_a(t_{NC,i})$ continuous: $\xi(f^M_a(t_{NC,i}))=1$.
\item[(b)] Similarly to (a),  $t_{NC,i}$ is never a non-Markovian discontiuity for $f^{(p)}(t)$. Since $f^M_a(t)$ has to change sign, the best strategy is to maximize the final distance from zero. Therefore, we impose $\xi(f^M_a(t_{NC,i}))=-1/(d^2-1)$.
\item[(c)] $\xi(f^M_a(t_{NC,i}))=1$ implies a non-Markovian discontinuity for $f^{(p)}(t_{NC,i})$ for any {$p<1$}. Since $\xi(f^M_a(t_{NC,i}))<1$ makes $f^M_a(t_{NC,i})$ and $f^{(p)}(t_{NC,i})$ closer to zero, we need the minimal intervention to make $f^{(p)}(t)$ Markovian and positive. {Due to this ambiguity}, we introduce the parameter $\Xi_i=\xi^{M}(f^{M}_a(t_{NC,i}))\in[0,1)$  \cite{foot3}.

\item[(d)]  $\xi(f^M_a(t))=-1/(d^2-1)$ implies  $\xi(f^{(p)}(t_{NC,i})) <-1/(d^2-1)$ for any {$p<1$}. In this case, we introduce the parameter $\Xi_i=\xi^{M}(f^{M}_a(t_{NC,i}))\in(-1/(d^2-1),0]$.
\end{itemize}

Therefore, these conditions fix  the behavior of $f^M_a(t)$ when  $f^{NM}(t)$ shows a Markovian discontinuity.  

\paragraph{Non-Markovian discontinuities:--}\label{secNMD}
In this section we define the behavior of the optimal $f^M_a(t)$ for those times when $f^{NM}(t)$ shows non-Markovian discontinuities, namely we consider times $t_{NC,i}\in W_{NC}^{NM}$ such that $\xi(f^{NM}(t_{NC,i}))\notin I_\mathcal{D}$. Having fixed $\boldsymbol{\sigma}_{a}=(\sigma_{a,1},\dots, \sigma_{a,i},\sigma_{a,i+1}, \dots)$, we know the sign of $f^{M}_a(t)$ and $f^{(p)}_a(t)$  before and after $t_{NC,i}$. Moreover, we need to decide what value has to assume $f^M_a(t_{NC,i}^+)$.

\begin{table}[t]
\centering
\begin{tabular}{l ||  l | l  }
$\boldsymbol{t\in W_{NC}^{NM}}$ &  $
\begin{array}{c}
\sigma_{a,i}=+1 \\ 
\sigma_{a,i+1}=+1
\end{array}
$ & $
\begin{array}{c}
\sigma_{a,i}=+1 \\ 
\sigma_{a,i+1}=-1
\end{array}
$  \\ \hline \hline
$
\begin{array}{c}
\sign{(f^{NM}(t_{NC,i}^-))}=+1 \\ 
\sign{(f^{NM}(t_{NC,i}^+))}=+1
\end{array}
$ & \hspace{0.6cm} (e) & \hspace{0.6cm} (g) \\ \hline
$
\begin{array}{c}
\sign{(f^{NM}(t_{NC,i}^-))}=+1 \\ 
\sign{(f^{NM}(t_{NC,i}^+))}=-1
\end{array}
$ & \hspace{0.6cm} (f) &  \hspace{0.6cm}  (h) \\ \hline
$
\begin{array}{c}
\sign{(f^{NM}(t_{NC,i}^-))}=-1 \\ 
\sign{(f^{NM}(t_{NC,i}^+))}=+1
\end{array}
$  & \hspace{0.6cm} (e)& \hspace{0.6cm} (g)\\ \hline
$
\begin{array}{c}
\sign{(f^{NM}(t_{NC,i}^-))}=-1 \\ 
\sign{(f^{NM}(t_{NC,i}^+))}=-1
\end{array}
$   &\hspace{0.6cm} (f) & \hspace{0.6cm}  (h)
\end{tabular}
\caption{Discontinuities of $f^M_a(t)$ depending of $\sigma_{a,i}$, $\sigma_{a,i+1}$, $\sign{(f^{NM}(t_{NC,i}^-))}$ and $\sign{(f^{NM}(t_{NC,i}^+))}$ in the case that $t_{NC,i}$ is a non-Markovian discontinuity for $f^{NM}(t)$. The remaining combinations are obtained by flipping all the signs of this table, where the optimal strategies are the same.}\label{fMaNCNM}
\end{table}

In order to illustrate the discontinuities that $f^M_a(t)$ has to show for any combination of $\sigma_{a,i}$, $\sigma_{a,i+1}$, $\sign{(f^{NM}(t_{NC,i}^-))}$ and $\sign{(f^{NM}(t_{NC,i}^+))}$, we follow the scheme of Table \ref{fMaNCNM}.

\begin{itemize}

\item[(e)] Similarly to case (c), we introduce the parameter $\Xi_i=\xi(f^{M}_a(t_{NC,i}))\in[0,1)$.

\item[(f)] Calculations show that the optimal $f^{M}_a(t)$ is obtained when $f^M_a(t)$ is continuous at time $t=t_{NC,i}$, namely by imposing $\xi(f^{M}_a(t_{NC,i}))=1$.

\item[(g)] Calculations show that the optimal $f^{M}_a(t)$ is obtained when $\xi(f^{M}_a(t_{NC,i}))=-1/(d^2-1)$.

\item[(h)] Similarly to case (d), we introduce the parameter $\Xi_i=\xi(f^{M}_a(t_{NC,i}))\in(-1/(d^2-1),0]$.
\end{itemize}
Therefore, these conditions fix  the behavior of $f^M_a(t)$ when  $f^{NM}(t)$ shows a non-Markovian discontinuity.

\paragraph{Evaluation of $p_{a,a}$:--}\label{abra}
 We show the procedure to define the optimal $f^M_a(t)$ until $t=t_{NC,2}$.  
 \begin{itemize}
\item First interval of continuity $[0,t_{NC,1})$: we start by imposing the condition of physicality $f^M_a(0)=1$. We have $\sign(f^{M}_a(t))=\sign(f^{(p)}(t))=+1$. The evolution of $f^M_a(t)$ for $t\in T_{C,1}=(0,t_{NC,1})$ is given in Table \ref{fMaC}. 
\item First time of discontinuity $t_{NC,1}$: the behavior of $f^M_a(t)$ for $t=t_{NC,1}$ is given by Table \ref{fMaNCM} if $t_{NC,1}$ is a Markovian discontinuity for $f^{NM}(t)$ and by Table \ref{fMaNCNM} if $t_{NC,1}$ is a non-Markovian discontinuity for $f^{NM}(t)$.
\item Second interval of continuity $T_{C,2}=(t_{NC,1},t_{NC,2})$: we have $\sign(f^{M}_a(t))=\sign(f^{(p)}(t))=\sigma_{a,2}$. The evolution of $f^M_a(t)$ is given in Table \ref{fMaC}. 
\end{itemize}
The definition of this characteristic function for any $t\geq t_{NC,2}$ is now obvious.

We saw that in order to define $f^M_a(t)$ for $t\in (t_{NC,i-1},t_{NC,i})$ it may be necessary to introduce a parameter $\Delta>0$ that allows to make $\dot{f}^{(p)}(t)=0$ when the cross-diagonal conditions of Table \ref{fMaC} occur (see Eq.~(\ref{afterDelta})). 
Moreover, for each time of discontinuity $t_{NC,i}\in W_{NC}$ we have to define $\xi(f^{M}_a(t_{NC,i}))$. 
For each discontinuity of type (a) or (f), we impose $\xi(f^{M}_a(t_{NC,i}))=1$. For each discontinuity of type (b) or (g), we impose $\xi(f^{M}_a(t_{NC,i}))=-1/(d^2-1)$. For each discontinuity of type (e) or (c), we introduce a parameter $\Xi_i\in[0,1)$. For each discontinuity of type (d) or (h), we introduce a parameter $\Xi_i=\xi(f^{M}_a(t_{NC,i}))\in (-1/(d^2-1),0]$. Therefore, in general, we introduce a set of parameters that defines $f^M_a(t)$:
\begin{equation}\label{afterdisc}
f^M_a(t)=f^M_a(t , \Delta,\{\Xi_i\}_i) \, .
\end{equation}

We seek a combination of $\Delta$ and $\{\Xi_i\}_i$  that {minimizes} the value of $p$ for which $f^{(p)}(t) \in \mathfrak{F}^M_a$.  Eq.~(\ref{pab}) becomes
\begin{equation}
p_{a,a}= {\min_{\Delta,\{\Xi_i\}_i}} \, \{ p \, | \, f^M_a(t,\Delta,\{\Xi_a\}_i) \mbox{ and }  f^{(p)}(t)\in\mathfrak{F}^M \} \, .
\end{equation}
Therefore, we obtained a drastic simplification of the {minimization} required in Eq.~(\ref{pab}). Indeed, to calculate $p_{a,b}$, we formally need to perform a {minimization} over the elements of $\mathfrak{F}_a^M$, which have infinite degrees of freedom. Instead, thanks to this procedure, we only need to perform a {minimization} over $\Delta$ and $\{\Xi_i\}_i$.  Notice that, if the discontinuities of type (c), (d), (e) and (h) are finite, the total number of parameters over which we need to optimize $p_{a,a}$ is finite.

\section{Dephasing evolutions}\label{DEPHASING}
{
In this section we show that the convex class of dephasing evolutions for qubits $\mathcal{Z}$ requires a method to evaluate the corresponding measure of non-Markovianity $p(Z^{NM}|\mathcal{Z}^M)$ which is very similar to the depolarizing case. A dephsing evolution $Z=\{Z_t\}_t \in \mathcal{Z}$ corresponds to a family of dynamical maps $Z_t$ that at any time $t\geq 0$ assumes the form
\begin{equation}\label{Dep}
Z_t (\cdot) = \phi(t) \, \mbox{id} (\cdot) + (1-\phi(t)) \sigma_z  \cdot \sigma_z \, ,
\end{equation}
with $\sigma_z = \mbox{diag}(1,-1)$ being the diagonal $z$-Pauli matrix . We have that
$\phi(t)\in [0,1]$ is a necessary and sufficient condition to ensure $Z_t$ to be CPTP. We rewrite Eq.~(\ref{Dep}) making use of $\varphi(t)\equiv 2\phi(t)-1$, namely considering
\begin{equation}\label{Dep2}
Z_t (\cdot) = \frac{1+\varphi(t)}{2} \, \mbox{id} (\cdot) + \frac{1-\varphi(t)}{2}  \sigma_z  \cdot \sigma_z \, ,
\end{equation}
where  $\varphi(t)$ belonging to
\begin{equation}\label{Ideph}
I_{\mathcal{Z}}\equiv [-1,1] \, ,
\end{equation}
is the necessary and sufficient condition to ensure $Z_t$ to be CPTP. 

In order to characterize Markovian dephasing evolutions, similarly to the case of depolarizing channels, if $\varphi(s)=0$ for some $s> 0$, then the intermediate map $Z_{t,s}$ from $s$ to $t\geq s$ of a dephasing channel can be CPTP if and only if $\varphi(t)=0$ for any $t\geq s$, i.e., $Z_{t,s}(\cdot)=\mbox{id}(\cdot) $ for any $t\geq s$. In the case of a non-zero value of $\phi(s)$, the parametrization given in Eq.~(\ref{Dep2}) allows us to write the intermediate map $Z_{t,s}$ for $t\geq s$  in the following convenient form
\begin{equation}\label{Dep2}
Z_{t,s} (\cdot) = \frac{1+\varphi(t)/\varphi(s)}{2} \, \mbox{id} (\cdot) + \frac{1-\varphi(t)/\varphi(s)}{2}  \sigma_z  \cdot \sigma_z \, ,
\end{equation}
which is a dephasing channel characterized by the value of $\varphi(t)/\varphi(s)$. As a consequence, $Z_{t,s}$ is CPTP if and only if $\varphi(t)/\varphi(s) \in I_{\mathcal{Z}}$.

From Eq.~(\ref{Dep2}) it is clear that we can use $\varphi(t)$ to uniquely characterize $Z$. We define the set of dephasing characteristic functions $\mathfrak{S}$ by requiring the same conditions of regularity considered in Sec.~\ref{secdep} for depolarizing evolutions. As a result, we have a one-to-one correspondence between  dephasing evolutions $Z \in \mathcal{Z}$ and ``regular'' (in general non-continuous) characteristic functions that take values in $I_\mathcal{Z}$, i.e.,  $\varphi(t) \in \mathfrak{S}$. 

In analogy to Eq.~(\ref{xi}), the non-continuous behavior of $\varphi(t) $ can be studied by considering the quantity
\begin{equation}\label{dephss}
\xi(\varphi(t))=\frac{\varphi(t^+)}{\varphi(t^-)} \, .
\end{equation}
Similarly to the depolarizing case, we have a Markovian discontinuity when $\xi(\varphi(t))\in I_\mathcal{Z}\setminus 1$, a non-Markovian discontinuity when $\xi(\varphi(t))\notin I_\mathcal{Z}$ and a time of continuity when $\xi(\varphi(t))=1$.

The similarities between the CPTP conditions for dephasing and depolarizing channels and the role of the corresponding characteristic functions allows to conclude that a dephasing evolution $Z$ with characteristic function $\varphi(t)$  exhibits a {\it Markovian behaviour} at time $\tau
\geq 0$ if one of the two conditions applies
\begin{eqnarray} 
\begin{array}{lclr} 
{ \mathbf{CM_1}(\tau):  } && { \mbox{$\xi(\varphi(\tau))=1$ and  $\frac{d}{d\tau}|\varphi(\tau)| \leq 0$}; }  \\
\mathbf{CM_2}(\tau):
&& \mbox{$\xi(\varphi(\tau)) \in I_{\mathcal{Z}} \setminus 1$} ; 
\end{array} \label{dephsss}
\end{eqnarray} 
where $\mathbf{CM_1}(\tau)$ has to be  replaced by $\dot{\varphi}(\tau^{\pm}) \varphi(\tau) \leq 0$ when $\dot{\varphi}(\tau)$ is non-continuous, i.e., $\dot \varphi(\tau^-)\neq \dot \varphi (\tau^+)$.
We define the set of Markovian dephasing characteristic functions as
\begin{equation}\label{dephsxs}
\mathfrak{S}^M = \left\{  \varphi(t) \in \mathfrak{S} \, | \, 
\mathbf{CM_1}(\tau)\; \mbox{or}\; \mathbf{CM_2}(\tau) = {\mbox{TRUE}, \forall\tau\geq 0 }
 \right\} \;,
\end{equation}
which  involves only local properties of  $\varphi(t)$. Consequently, we can define $\mathfrak{S}^{NM}\equiv \mathfrak{S} \setminus \mathfrak{S}^M$, $\mathcal{Z}^M$ and $\mathcal{Z}^{NM}$.

We can summarize the behavior of Markovian dephasing functions as follows. $\varphi^M(t)\in \mathfrak{S}^M$, when continuous ($\xi(\varphi(t))=1$), does not increase its distance from zero, i.e., its modulus is non-increasing. Therefore, in the time intervals where it is positive (negative) and it is continuous, it is monotonically non-increasing (non-decreasing). As a consequence, $\varphi^M(t)$ cannot change sign while being continuous, i.e., if $\varphi^M(s)=0$ for some  $s\geq 0$, then $\varphi^M(t)=0$ for any $t \geq s$. Discontinuities of Markovian characteristic functions cannot make $\varphi^M(t)$ increase its modulus. Therefore, $\varphi^M(t)$ can change its sign at a generic time $\tau$ (only) with a discontinuity, where  $|\varphi^M(\tau^+)|\leq |\varphi^M(\tau^-)|$. Non-Markovian characteristic functions $\varphi^{NM}(t)\in \mathfrak{S}^{NM}$, instead, can show any  discontinuity and non-monotonic behavior, with the only constraint of assuming values in $I_{\mathcal{Z}}=[-1,1]$ at any time.

We notice that the characterizations of Markovian dephasing evolutions  and depolarizing evolutions are analogous. Given the similarities between the Markovian conditions (\ref{MarkD}) and (\ref{dephsss}) and the dependence of the intermediate maps (\ref{V}) and (\ref{dephss}) from the respective characteristic functions $f(t)$ and $\varphi(t)$, we obtain a very similar procedure needed to evaluate the measure of non-Markovianity $p(Z^{NM} | \mathcal{Z}^M)$. Indeed, in this case we need to find a $Z^M \in \mathcal{Z}^M$ that allows to make $Z^{(p)}=(1-p) Z^{NM} + p Z^M$ Markovian for the smallest value of $p\in [0,1]$, where the Markovian condition for $Z^{(p)}$ can be studied by imposing $\varphi^{(p)}=(1-p) \varphi^{NM} (t) + p \varphi^M(t)$ to satisfy the Markovian conditions  (\ref{dephsss}). The main difference between the evaluations of $p(Z^{NM} | \mathcal{Z}^M)$ and $p(D^{NM} | \mathcal{D}^M)$ for generic $Z^{NM}\in \mathcal{Z}^{NM}$ and $D^{NM} \in \mathcal{D}^{NM}$ is given by the fact that $I_{\mathcal{D}}\neq  I_{\mathcal{Z}}$, which in particular implies that Markovian and non-Markovian characteristic functions of dephasing and depolarizing evolutions have different freedoms to assume values and show discontinuities (compare Eqs. (\ref{intervalI}) and (\ref{Ideph}) for the values of physicality of characteristic functions and $\mathbf{CM_2}(\tau)$ of Eqs. (\ref{MarkD}) and (\ref{dephsss}) for the definition of Markovian discontinuities). Nonetheless, the evaluation of $p(Z^{NM} | \mathcal{Z}^M)$ does not require any particular additional technique compared to the depolarizing case. 

Generalizing this approach to convex set of dynamics of similar forms is straightforward. Some examples are (i) $\mathcal{X}$ and $\mathcal{Y}$ obtained by replacing in Eq.~(\ref{Dep}) $\sigma_z$ with the Pauli matrix, respectively, $\sigma_x$ and $\sigma_y$ and, more in general, (ii) $\mathcal{N}$ obtained by replacing in Eq.~(\ref{Dep}) $\sigma_z$ with any $\sigma_n=n_x \sigma_x+n_y \sigma_y + n_z \sigma_z$ where $(n_x,n_y,n_z)$ is a unit real vector. 
}

\section{Conclusions}\label{CONCL}

We introduced a measure of non-Markovianity inspired by the  intuitive concept for which, in order to consider an evolution highly non-Markovian, it has to be difficult to make it Markovian via incoherent mixing with Markovian dynamics. We showed how to evaluate this measure in the case of depolarizing evolutions in arbitrary dimensions and we discussed the case of dephasing evolutions for qubits. Analytical results are derived for evolutions that satisfy precise continuity and regularity criteria, while we proposed a numerical approach for generic depolarizing evolutions.  
It would be interesting to generalize this analysis to other (even non-convex) classes of evolutions with particular symmetries, e.g. generalized amplitude damping channels and higher-dimensional pure dephasing evolutions. Moreover, conjecture (\ref{conj}) necessitates a valid proof to be enforced.

\section{Acknowledgments}\label{ACK}

We thank Matteo Rosati for helpful feedbacks.
D.D.S acknowledges support from the  
Spanish MINECO (QIBEQI FIS2016-80773-P and Severo Ochoa SEV-2015-0522), the
Fundaci\'o Privada Cellex, the Generalitat de Catalunya (CERCA Program and SGR1381), the ICFOstepstone programme, funded by the Marie Sk\l odowska-Curie COFUND action (GA665884) and  the European Union's Horizon 2020 research and innovation programme under the
Marie Sk\l odowska-Curie grant agreement No 665884. 
VG acknowledges support by MIUR via PRIN 2017 (Progetto di Ricerca di Interesse Nazionale): project QUSHIP (2017SRNBRK).

\appendix

\section{Derivation of Eq.~(\ref{COND4441})}  \label{DERIAPP} 
To cast inequality (\ref{COND444}) into the equivalent form (\ref{COND4441}) let us first consider the case where 
$f(s)> 0$. Under this condition  (\ref{COND444}) forces $f(t)$ to belong to the interval $[-\frac{f(s)}{d^2-1}, f(s)]$ which 
is centred on the point 
\begin{eqnarray}f_{M} \equiv \frac{1}{2} \left(f(s) -\frac{f(s)}{d^2-1}\right)= \frac{d^2 -2}{2(d^2-1)} f(s) \;, \label{MID0} 
\end{eqnarray} 
and has width 
\begin{eqnarray}W \equiv  f(s) +\frac{f(s)}{d^2-1}= \frac{d^2}{d^2-1} f(s) \;.
\end{eqnarray} 
Accordingly imposing $f(t)\in[-\frac{f(s)}{d^2-1}, f(s)]$ is equivalent to require
\begin{eqnarray} \label{MID} 
|f(t) - f_M|   \leq W/2\;,
\end{eqnarray} 
that is \begin{eqnarray} 
|2 ({d^2 -1}) f(t) -  ({d^2 -2}) f(s)|   \leq  {d^2} f(s)\;,
\end{eqnarray} 
which corresponds to (\ref{COND4441}). Similarly 
if $f(s)\leq 0$, Eq.~(\ref{COND444}) forces $f(t)$ to belong to the interval $[f(s),-\frac{f(s)}{d^2-1}]$
which can still be expressed as in (\ref{MID}) by observing that $f_M$ is still as in~(\ref{MID0})
while $W$ becomes 
\begin{eqnarray}W \equiv -\frac{f(s)}{d^2-1}-f(s)= - \frac{d^2}{d^2-1} f(s) \;.
\end{eqnarray} 
In this case hence 
we get 
\begin{eqnarray} 
|2 ({d^2 -1}) f(t) -  ({d^2 -2}) f(s)|   \leq - {d^2} f(s)\;,
\end{eqnarray} 
which corresponds to (\ref{COND4441}) for nonpositive values of $f(s)$.

\section{
Non convexity of the Markovian and non-Markovian subsets of  depolarizing evolutions} \label{NONCONV} 
From the results of  Ref.~\cite{assessing} it follows that neither the Markovian subset $\mathcal{E}^{M}$ nor
its complement $\mathcal{E}^{NM}$ are convex (or equivalently that  $\mathcal{E}^{M}$ is neither 
convex nor concave). In Subsec.~\ref{NCNM} and \ref{example1}
 we show that the same property holds also for the Markovian and non-Markovian parts of the depolarizing 
trajectories  $\mathcal{D}$. 
\subsection{Non-convexity of $\mathcal{D}^{NM}$}\label{NCNM}

Consider 
 the pair of non-Markovian depolarizing evolutions $D^{NM,1}$ and $D^{NM,2}$ with characteristic functions 
\begin{eqnarray}
f^{NM,1}(t)&\equiv \theta_H(1-t)+\theta_H(t-1)\cos^2(t-1) \, , \\ 
f^{NM,2}(t)&\equiv \theta_H(1-t)+\theta_H(t-1)\sin^2(t-1) \, ,
\end{eqnarray}
where $\theta_H(\tau)=1$ for $\tau\geq 0$ and $\theta_H(\tau)=0$ for $\tau<0$. The characteristic functions $f^{NM,1}(t)$ and $f^{NM,2}(t)$ belong to $\mathfrak{F}$
but fail to fulfil the conditions (\ref{MarkD}) for all $t$, hence they are elements of $\mathfrak{F}^{NM}$.
{Interestingly these two evolutions are maximally non-Markovian. Indeed, they show infinitely many non-Markovian gaps $\Delta_k^{NM}=1$ while being positive and continuous. $f^{NM,1}(t)$ is continuous at any time and $f^{NM,2}(t)$, even if it is not continuous at $t=1$, belongs to the family described in Eq.~(\ref{genmulttext}). Hence, since for both of them we have $\Delta^{NM}=\sum_k \Delta_k^{NM}=+\infty$, they assume the maximal value for the measure of non-Markovianity $p(D^{NM,1}|\mathcal{D}^M)=p(D^{NM,2}|\mathcal{D}^M)=1$ (see Eq.~(\ref{manypos})).
Nonetheless,} the convex combination {$f^{(p)}(t)=(1-p)f^{NM,1}(t) + pf^{NM,2}(t)$} is Markovian for $p=1/2$. Indeed, we have 
\begin{equation}
f^{(1/2)}(t)=\theta_H(1-t)+\frac{\theta_H(t-1)}{2}= \left\{
\begin{array}{ccc}
1 & t\in[0,1]\;, \\ \\
\frac{1}{2} & t>1\;,
\end{array} \right. 
\end{equation}
which 
 is an element of $\mathfrak{F}^M$ with  a Markovian discontinuity at $t=1$ (indeed $\xi(f^{(1/2)}(1))=1/2\in I_{\mathcal{D}}$). Accordingly the process $(D^{NM,1}+D^{NM,2})/2$ is an element of $\mathcal{D}^{M}$
 proving that $\mathcal{D}^{NM}$ is not closed under convex combination.

\subsection{Non-convexity of $\mathcal{D}^{M}$}\label{example1}

Focusing on the qubit case $d=2$, we show an example where any non-trivial convex combination of two Markovian depolarizing evolutions provide a non-Markovian depolarizing evolution (generalization for $d>2$ being trivial). Therefore, this proves that the Markovian set of depolarizing channels is non-convex and that the two Markovian evolutions used in this example belong to  the border of the Markovian set $\mathcal{E}^M$. 

\begin{figure}
\includegraphics[width=0.5\textwidth]{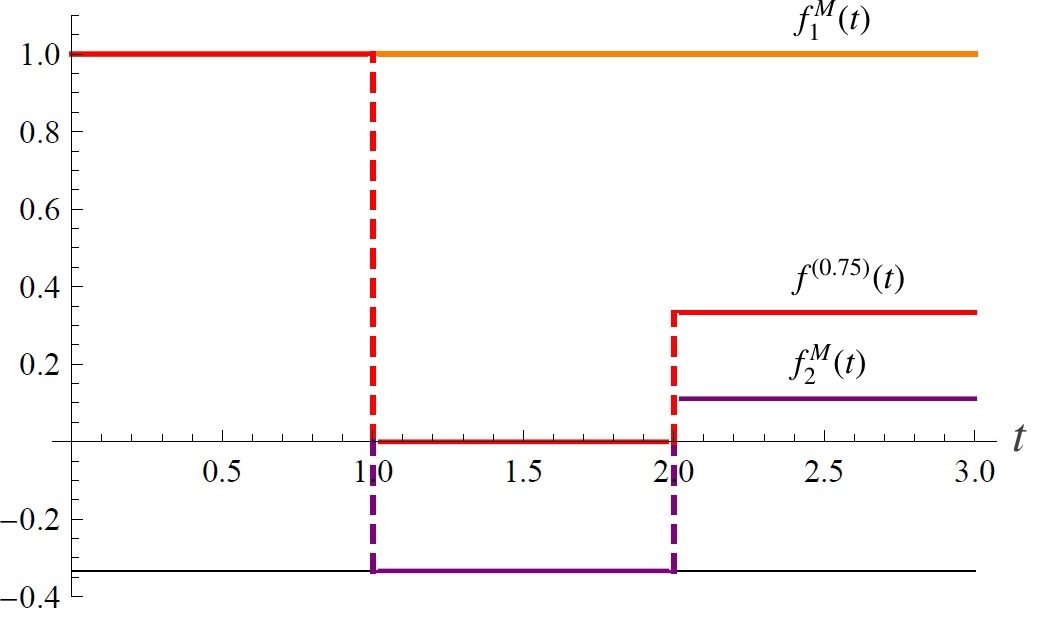}
\caption{Plots of $f^M_1(t)=1$ (orange), $f^M_2(t)$ (purple) and {$f^{(p)}(t)=p f^M_1(t) + (1-p) f^M_2(t)$} for $p=0.75$ (red){, where $t_{NC,1}=1$ and $t_{NC,2}=2$}. While $f^M_{1,2}(t)$ satisfy the conditions (\ref{MarkD}) at any time, this is not the case for $f^{(0.75)}(t)$: it  shows a non-Markovian discontinuity at $t_{NC,2}=2$   when  $\mathbf{CM_2}(2)$ is violated. Indeed, $f^{(0.75)}(2^-)=0$, $f^{(0.75)}(2^+)>0$ and $\xi(f^{(0.75)}(2))=+\infty$.}\label{nonconv}
\end{figure}

Consider two Markovian qubit evolutions $D^{M,1}$ and $D^{M,2}$ defined by the characteristic functions $f^{M,1}(t)$ and $f^{M,2}(t)$, respectively. First, we define $f^{M,1}(t)=1$ for any $t$, {noticing}   that  $D^{M,1}_t (\cdot) = \mbox{id} (\cdot)$ is the identical map for any $t\geq 0$. Secondly we take
\begin{equation}\label{ofMnc}
 f^{M,2}(t) \equiv \left\{
\begin{array}{cc}
 1 & t \leq t_{NC,1} \\ 
  -1/3 & t\in (t_{NC,1},t_{NC,2}] \\ 
1/9 & t > t_{NC,2}\;, 
\end{array}
\right. 
\end{equation}
which exhibits Markovian discontinuities 
\begin{equation}\label{genthis}
\xi(f^{M,2}(t_{NC,1}))=\xi(f^{M,2}(t_{NC,2}))=-\frac{1}{d^2-1}=-\frac{1}{3} \, .
\end{equation}
 The convex combination {$D^{(p)}= (1-p) D^{M,1} + p D^{M,2} $} is characterized by {$f^{(p)}(t) =(1-p)f^{M,1}(t) + p f^{M,2}(t)$}. While the discontinuity that $f^{(p)}(t)$ shows at $t=t_{NC,1}$ is always Markovian, at $t=t_{NC,2}$ we have
\begin{equation}\label{Mjump2}
\xi(f^{(p)}(t_{NC,2})) = \frac{9-8p}{9-12p} \notin I_{\mathcal{D}} \, , \,\, \forall p\in(0,1) \, .
\end{equation}
Indeed, $\xi(f^{(p)}(t_{NC,2}))>1$  for any $p\in(0,3/4)$, $\xi(f^{(p)}(t_{NC,2}))<-1/3$  for any $p\in(3/4,1)$ and it diverges for $p=3/4$, i.e., $\lim_{p\rightarrow 3/4^\mp} \xi(f^{(p)}(t_{NC,2})) =\pm \infty$ (see Fig. \ref{nonconv}).
Therefore, \textit{any} depolarizing evolution $D^{(p)}$ obtained by the non-trivial convex combination of the Markovian depolarizing evolutions $D^{M,1}$ and $D^{M,2}$ is non-Markovian. 

\section{Markovian and non-Markovian positive depolarizing evolutions}\label{deppos}

{ We define $\mathcal{D}_+\subset \mathcal{D}$ to be the class of positive depolarizing evolutions which is defined by non-negative characteristic functions, namely the set $\mathfrak{F}_+ \subset \mathfrak{F}$ made by the elements of $\mathfrak{F}$ that are non-negative for any $t\geq 0$. Given the defining feature of the elements of the $\mathfrak{F}_+$, it is clear that the positive depolarizing evolutions form a convex set. We define $\mathcal{D}_+^M$ to be the Markovian subset of $\mathcal{D}_+$ which is in one-to-one correspondence with the set of characteristic functions $\mathfrak{F}^M_+\subset \mathfrak{F}^M$. Similary, we define  $\mathcal{D}_+^{NM}$ and $\mathfrak{F}_+^{NM}$. 

The characteristic functions of $\mathfrak{F}_+$ are in general non-continuous. Indeed, we require that $\xi(f^M(t))\in[0,1]$ for any $f^M_+(t)\in \mathfrak{F}^M_+$ and $t\geq 0$. Analogously,  $\xi(f^{NM}(t))\in[0,+\infty]$ for any  $f^{NM}_+(t)\in \mathfrak{F}_+^{NM}$ and $t\geq 0$. A negative value of $\xi(f(t))$ implies that $f(t)$ changes sign at time $t$ and this circumstance cannot occur for positive $f(t)$. A straightforward calculation shows that $f^{(p)}(t)=(1-p)f^{M,1}_+ (t) + p f^{M,2 }_+ (t) $ cannot show non-Markovian discontinuities and more in general cannot be non-Markovian. Hence,
 \begin{itemize}
\item $\mathcal{D}_+$ is convex,
\item $\mathcal{D}^M_+$ is closed and convex,
\item $\mathcal{D}^{NM}_+$ is open and non-convex. 
\end{itemize}
 
We remember that the set of continuous depolarizing evolutions $\mathcal{D}_C$ has a convex Markovian subset that we called $\mathcal{D}_C^M$. The set of characteristic functions that corresponds to $\mathcal{D}_C^M$ is $\mathfrak{F}^M_C$, which is the collection of non-increasing continuous non-negative functions $f^M_C(t)$. Therefore, we can conclude that
\begin{equation}
\mathcal{D}^M_C \subset \mathcal{D}^M_+  \; ,
\end{equation}
where $\mathcal{D}^M_+  \setminus  \mathcal{D}^M_C $ is given by those evolutions of $\mathcal{D}^M_+$ that show at least one (Markovian) discontinuity. Moreover, since no $f^{NM}_+(t)\in \mathfrak{F}^{NM}_+$ can assume negative values, it is easy to see that $\mathcal{D}^{NM}_C \nsubset \mathcal{D}^{NM}_+$, $\mathcal{D}^{NM}_+ \nsubset \mathcal{D}^{NM}_C $ and $\mathcal{D}^{NM}_+ \cap \mathcal{D}^{NM}_C\neq \emptyset$, namely the intersection is not empty. 
}

\subsection{Positive non-Markovian characteristic functions}\label{posdepB}
{
We discuss the value of $p(D^{NM}_+ | \mathcal{D}^M_+)$ when $D^{NM}_+\in \mathcal{D}^M_+$, namely a non-Markovian depolarizing evolution with a positive characteristic function $f^{NM}_+(t) \in \mathfrak{F}^{NM}_+$. Therefore we have to consider the convex combination $f^{(p)}(t)=(1-p) f^{NM}_+(t) + p f^{M}_+ (t)$ and evaluate the smallest $p$ for which there exists a $f^M_+(t) \in \mathfrak{F}^M_+$ that makes $f^{(p)}(t)$ Markovian, more precisely an element of $\mathfrak{F}^M_+$.

Similarly to the previous sections, we define $T^{NM}_+$ and $\Delta^{NM}$ exactly as in the continuous case, i.e., the collection of the time intervals $T_k^+=(t_k^{(in)},t_k^{(fin)})$ where a non-Markovian gap $\Delta_k^{NM}>0$ is shown while the non-negative $f^{NM}_+(t)$ is continuous. Analogously to Sec.~\ref{npos}, we introduce
\begin{eqnarray}
\label{Deltas2lim}
 \Delta_k^{M} \in[-1,0] \, \;, \qquad 
\Delta^{M}\equiv  \sum_{k=1}^L \Delta_k^{M} \in[-1,0] \; ,
\end{eqnarray}
where $\Delta_k^M=f^M_+(t_k^{(fin)}) - f^M_+(t_k^{(in)}) \leq 0$ is the gap that $f^M_+(t)$ describes when $f^{NM}_+(t)$ is increasing.

Moreover, we introduce $W_+^{NM} \equiv \{\tau_i \}_i$
as the discrete set of times when $f^{NM}_+(t)$ shows a non-Markovian discontinuity, namely such that $\xi(f^{NM}_+(\tau_i))\in (1,\infty]$ (remember that $\xi(f^{NM}_+(t))$ and $f^{NM}_+(t)$ itself cannot be negative). Analogously to $\Delta_k^{NM}$, we introduce the quantities 
\begin{eqnarray}
&\pi^{NM}_i& \equiv f_+^{NM}(\tau_i^+) - f_+^{NM}(\tau_i^-)>0 \, , \label{gnm} \\ 
&\pi^{M}_i& \equiv f_+^{M}(\tau_i^+) - f_+^{M}(\tau_i^-)< 0 \, , \label{gm} 
\end{eqnarray}
respectively the non-Markovian gaps shown by $f^{NM}_+(t)$ and the Markovian gaps shown by $f^M_+(t)$ at the times when $f^{NM}_+(t)$ shows non-Markovian discontinuities. Moreover,
\begin{eqnarray}
&\pi^{NM}&=\sum_i \pi_i^{NM} >0 \, , \label{nmjumps} \\ 
&\pi^{M}&=\sum_i \pi_i^{M} \in[-1,0] \, , \label{mjumps} 
\end{eqnarray}
are respectively the sum of all the non-Markovian jumps shown by $f^{NM}_+(t)$ and all the Markovian jumps shown by $f^M_+(t)$. Notice that, since $f^M_+(t)$ is non-increasing, $\Delta^M + \pi^M \in [-1,0]$ \cite{foot4}.  Indeed, it is easy to show that, in order to calculate  $p(D^{NM}_+ | \mathcal{D}^M_+)$, by considering  $f^M_+(t)$ with (Markovian) discontinuities for some $t\notin W_+^{NM}$ we do not obtain an advantage. More precisely, we have to consider $f^M_+(t)$ that show Markovian discontinuities if and only if $t \in W_+^{NM}$.

A necessary condition to make $f^{(p)}(t)$ Markovian is that $p\geq p_+\equiv (\Delta^{NM} + \pi^{NM})/(1 + \Delta^{NM} + \pi^{NM})$. This relation is obtained as Eq.(\ref{abrakadabra}), where we also require that $\pi^{(p)}_i=(1-p) \pi^{NM}_i + p \pi^M_i \leq 0$, namely that the discontinuities of $f^{(p)}(t)$ are Markovian.

In order to evaluate the measure of non-Markovianity of $f^{NM}_+(t)$, we adapt the tools introduced in Sec.~\ref{npos} (where we studied non-negative continuous non-Markovian characteristic functions) to implement the cases where $f^{NM}_+(t)$ shows non-Markovian discontinuities.  We define $g_+^{M}(t) $ as the following function
\begin{equation}\label{gplus}
 \left\{
\begin{array}{cc}
1 & \,\,\, t\leq t_1 \, , \\ 
\cdots \\ 
g_+^{M} (t_{k-1}^{(fin)}) -\left( f^{NM}(t) - f^{NM}(t_{k}^{(in)}) \right)/(\Delta^{NM} + \pi^{NM}) & t\in T_k^+ \, , \\ 
\cdots  \\ 
g_+^M(\tau_i^-) - \pi^{NM}_i / (\Delta^{NM} + \pi^{NM}) & t =  \tau_i \, .
\end{array}
\right. 
\end{equation}
where $t_1\equiv\min\{\tau_1,t_1^{(in)}\}$.
It is easy to see that Eq.~(\ref{gplus}) is obtained from Eq.~(\ref{ofMgen}) by replacing $\Delta^{NM}$ with $\Delta^{NM} + \pi^{NM}$ and by implementing the Markovian gaps $\pi^M_i \equiv -\pi^{NM}_i / (\Delta^{NM} + \pi^{NM})$ that $g^M_+(t)$ shows when $f^{NM}_+(t)$ shows a non-Markovian discontinuity. Moreover, we notice that by considering $g_+^M(t)$ we have $\Delta^{M}+ \pi^M=-1$. The function $f^{(p)}(t)=(1-p)f^{NM}_+(t) + p g_+^M(t)$ belongs to $\mathfrak{F}^M_+$ for any $p\geq p_+$, where $f^{(p_+)}(t)$ is constant for any $t\in T^{NM}$ and continuous for any $t\in W_+^{NM}$. Finally, we can state that 
\begin{equation}\label{pplus}
p(D^{NM}_+ | \mathcal{D}^M_+) = \frac{\Delta^{NM} + \pi^{NM}}{1 + \Delta^{NM} + \pi^{NM}} \, ,
\end{equation}
and therefore this measure of non-Markovianity depends on the non-Markovian gaps shown by the non-Markovian characteristic function (in this case $\Delta^{NM} + \pi^{NM}$) as in the continuous case.

We notice that, while for continuous depolarizing evolutions we have $p(D_C^{ NM}| \mathcal{D}^M)=p(D_C^{ NM}| \mathcal{D}_C^M)$ (see Sec.~\ref{contnoncont}), in the case of positive depolarizing evolutions we have $p(D_+^{ NM}| \mathcal{D}^M)\leq p(D_+^{ NM}| \mathcal{D}_+^M)$. We present a simple example that shows this feature. Consider $f^{NM}_+(t)$ with (i) a single non-Markovian discontinuity $W_+^{NM}=\{\tau\}$, (ii) no non-Markovian intervals of non-Markovianity $T_k^{NM}$ and (iii) such that $\dot{f}^{NM}_+(t)=0$ for any $t\geq \tau$. In this case we have that $g_+^M(t)=1$ for any $t\leq \tau$ and $g_+^M(t)=0$ for any $t>\tau$. Therefore, we obtain $p(D_+^{ NM}| \mathcal{D}_+^M)=\pi^{NM}/(1+\pi^{NM})$, where $\pi^{NM}=f^{NM}_+(\tau^+) - f^{NM}_+(\tau^-)$. By considering the non-positive Markovian characteristic function $g^M(t)=1$ for $t\leq \tau$ and $g^M(t)=-1/(d^2-1)$  for $t\geq \tau$ we have \cite{foot5}
$$p(D_+^{ NM}| \mathcal{D}^M)=\frac{\pi^{NM}}{1+\frac{1}{d^2-1}+\pi^{NM}} < p(D_+^{ NM}| \mathcal{D}^M_+) \, .$$}

\section{Uniqueness of the optimal continuous Markovian characteristic function} \label{UNIQUE} 

We consider the evaluation of $p(D^{NM}_C|\mathcal{D}_C^M)$ when $D_C^{NM}\in \mathcal{D}_C^{NM}$ and $f^{NM}_C(t)$ is the corresponding continuous characteristic function.
For the purpose of evaluating this quantity, in Sec.~\ref{nposnegtech} we saw that its value is given by {$\Gamma^{NM}/(1+\Gamma^{NM})$} and a continuous characteristic function that makes the corresponding $f^{(p)}(t)$ Markovian for $p=p(D^{NM}_C|\mathcal{D}_C^M)$ is $h_C^M(t)$ (see Eq.~(\ref{hCMapp})). In this section we show that $h_C^M(t)$ is the \textit{only} continuous Markovian characteristic function that makes $f^{(p)}(t)$ Markovian for any {$p\geq p(D^{NM}_C|\mathcal{D}_C^M)$.}

Any continuous Markovian characteristic function $f_C^M(t)$ assumes values in $[0,1]$ and is non-increasing. Let start noticing that, if $f_C^M(t)$ decreases while $h_C^M(t)$ is constant, {given what we discussed in Sec.~\ref{measureofnmc}  we conclude that} the former has no chance to perform better than the latter. Therefore, consider a time interval $(t_1,t_2)$ of non-Markovianity where $\dot{h}_C^M(t)<0$ and $\dot{f}^{NM}(t)>0$. If for some $t\in (t_1,t_2)$ we have $\dot{h}_C^M(t)< \dot{f}_C^M(t)\leq 0$, the $f^{(p)}(t)$ obtained with $f^M_C(t)$ has a time derivative that can be made non-positive for {larger} values of $p$ if compared with the $f^{(p)}(t)$ obtained with $h_C^M(t)$. Therefore, in this situation $f^M_C(t)$ is less efficient than $h_C^M(t)$ to make $f^{(p)}(t)$ Markovian.

Consider a time interval of non-Markovianity $(t_k^{(in)},t_k^{(fin)})$ where $\dot{f}_C^M(t)<\dot{h}_C^M(t)<0$ for some  $t\in (t_k^{(in)},t_k^{(fin)})$ and $\dot{f}_C^M(t)\leq \dot{h}_C^M(t)<0$ for every $t\in (t_k^{(in)},t_k^{(fin)})$. Assume that $f^{NM}(t)\geq 0$ and therefore $\lim_{t\rightarrow\infty} h_C^M(t)=0$. Using the notation introduced in Eqs. (\ref{Deltask1}), (\ref{Deltask2}) and (\ref{Deltask3}), we see that by using $h_C^M(t)$ all the $\Delta^{(p)}_k$ are non-positive for {$p\geq p(D^{NM}_C|\mathcal{D}_C^M)$}, while they are all positive for {$p<p(D^{NM}_C|\mathcal{D}_C^M)$}. In the case of the $f^M_C(t)$ described above, we may have that {some}$\Delta_k^{(p)}(t)$ can be made non-positive for some {$p < p(D^{NM}_C|\mathcal{D}_C^M)$}. Since $\sum_k \Delta_k^M\in [-1,0]$ and by considering that with $h_C^M(t)$ we have $\sum_k \Delta_k^M=-1$, there must be a $k'\neq k$ such that the value of $|\Delta^M_{k'}|$ obtained with $f^{M}_C(t)$ is smaller than the one obtained with $h_C^M(t)$. Hence, while $h_C^M(t)$ can make $f^{(p)}(t)$ Markovian for $p=p(D^{NM}_C| \mathcal{D}_C)$, $f^M_C(t)$ cannot do the same. A similar argument can be used for $f^{NM}(t)$ that assume positive and negative values.

Finally, since any $f^M_C(t)$ that make $f^{(p)}(t)$ Markovian for $p=p(D^{NM}|\mathcal{D}_C^M)$ cannot have a time derivative different from $\dot{h}_C^M(t)$,  $h_C^M(t)$ is the only continuous Markovian characteristic function that is optimal to make $f^{(p)}(t)$ Markovian, i.e., $f^{(p)}(t)$ can be made Markovian for $p=p(D^{NM}_C| \mathcal{D}_C)$ with a continuous characteristic function if and only if we consider $h_C^M(t)$. {In particular, given the results of Sec.~\ref{contnoncont}, we can state that the optimal Markovian evolution needed to evaluate $p(D^{NM}_C|\mathcal{D}^M)$ is unique and defined by $h^{M}_C(t)$.}

\section{Different vectors of signs for $f^M_a(t)$ and $f^{(p)}(t)$}\label{aneqb}

We consider $f^{(p)}(t)\in \mathfrak{F}_b^M$ for some $\boldsymbol{\sigma}_b$, where $f^M_a(t)\in \mathfrak{F}^M_a$ for some $\boldsymbol{\sigma}_a\neq \boldsymbol{\sigma}_b$. In the following, $b$ is always the index attached to $f^{(p)}(t)$ and $a$ is always the index attached to $f^M_a(t)$.

First, we make the following consideration. In the case that $f^{NM}(t)$ violates the conditions of Markovianity given in Eq.~(\ref{MarkD}) in a time interval in $ T_{C,i}$, then we must consider $\sigma_{a,i}=\sigma_{b,i}$. Indeed, if $\sign(f^M(t))=-\sign(f^{NM}(t))$ and $f^{NM}(t)$ shows a non-Markovian behavior while being continuous at time $t$, $f^{(p)}(t)$ can be made Markovian at time $t$ if and only if $\sign(f^{(p)}(t))=\sign(f^{M}(t))$. Therefore:
\begin{itemize}
\item[(A)] If $T_{C,i}$ is a time interval when $f^{NM}(t)$ behaves as a non-Markovian characteristic function, then ${\sigma}_{a,i}={\sigma}_{b,i}$.
\end{itemize}
Therefore, if $\boldsymbol{\sigma}_a$ and $\boldsymbol{\sigma}_b$ do not satisfy (A) for at least one time interval $T_{C,i}$ we set {$p_{a,b}=1$} because $f^M_a(t)$ cannot make $f^{(p)}(t)\in \mathfrak{F}_b^M$.

\subsection{Times of continuity}\label{ToCapp}

Let consider $\boldsymbol{\sigma}_a$ and $\boldsymbol{\sigma}_b$ that satisfy (A) for each $T_{C,i}$ and define the optimal $f^M_a(t)$ for a generic time interval $T_{C,i}$ when $\sigma_{a,i}\neq \sigma_{b,i}$. During this time interval  $f^{NM}(t)$  behaves as a continuous Markovian characteristic function and therefore cannot change its sign. As a consequence, we must be in a situation where $\sigma_{a,i}= - \sigma_{b,i}$ while $\sign(f^M_a(t)) = \sigma_{a,i}$ and $\sign(f^{NM} (t)) = \sign(f^{(p)}(t))=\sigma_{b,i}$. Therefore, with opposite signs, $f^M_a(t)$ and $f^{NM}(t)$ are approaching continuously zero and we need to make their convex combination be of the same sign of $f^{NM}(t)$. 

{ We analyze the situation $\sigma_{a,i}=+1$ and $\sigma_{b,i}=-1$, where $\sign(f^M_a(t))=+1$ and $\sign(f^{NM}(t))=\sign (f^{(p)}(t))=-1$ for any $t\in [t_{NC,i-1},t_{NC}]$ (the same results can also be obtained for $\sign(f^M_a(t))=-1$ and $\sign(f^{NM}(t))=\sign (f^{(p)}(t))=+1$). We write $f^M_a(t_{NC,i})=|f^M_a(t_{NC,i})|=|f^M_a(t_{NC,i-1})|-\delta_{M,i}$ and $f^{NM}(t_{NC,i})=-|f^{NM}(t_{NC,i})|=-|f^{NM}(t_{NC,i-1})|+\delta^{NM}_{i}$, where $\delta^M_{i},\delta^{NM}_{i}\geq 0$. Indeed, between $t_{NC,i-1}$ and $t_{NC,i}$, $f^M_a(t)\geq 0$ decreases and $f^{NM}(t)\leq 0$ increases. Therefore, if we consider the value of $f^M_a(t_{NC,i-1})$ fixed by the study of the time interval $[t_{NC,i-2},t_{NC,i-1}]$, we have to study for which values of $p$ the function $f^{(p)}(t)$ in negative and non-decreasing in $[t_{NC,i-1},t_{NC,i}]$, when $\delta_i^M$ varies. These two conditions can be respectively written as:
\begin{equation}\label{condix1}
p\leq \frac{|f^{NM}(t_{NC,i})|}{|f^{NM}(t_{NC,i})|+|f^{NM}(t_{NC,i})|-\delta^M_i}\leq 1 \, ,
\end{equation}
\begin{equation}\label{condix2}
p\leq \frac{\delta^{NM}_i}{\delta^{NM}_i+\delta^M_i}\leq 1 \, ,
\end{equation}
which are upper bounds for $p$. This is the first time that we obtain upper bounds on $p$ rather than lower bounds. The reason of this new situation is given by the fact that we impose $f^{(p)}(t)$ to have the same sign of $f^{NM}(t)$ and the opposite sign of 	$f^M_a(t)$. Hence, this condition cannot be satisfied if $p$ is too large and it is surely verified when $p$ is small enough.
Notice that (\ref{condix1}) provides the largest interval of validity when  $\delta^M_i$ is as large as possible, i.e., $|f^{NM}(t_{NC,i})|-\delta^M_i=0$, while for (\ref{condix2}) we have the opposite situation. Since they are both upper bounds, $\delta^M_i=0$ may seem the best choice. Nonetheless, we have to consider that  (\ref{condix1}) have to be consistent with the lower bounds on $p$ that we obtain when we impose Markovianity for $f^{(p)}(t)$ in the other time intervals and times of discontinuity.
As a consequence, the choice of $\delta^M_i$ is not obvious, and we have to implement a variable $\delta_i^M$ that we fix when we calculate $p_{a,b}$. Therefore, for each time interval where $\sigma_{a,i}\neq \sigma_{b,i}$ we introduce a parameter $\delta^M_i$ and $f^M_a(t)$ has to be parametrized by this set, namely $f^M_a(t)=f^M_a(t,\{\delta^M_i\}_i)$.}

Notice that, for the time intervals when instead we have $\sigma_{a,i}=\sigma_{b,i}$, we use the conditions introduced in Table \ref{fMaC}. Therefore, we defined the behavior of $f^M_a(t)$ for all the times $t\in W_C$.

\subsection{Discontinuities}\label{ToDapp}

Let consider those discontinuities that cannot be described by Tables \ref{fMaNCM} and \ref{fMaNCNM}, namely those times $t_{NC,i}$ such that: 
\begin{equation}\label{BCD}
\left\{
\begin{array}{ccccc}
 \sigma_{a,i-1}=&-\sigma_{b,i-1} & \mbox{ and  } &  \sigma_{a,i}=&\sigma_{b,i} \\
 \sigma_{a,i-1}=&\sigma_{b,i-1}  & \mbox{ and  } & \sigma_{a,i}=&-\sigma_{b,i} \\ 
\sigma_{a,i-1}=&-\sigma_{b,i-1}  & \mbox{ and  } &   \sigma_{a,i}=&-\sigma_{b,i} 
\end{array}
\right. .
\end{equation}
These discontinuities, analogously to (\ref{condix1}) and (\ref{condix2}), often provide {upper}-bounds for $p$. Consider that, if there is just one time interval where the cross-diagonal conditions of Table \ref{fMaC} occur, we obtain a {lower-bound $p\geq p_{up}= \Delta/(1+\Delta)$}. Moreover, {lower}-bound conditions are obtained when we consider Tables \ref{fMaNCM} and \ref{fMaNCNM} (while (\ref{BCD}) does not apply). 
Moreover we notice that we must have at least one {lower}-bound condition, otherwise {$p=0$} would be consistent with $f^{(p)}(t)$ being Markovian, which is a contradiction.
Therefore, we may be interested to {maximize} $p_{lim}$ as a function of $ \xi(f^M_a(t))$ in order to make it compatible with one or more {lower}-bound conditions. In several cases given by Eq.~(\ref{BCD}) this result is obtained for $\xi(f^M_a(t))=0$. Therefore, there is a trade-off between $p_{lim}$ and the ability of $f^M_a(t)$ to make $f^{(p)}(t)$ for later times. We conclude that, for each Markovian and non-Markovian discontinuity of type (\ref{BCD}), we introduce a parameter $\overline{\Xi}_i$ that defines the value of $\xi(f^M_a(t_{NC,i}))$. 

\subsection{Evaluation of $p_{ab}$}

Therefore, having $\boldsymbol{\sigma}_a\neq \boldsymbol{\sigma}_b$ such that condition (A) is satisfied, in general we need to consider an $f^M_a(t)$ that depends on the parameters $\Delta$ (see Section~\ref{ToC}), $\{\Xi_i\}_i$ (see Sections \ref{secMD} and \ref{secNMD}), $\delta^M_i$ (see Section~\ref{ToCapp}) and $\{\overline \Xi_i\}_i$ (see Section~\ref{ToDapp}) and $p_{a,b}$ is obtained by the optimization
\begin{equation}
p_{a,b}=\!\!\!\!\!\!\! \min_{\Delta,\{\delta_i^M\}_i,\{\Xi_i\}_i,\{\overline{\Xi}_i\}_i} \!\!\!\!\! \{ p \, | f^M_a(\Delta,\{\delta_i^M\}_i,\{\Xi_i\}_i,\{\overline{\Xi}_i\}_i) \in \mathfrak{F}_a^M, f^{(p)}(t)\in\mathfrak{F}_b^M \}.
\end{equation}
It is plausible that, even if condition (A) holds, the maximization required for $p_{a,b}$ has no solution. Indeed, the upper-bound and lower-bound conditions discussed above may not be made compatible for any $f^M_a(t)\in \mathfrak{F}_a^M$.


\begin{thebibliography}{xx}
\bibitem{foot1} We introduce this parameter in analogy with Eq.~(\ref{hCM}). If $f^{NM}(t)$ does not show any discontinuity, $\Delta=\Gamma^NM$.
\bibitem{foot2} We remember that $\xi(f^M_a(t_{NC,i}))$ if and only if $f^{M}_a(t_{NC,i}^+)=0$ and $f^{M}_a(t)=0$ for any $t\geq t_{NC,i}$. Therefore, we can pick this value if and only if $f^{NM}(t)$ does not show any non-Markovian behavior for $t\geq t_{NC,i}$.
\bibitem{foot3} For each value of $\Xi_i$ we have a different interval of $p$ such that $f^{(p)}(t)$ is Markovian and with the same sign of $f^M_a(t)$. If $\xi(f^{NM}(t_{NC,i}))>0$, the choice $\Xi_i=\xi(f^{NM}(t_{NC,i}))$ allows the largest value of $p$ for which we can make $f^{(p)}(t)$ Markovian and with the same sign of $f^{M}_a(t)$, but it implies that $f^{(p)}(t_{NC,i}^+)=0$ and it denies any further possibility to make $f^{(p)}(t)$ Markovian for $t> t_{NC,i}^+$. Therefore, chosen a value of $\Xi_i$, we obtain some conditions $p\leq p(\Xi_i)$ for which $f^{(p)}(t)$ is Markovian and with the same sign of $f^{M}_a(t)$.
\bibitem{foot4} This quantity can be equal to -1 if and only if $f^M_+(t)$ is constant for those times when $f^{NM}_+(t)$ behaves as a Markovian characteristic function.
\bibitem{foot5}  We think that it is not necessary to prove that there are no $f^M(t)\in \mathfrak{F}^M$ that are able to make $f^{(p)}(t)$ Markovian for smaller values of $p$.
\end{thebibliography}
\end{document}